\documentclass[aps,prd,reprint,amsmath,amssymb,nofootinbib]{revtex4-1}

\usepackage{graphicx}
\usepackage{xcolor}
\usepackage{amsmath, slashed}
\usepackage{subcaption}
\usepackage{url}
\usepackage[colorlinks,linkcolor=blue,urlcolor=blue,citecolor=blue]{hyperref}

\allowdisplaybreaks

\makeatletter
\renewcommand\@makecaption[2]{%
  \par
  \vskip\abovecaptionskip
  \begingroup
   \small\rmfamily
    \begingroup
     \samepage
     \flushing
     \let\footnote\@footnotemark@gobble
     \@make@capt@title{#1}{#2}\par
    \endgroup
  \endgroup
  \vskip\belowcaptionskip
}
\makeatother

\begin{document}

%%%%%%%%%%%%%%%%%%%%%%%%%%%%%%%%%%%%%%%%%%%%%%%%%%%%%%%%%
\title{\texorpdfstring{Addressing $R_{D^{(*)}}$, $R_{K^{(*)}}$, muon $g-2$ and ANITA anomalies in a minimal $R$-parity violating supersymmetric framework}{Addressing R(D(*)), R(K(*)), muon g-2 and ANITA anomalies in a minimal R-parity violating supersymmetric framework}}
%%%%%%%%%%%%%%%%%%%%%%%%%%%%%%%%%%%%%%%%%%%%%%%%%%%%%%%%%

\author{Wolfgang Altmannshofer$^{1}$, P. S. Bhupal Dev$^{2}$, Amarjit Soni$^3$, Yicong Sui$^2$}

\affiliation{$^1$Department of Physic and Santa  Cruz  Institute  for  Particle  Physics,
University  of  California,  Santa  Cruz,  CA  95064,  USA}
\affiliation{$^2$Department of Physics and McDonnell Center for the Space Sciences,
Washington University, St. Louis, MO 63130, USA}
\affiliation{$^3$Physics Department, Brookhaven National Laboratory, Upton, NY 11973, USA}

%%%%%%%%%%%%%%%%%%%%%%%%%%%%%%%%%%%%%%%%%%%%%%%%%%%%%%%%%
\begin{abstract}
We analyze the recent hints of lepton flavor universality violation in both charged-current and neutral-current rare decays of $B$-mesons in an $R$-parity violating supersymmetric scenario. Motivated by simplicity and minimality, we had earlier postulated the third-generation superpartners to be the 
lightest (calling the scenario ``RPV3") and explicitly showed that it preserves gauge coupling unification and of course has the usual attribute of naturally addressing the Higgs radiative stability. Here we show that both $R_{D^{(*)}}$ and $R_{K^{(*)}}$ flavor anomalies can be addressed in this RPV3 framework. Interestingly, this scenario may also be able to accommodate two other seemingly disparate anomalies, namely, the longstanding discrepancy in the muon $(g-2)$, as well as the recent anomalous upgoing ultra-high energy ANITA events. Based on symmetry arguments, we consider three different benchmark points for the relevant RPV3 couplings and carve out the regions of parameter space where all (or some) of these anomalies can be simultaneously explained. We find it remarkable that such overlap regions exist, given the plethora of precision low-energy and high-energy experimental constraints on the minimal model parameter space. The third-generation superpartners needed in this theoretical construction are all in the 1--10 TeV range, accessible at the LHC and/or next-generation hadron collider. We also discuss some testable predictions for the lepton-flavor-violating decays of the tau-lepton and $B$-mesons for the current and future $B$-physics experiments, such as LHCb and Belle II. Complementary tests of the flavor anomalies in the high-$p_T$ regime in collider experiments such as the LHC are also discussed. 
\end{abstract}
%%%%%%%%%%%%%%%%%%%%%%%%%%%%%%%%%%%%%%%%%%%%%%%%%%%%%%%%%

%\date{\today}
\maketitle
%\tableofcontents

%%%%%%%%%%%%%%%%%%%%%%%%%%%%%%%%%%%%%%%%%%%%%%%%%%%%%%%%%
\section{Introduction}  \label{sec:intro}
%%%%%%%%%%%%%%%%%%%%%%%%%%%%%%%%%%%%%%%%%%%%%%%%%%%%%%%%%
A very likely way new physics beyond the Standard Model (SM) could show up in experiments is through anomalous features in the data that cannot be explained by any known SM physics. While some of these anomalies may well be due to statistical fluctuations and/or systematic or theoretical issues that need further understanding,
it is also possible that one (or more) such deviation(s) from the SM may well be a genuine signal of some new beyond the SM (BSM) physics. Moreover,
given their possible impact and global ramifications, it is worthwhile to carefully scrutinize them in light of possible underlying BSM scenarios.  

Of the existing statistically persistent anomalies, particularly prominent ones are the hints of Lepton Flavor Universality Violation (LFUV) in both charged current tree-level  and neutral current one-loop rare decays of $B$-mesons,  based on the $b \to c \ell^- \bar{\nu}$ (with $\ell=e,\mu,\tau$) and $b \to s \ell^+ \ell^-$ (with $\ell=e,\mu$) transitions, respectively. In particular, hints for LFUV are seen in the following ratios of branching ratios (BRs): 
\begin{eqnarray}
    R_{D^{(*)}} & \ = \ & \frac{\text{BR}(B \to D^{(*)} \tau \nu)}{\text{BR}(B \to D^{(*)} \ell \nu)}~ \, \, ({\rm with}~\ell=e,\mu)\, ,
    \label{eq:RD} \\
    R_{K^{(*)}} & \ = \ & \frac{\text{BR}(B \to K^{(*)} \mu^+\mu^-)}{\text{BR}(B \to K^{(*)} e^+e^-)} ~, \label{eq:RK}
\end{eqnarray}
where $D^{*}$ and $K^{*}$ denote excited states of $D$ and $K$ mesons, respectively. 
These ratios of BRs are interesting observables due to several reasons: 
\begin{enumerate}
\item [(i)] Different experiments with completely independent data sets, namely, BaBar~\cite{Lees:2012xj}, Belle~\cite{Huschle:2015rga,Hirose:2016wfn,Abdesselam:2019dgh} and LHCb~\cite{Aaij:2015yra,Aaij:2017uff} for $R_{D^{(*)}}$, 
as well as LHCb~\cite{Aaij:2017vbb, Aaij:2019wad} and Belle~\cite{Abdesselam:2019wac,Abdesselam:2019lab} for $R_{K^{(*)}}$, have reported results for these observables, thus reducing the chances of a statistical fluctuation. 

\item [(ii)] Such ratios are theoretically clean observables, {\it i.e.}, with strongly suppressed hadronic uncertainties, thus making them less vulnerable to higher-order quantum corrections~\cite{Bordone:2016gaq, Bernlochner:2017jka}. 

\item [(iii)] LFUV is intimately linked with lepton flavor violation (LFV)~\cite{Glashow:2014iga}, which is another `clean' signal of BSM physics.

\item [(iv)] There are only a few BSM candidates discussed in the literature, typically involving scalar or vector leptoquarks~\cite{Alonso:2015sja, Bauer:2015knc, Fajfer:2015ycq, Barbieri:2015yvd, Das:2016vkr, Becirevic:2016yqi, Sahoo:2016pet, Bhattacharya:2016mcc,  Chen:2017hir, Crivellin:2017zlb, Alok:2017jaf, Buttazzo:2017ixm, Assad:2017iib, DiLuzio:2017vat, Bordone:2017bld, 
Becirevic:2018afm, Crivellin:2018yvo, Faber:2018qon, Angelescu:2018tyl, Chauhan:2018lnq, Cornella:2019hct, Popov:2019tyc, Bigaran:2019bqv, Hati:2019ufv, Datta:2019bzu, Balaji:2019kwe, Crivellin:2019dwb, Altmannshofer:2020ywf} (see however Refs.~\cite{Bhattacharya:2014wla, Greljo:2015mma, Calibbi:2015kma, Boucenna:2016qad, Blanke:2018sro, Kumar:2018kmr,  Li:2018rax, Matsuzaki:2018jui, Marzo:2019ldg}) for other plausible BSM explanations) that can simultaneously account for both $R_{D^{(*)}}$ and $R_{K^{(*)}}$ anomalies, while being consistent with all other theoretical and experimental
constraints~\cite{Tanabashi:2018oca,Amhis:2019ckw}. 
\end{enumerate}

In this paper, we take the $B$-anomalies at face value and address them in a minimal $R$-parity violating (RPV) supersymmetric (SUSY) framework, with the third generation superpartners lighter than the other two generations (hence dubbed as `RPV3'). The RPV3 framework was earlier proposed~\cite{Altmannshofer:2017poe} to explain the $R_{D^{(*)}}$ anomaly and its possible interconnection with the radiative stability of the SM Higgs boson. The basic idea behind this suggestion comes from the simple observation that the $R_{D^{(*)}}$ anomaly involves $b$ and $\tau$, both members of the third fermion family. On the other hand, it is again another third-family fermion, namely, the top quark, that is primarily responsible for the Higgs naturalness problem in the SM. The best known candidate theory for addressing the naturalness problem is (still) low-scale SUSY. However, given the null results in direct SUSY searches at the LHC so far~\cite{Tanabashi:2018oca, ATLAS_SUSY, CMS_SUSY}, SUSY solutions to naturalness have become less appealing. As argued in Ref.~\cite{Altmannshofer:2017poe} (see also Ref.~\cite{Brust:2011tb}), the RPV3 framework which assumes only the third-family fermion to be effectively supersymmetric at the low-scale, while the sfermions belonging to the first two families are decoupled from the low-energy spectrum, provides a simple and minimal solution to the naturalness issue, while being consistent with the LHC constraints so far~\cite{Papucci:2011wy, Buckley:2016kvr}, as well as preserving the attractive features of SUSY, such as gauge-coupling unification\footnote{As shown in Ref.~\cite{Altmannshofer:2017poe}, gauge coupling unification occurs regardless of whether only one, two, or all three fermion families are supersymmetrized at the TeV scale.} and the existence of dark matter candidate(s). 

Here we extend our previous analysis to address both $R_{D^{(*)}}$ and $R_{K^{(*)}}$ anomalies simultaneously in the RPV3 framework. In addition, we also examine the possibility of addressing two other intriguing and seemingly unrelated anomalies within the same RPV3 framework, namely, (i) the longstanding discrepancy between the SM prediction and experimental measurement of the muon anomalous magnetic moment~\cite{Bennett:2006fi, Tanabashi:2018oca}, and (ii) the recent anomalous upgoing ultra-high energy (UHE) air showers seen by the ANITA balloon experiment~\cite{Gorham:2016zah, Gorham:2018ydl}. Our goal in this paper is to see if we can carve out a common allowed parameter space within our RPV3 framework where the regions favored by the $B$-anomalies can overlap with the muon $(g-2)$ and ANITA-favored regions, while being consistent with all relevant theoretical and experimental constraints. For simplicity, we consider three different versions of our scenario enumerated later, based on certain symmetry arguments, and in each case, we investigate whether there is any available  parameter space where all these anomalies can coexist. In one of the three scenarios we find a common overlap region at $3\sigma$ confidence level (CL) satisfying all the anomalies, while in the other two simpler scenarios not all the four anomalies could be accounted for, but a combination of either two or three of them could coexist. To the best of our knowledge, this is the first analysis of its kind to unify the $B$-anomalies with the muon $g-2$ and ANITA anomalies in a single testable framework. In passing, let us also mention that while in the past few years many papers~\cite{Alonso:2015sja, Bauer:2015knc, Fajfer:2015ycq, Barbieri:2015yvd, Das:2016vkr, Becirevic:2016yqi, Sahoo:2016pet, Bhattacharya:2016mcc,  Chen:2017hir, Crivellin:2017zlb, Alok:2017jaf, Buttazzo:2017ixm, Assad:2017iib, DiLuzio:2017vat, Bordone:2017bld, 
Becirevic:2018afm, Crivellin:2018yvo, Faber:2018qon, Angelescu:2018tyl, Chauhan:2018lnq, Cornella:2019hct, Popov:2019tyc, Bigaran:2019bqv, Hati:2019ufv, Datta:2019bzu, Balaji:2019kwe, Crivellin:2019dwb, Altmannshofer:2020ywf, Bhattacharya:2014wla, Greljo:2015mma, Calibbi:2015kma, Boucenna:2016qad, Blanke:2018sro, Kumar:2018kmr,  Li:2018rax, Matsuzaki:2018jui, Marzo:2019ldg} jointly discuss both $R_{D^{(*)}}$ and $R_{K^{(*)}}$ anomalies, only a few~\cite{Bauer:2015knc, Das:2016vkr, Chen:2017hir, Datta:2019bzu, Crivellin:2019dwb} also simultaneously address the muon $(g-2)$ or ANITA~\cite{Chauhan:2018lnq}, but not both together.

In this work while we are using our RPV3 scenario to understand several of the anomalies
because we think it has considerable theoretical appeal for such issues, we will in the following section also voice our concerns regarding experiments and theory pertaining to the results of interest. 

The rest of the paper is organized as follows: In Section~\ref{sec:exp}, we give a brief description of the anomalies under consideration. In Section~\ref{sec:RPV}, we discuss how these anomalies can be explained in our RPV3 setup. Our main numerical results are presented in Section~\ref{sec:results}. The low-energy experimental constraints used in our analysis are discussed in Section~\ref{sec:constraints}. In Section~\ref{sec:LFV}, we make predictions for the LFV decays of $\tau$ and $B$-mesons. Complementary high-$p_T$ tests of the $B$-anomalies at the LHC are discussed in Section~\ref{sec:LHC}. We conclude in Section~\ref{sec:conclusions}. In Appendix~\ref{app:bino}, we calculate the extra contribution to $R_{D^{(*)}}$ from a light bino in the final state. In Appendix~\ref{app:anita}, we show the bino mean free path as a function of its energy for some benchmark values of the RPV3 parameters.

%%%%%%%%%%%%%%%%%%%%%%%%%%%%%%%%%%%%%%%%%%%%%%%%%%%%%%%%%
\section{The Anomalies}  \label{sec:exp}
%%%%%%%%%%%%%%%%%%%%%%%%%%%%%%%%%%%%%%%%%%%%%%%%%%%%%%%%%
In this section, we critically assess the status of each of the experimental anomalies to be subsequently addressed in our RPV3 framework. Although we indulge in a BSM explanation of the anomalies using our RPV3 scenario, 
%we also voice our reservations and %concerns regarding experiment and %theory pertaining to the anomalies. %Thus, 
and even though the global pull of the $B$ anomalies against the SM 
%may 
appears to be over 5$\sigma$~\cite{Aebischer:2019mlg} (see Table~\ref{tab:sigmas}), its interpretation as robust evidence of LFUV does not seem compelling to us at this point. It is quite plausible that the resolution of some of these anomalies may well lie in fluctuation of one or more  of these experimental results by a few $\sigma$. We discuss how the remaining experimental and theoretical issues may be addressed. 

In Table~\ref{tab:sigmas} we summarize the anomalies and their pulls. When combining the pulls of several observables we treat all observables as independent degrees of freedom.
We do not include ANITA in this table, as it is difficult to reliably estimate the associated systematic errors and therefore the precise significance of the ANITA anomaly is hard to quantify.

\renewcommand{\arraystretch}{2.0}
%%%%%%%%%%%%%%%%%%%%%%%%%%%%%%%%%%%%%%
\begin{table*}[tbh]
    \centering
    \begin{tabular}{cccccc} 
    \hline\hline
         ~~~Observable~~~ & ~~~$R_{D^{(*)}}, R_{J/\psi}$~~~ & ~~~$R_{K^{(*)}}$~~~ & ~~~$(g-2)_\mu$~~~ & ~~~All but $(g-2)_\mu$~~~ & ~~~~~All~~~~~ \\ \hline
         Pull & $3.3\sigma$ ($2.2\sigma$) & $3.4\sigma$ & $3.3\sigma$ & $4.5\sigma$ ($3.7\sigma$) & $5.3\sigma$ ($4.6\sigma$) \\
    \hline\hline
    \end{tabular}
    \caption{Summary of the anomalies in the observables $R_{D^{(*)}}$, $R_{J/\psi}$, $R_{K^{(*)}}$, and $(g-2)_\mu$. Listed are the pulls of various subsets of observables. The pulls are combined assuming the observables are independent from each other. The values in parenthesis exclude the BaBar results for $R_{D^{(*)}}$.}
    \label{tab:sigmas}
\end{table*}
%%%%%%%%%%%%%%%%%%%%%%%%%%%%%%%%%%%%%%

\subsection{\texorpdfstring{Hints for LFUV in $B$ Decays}{Hints for LFUV in B Decays}}
As alluded to in Section~\ref{sec:intro}, 
multiple experimental results from BaBar, Belle and LHCb are pointing to non-standard sources of LFUV in charged-current and in flavor-changing neutral current (FCNC) decays of $B$-mesons, based on the $b \to c \ell^- \bar{\nu}$ and $b \to s \ell^+ \ell^-$ transitions, respectively, as measured by the ratios of the BRs, $R_{D^{(*)}}$ and $R_{K^{(*)}}$  [cf.~Eqs.~\eqref{eq:RD} and \eqref{eq:RK}]. We briefly review the current experimental results on these observables and the significance of the discrepancies with respect to the SM predictions.

\subsubsection{\texorpdfstring{$R_D$, $R_{D^*}$ and $R_{J/\psi}$}{R(D), R(D*) and R(J/psi)}}  \label{sec:2A1}
\begin{figure*}[tbh]
  \centering
  \includegraphics[width=0.47\textwidth]{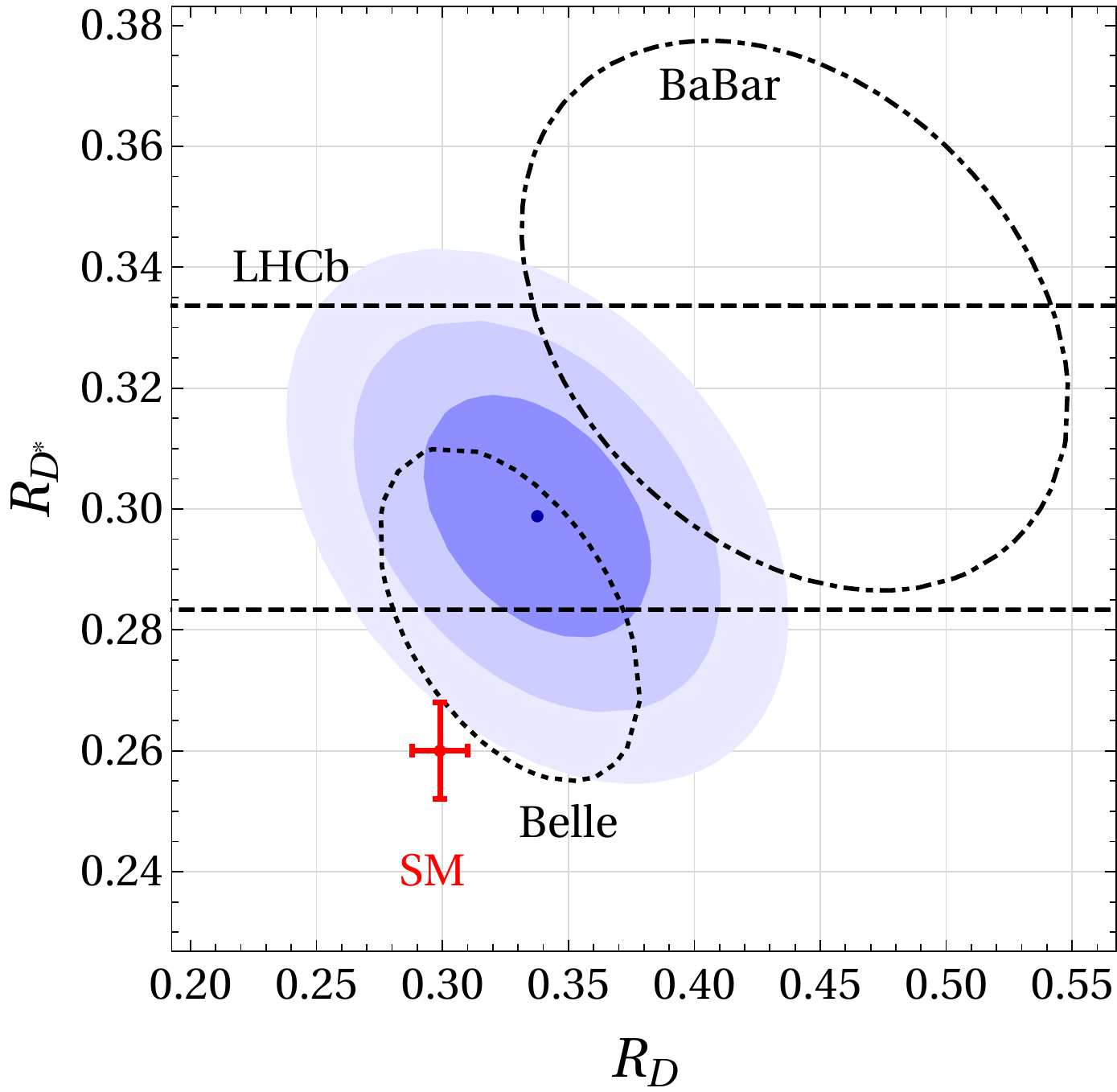} ~~~~~
  \includegraphics[width=0.46\textwidth]{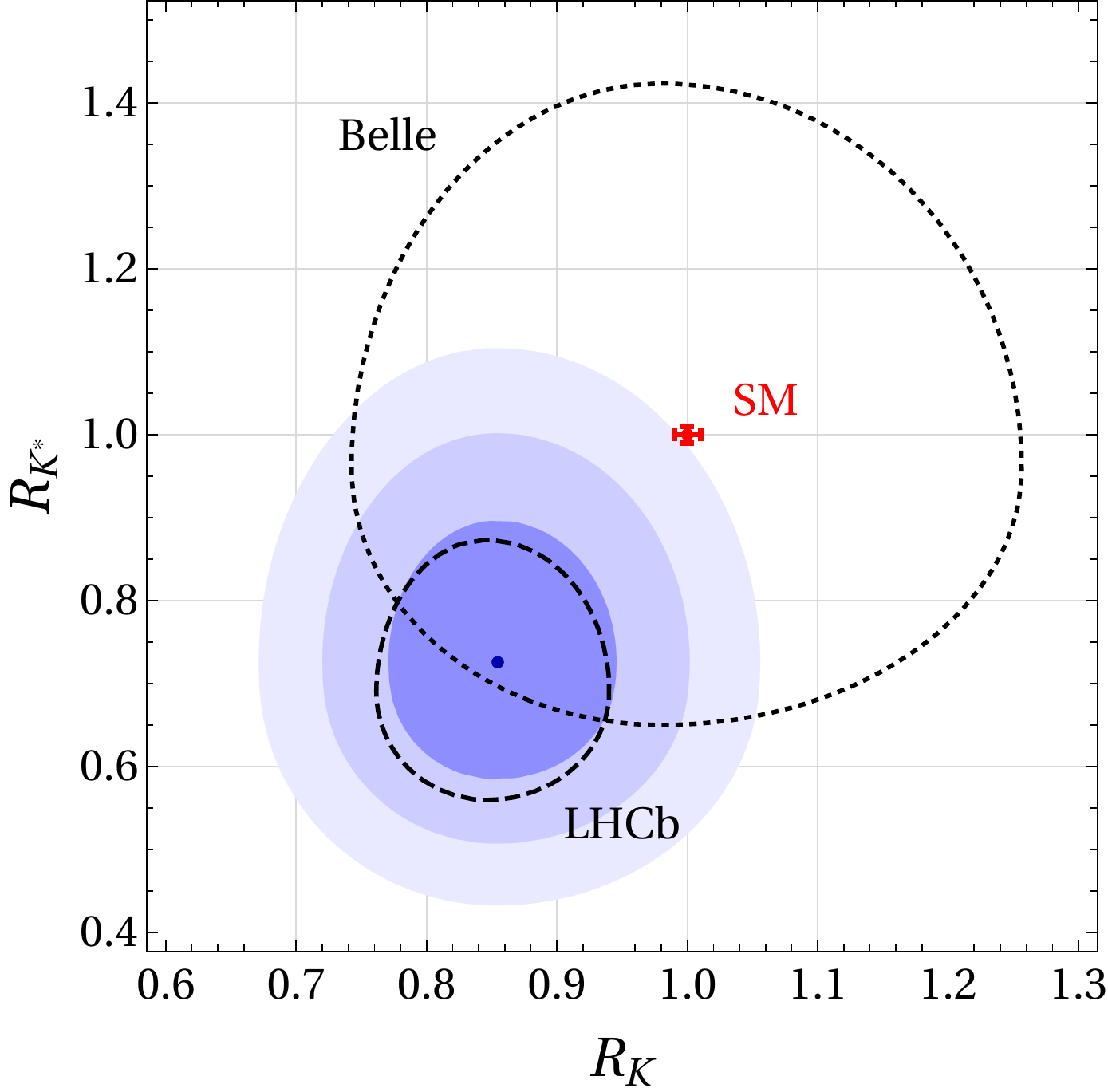}
  \caption{Experimental averages (shown by the  blue dot for the best-fit and darker-to-lighter shaded regions for $1\sigma, 2\sigma, 3\sigma$) and SM predictions (shown by red error bars) for the LFUV observables $R_D$ and $R_{D^*}$ (left), as well as $R_K$ and $R_{K^*}$ (right). The values for $R_{K^{(*)}}$ correspond to a dilepton invariant mass squared of 1.1\,GeV$^2 < q^2 < 6$\,GeV$^2$. Individual 1$\sigma$ regions from Belle, LHCb, and BaBar are also shown by the dotted, dashed, and dash-dotted contours, respectively.}
  \label{fig:R}
\end{figure*}

Measurements of $R_{D^{(*)}}$ exist from BaBar~\cite{Lees:2012xj}, Belle~\cite{Huschle:2015rga,Hirose:2016wfn,Abdesselam:2019dgh}, and LHCb~\cite{Aaij:2015yra,Aaij:2017uff}. Combining all these, we find
\begin{eqnarray}
    R_D &\ = \ & 0.337 \pm 0.030 ~, \\
    R_{D^*} & \ = \ & 0.299 \pm 0.013 ~,
\end{eqnarray}
with an error correlation between $R_{D}$ and $R_{D^*}$ of $\rho = -38\%$.
This is in very good agreement with the average from the Heavy Flavor Averaging Group (HFLAV)~\cite{Amhis:2019ckw}.
Our $R_{D^{(*)}}$ combination is shown in the left plot of Fig.~\ref{fig:R}. For the SM predictions we use in our analysis
\begin{eqnarray}
 R_D^\text{SM} &\ = \ & 0.299 \pm 0.011 \text{~\cite{Lattice:2015rga}} ~, \label{eq:RDSM} \\
 R_{D^*}^\text{SM} & \ = \ & 0.260 \pm 0.008\text{~\cite{Bigi:2017jbd}} ~. \label{eq:RDsSM}
 \end{eqnarray}
 Note that the above uncertainties are somewhat larger than those quoted in {\it e.g.}~Refs.~\cite{Bernlochner:2017jka, Jaiswal:2017rve,Jaiswal:2020wer}, but we prefer to be conservative for reasons described below.

%%%%%%%%%%%%%%%%%%%%%%%%%%%%%%%%%%%%%%%%%%
%%%%%%%%%%%%%%%%%%%%%%%%%%%%%%%%%%%%%%%%%%

LFUV in the same quark-level transition can also be probed by the decay $B_c \to J/\psi\: \ell \nu$. The corresponding experimental result from LHCb reads~\cite{Aaij:2017tyk}
\begin{equation}
 R_{J/\psi} \ = \ \frac{{\rm BR}(B_c \to J/\psi \: \tau\nu)}{{\rm BR}(B_c \to J/\psi\: \ell\nu)} = 0.71 \pm 0.17 \pm 0.18 ~,
\end{equation}
whereas the SM prediction is~\cite{Murphy:2018sqg, Dutta:2017xmj,Issadykov:2018myx,Watanabe:2017mip,Cohen:2018dgz,Berns:2018vpl}
\begin{equation}
 R_{J/\psi}^\text{SM}  \ = \  0.26 \pm 0.02\text{~} ~.
\end{equation}

The SM predictions of the individual observables disagree with the experimental results by $1.4\sigma$ ($R_D$), $2.5\sigma$ ($R_{D^*}$), and $1.7\sigma$ ($R_{J/\psi}$). 
The combined discrepancy between the quoted SM predictions and our experimental average is $3.3 \sigma$, as shown in Table~\ref{tab:sigmas}.

A few remarks are in order on the theoretical and experimental errors. 
\begin{enumerate}
\item [(i)] Lattice calculations for $B \to D$ semileptonic decay are fairly mature by now with stated errors up to around 4\%
~\cite{Bailey:2014tva, Lattice:2015rga, Harrison:2017fmw}. \
%blue{check if these are the correct refs?} 
%lattice B2D.
However,
these quoted errors so far do not include corrections due to soft photons with energy 
below the experimental threshold; these corrections could be around a few \%~\cite{deBoer:2018ipi}. These calculations may also need to be corrected for electromagnetic and isospin effects, {\it e.g} difference between charged and neutral $B$ decays etc. 

\item [(ii)] For $B \to D^*$ semileptonic case there appear to be more serious issues with the theory calculations. An important point that needs to be considered seriously is that since $D^*$ carries spin, its production and decay cannot be rigorously factorized. In fact in a construction of the quantum amplitude the production from $B \to D^* \ell \nu$ must be correlated with the final decay, say 
$D^* \to D \pi$, with an appropriate spin-1 $D^*$ propagator with its width. It is quite likely that unless this effect is correctly
taken into account both the extraction of $V_{cb}$ and $R_{D^*}$ suffer from some inaccuracies. 

\item [(iii)] Moreover, for $B \to D^*$ transition, a complete lattice calculation with the full
$q^2$-dependent form-factors does not exist yet and from the lattice perspective given that for a vector final state there are four and not two form-factors
(unlike the case of a pseudoscalar final state), it is difficult to see why the theory errors for the case of $R_{D^{(*)}}$ should not be appreciably bigger than for
$R_D$ [cf.~Eqs.~\eqref{eq:RDSM} and \eqref{eq:RDsSM}]. 

\item [(iv)] There may also be a rather serious concern, at present, on the experimental side, namely, most of the experimental results 
so far using the leptonic $\tau \to \mu \nu \nu$ decays involving two neutrinos seem to indicate somewhat larger deviations from theoretical expectations based on the SM compared to two recent measurements, one from LHCb~\cite{Aaij:2017uff} and the other from Belle~\cite{Hirose:2016wfn} which use $\tau \to {\rm hadron(s)} + \nu$; see Table~\ref{tab:semi-leptonic-data}. Although the error in each class of measurements is rather large so that the difference in the central values is not different by a significant amount, this difference needs to be better understood as it may originate from some important experimental systematics. Superficially, for example, $\tau$ decays involving two neutrinos in the final state
appear more vulnerable to backgrounds from higher $D^*$ resonances. Theoretical estimates on such contaminations are quite unreliable and they should  be subtracted by using experimental measurements, which can be quite challenging.

\item [(v)] Another issue on the experimental side that is somewhat disconcerting is that the very first experimental results on the charged-current anomaly came from BaBar~\cite{Lees:2012xj} and they seem to indicate the most significant deviations from the SM; in contrast, all the Belle results seem to show only mild deviations [cf.~Table~\ref{tab:semi-leptonic-data}]. That is why excluding the BaBar results leads to a smaller pull of only $2.2\sigma$ for $R_{D^{(*)}}$, as shown in Table~\ref{tab:sigmas}.  
\end{enumerate}

The concerns regarding theory errors voiced above in (i)--(iii) on the charged-current anomaly not withstanding, we also want to stress that at this point the theory errors
are subdominant and unlikely to be the sole cause of the discrepancy.

%table with all results of RD, RD* and R_psi
%%%%%%%%%%%%%%%%%%%%%%%%%%%%%%%%%%%%%%%%%%%%%%%%
\begin{table*}[tbh!]
\begin{tabular}{||c|c|c|c|c|c||}\hline\hline
Experiment & Tag method & $\tau$ decay mode & $R_D$ & $R_{D^{*}}$ & $R_{J/\psi}$   \\ 
           
\hline

Babar (2012)~\cite{Lees:2012xj} & hadronic & $\ell \nu \nu$ & $0.440 \pm 0.058 \pm 0.042$ & $0.332 \pm 0.024 \pm 0.0.018$ &\\

Belle (2015)~\cite{Huschle:2015rga}& hadronic & $\ell \nu \nu$ & $0.375 \pm 0.064 \pm 0.026 $ & $0.293 \pm 0.038 \pm 0.015$ &\\

LHCb (2015)~\cite{Aaij:2015yra}  & hadronic &  $\ell \nu \nu$ & - & $0.336 \pm 0.027 \pm 0.030$ &\\

Belle (2016)~\cite{Huschle:2015rga}& semileptonic &  $\ell \nu \nu$  & - &  $0.302 \pm 0.030 \pm 0.011$ &  \\

Belle (2017)~\cite{Hirose:2016wfn}& hadronic & $\pi (\rho) \nu$ & - &  $0.270 \pm 0.035 \pm 0.027$ & \\

LHCb (2017)~\cite{Aaij:2017uff} & hadronic & $ 3 \pi  \nu$ & - &  $0.291 \pm 0.019 \pm 0.029$ &  \\

Belle (2019)~\cite{Abdesselam:2019dgh}& semileptonic & $\ell \nu \nu$ & $0.307 \pm 0.037 \pm 0.016 $ & $0.283 \pm 0.018 \pm 0.014$& \\

LHCb (2016)~\cite{Aaij:2017tyk} & hadronic &  $\ell \nu \nu$ & - & - & $0.71 \pm 0.17 \pm 0.18 $  \\ \hline
SM   & -  & -& $  0.299 \pm 0.011$~\cite{Lattice:2015rga}   &$0.260 \pm 0.008$~\cite{Bigi:2017jbd}
 & $0.26\pm 0.02$~\cite{Murphy:2018sqg}\\
\hline\hline
\end{tabular}
%\caption{\label{tab:Higgs-filters}
\caption{All experimental results announced to date on $R_D$, $R_{D^*}$ and $R_{J/\psi}$ versus the predictions of those in the SM.} 
\label{tab:semi-leptonic-data}
\end{table*}
%%%%%%%%%%%%%%%%%%%%%%%%%%%%%%%%%%%%%%%%%%%%%%%%

Moreover, there is also an intriguing aspect of data from all three experimental groups on these semileptonic decays that is quite interesting and deserves attention. Table~\ref{tab:semi-leptonic-data} shows all available results to date indicating whether the other $B$ in the event was tagged hadronically or semileptonically and whether the $\tau$ decayed leptonically or hadronically. Table~\ref{tab:semi-leptonic-data} also includes the $R_{J/\psi}$ ratio from similar semileptonic decays of $B_c$ to $J/\psi\: \tau (\ell) \nu$. Altogether there are 11 entries and it is quite remarkable that the experimental central value of the $R$-ratio for each of these is {\it always}  without exception above the central value predicted by the SM. Note that the 11 experimental results in Table~\ref{tab:semi-leptonic-data} are not all completely independent. In fact in some cases, these are 
just updates of ongoing analyses with more data. Nevertheless,  
 many among these are independent and so the fact that so many experimental measurements are above the SM predictions is quite noteworthy.\footnote{For an important famous reminder from our past history that sometimes  many early experimental results can be somewhat incompatible with theoretical expectations, see Ref.~\cite{Lee:1965js}, in particular their discussion of the ``Michel parameter" in muon decay on p. 448, Fig.~6.}

\subsubsection{\texorpdfstring{$R_K$ and $R_{K^*}$}{R(K) and R(K*)}} 
The most precise measurement of the LFUV observable $R_K$ comes from LHCb~\cite{Aaij:2019wad}:
\begin{equation}
 R_K \ = \ 0.846^{+0.060}_{-0.054}{}^{+0.016}_{-0.014} ~,
\end{equation}
with the dilepton invariant mass-squared in the range 1.1\,GeV$^2 < q^2 < 6$\,GeV$^2$.
The SM predicts $R_K^\text{SM} \simeq 1$ with \%-level accuracy~\cite{Bordone:2016gaq}, corresponding to a $\sim 2.5\sigma$ tension with the experimental result.

The most precise measurement of $R_{K^*}$ is from a Run-1 LHCb analysis~\cite{Aaij:2017vbb} that finds
\begin{align}
 R_{K^*} \ = \ \begin{cases} 0.66^{+0.11}_{-0.07}\pm0.03 \, , \\ 0.69^{+0.11}_{-0.07}\pm0.05 \, ,
 \end{cases}
\end{align}
where the first and second values correspond to a $q^2$ range of 0.045\,GeV$^2 < q^2 < 1.1$\,GeV$^2$ and 1.1\,GeV$^2 < q^2 < 6$\,GeV$^2$, respectively.
The result for both $q^2$ bins are in tension with the SM prediction, $R_{K^*}^\text{SM} \simeq 1$~\cite{Bordone:2016gaq}, by $\sim 2.5\sigma$ each. Since the systematic errors here are subdominant, it is reasonable to add the deviations in these two bins in quadrature. Treating the two bins as independent observables we thus find that the deviations from the SM
in $R_{K^*}$ amounts to about 2.9$\sigma$.

Recent results for $R_{K^*}$ and $R_K$ by Belle
have sizable uncertainties and are compatible with both the SM predictions and the LHCb results. For the 1.1\,GeV$^2 < q^2 < 6$\,GeV$^2$ bin, Belle finds~\cite{Abdesselam:2019wac,Abdesselam:2019lab}
\begin{eqnarray}
  R_K & \ = \ & 0.99^{+0.27}_{-0.23}\pm 0.06 ~, \\
  R_{K^*} & \ = \ & 0.96^{+0.45}_{-0.29}\pm 0.11 ~.
\end{eqnarray}
In the right plot of Fig.~\ref{fig:R} we show the combination of the LHCb and Belle results for $R_{K^{(*)}}$ in the 1.1\,GeV$^2 < q^2 < 6$\,GeV$^2$ bin compared to the SM prediction. Combining the Belle and LHCb results, we get a net pull of $3.4\sigma$ in $R_{K^{(*)}}$ as shown in Table~\ref{tab:sigmas}. 

Unlike the charged-current semileptonic decays, in the case of FCNC decays $B \to K^{(*)} \ell^+ \ell^-$, there are hardly any nagging 
theoretical issues. So long as the lepton pair invariant mass is larger than about 500 MeV, the SM prediction for the ratio is rather 
clean and unambiguous. The reservation one may have is only about light lepton invariant mass, say below 500 MeV.
Then there is a concern that the electron pair may receive appreciably different radiative corrections from the muon 
pair~\cite{Bordone:2016gaq}.

The primary concerns about $\mu-e$ universality violation in FCNC is experimental. Of course the effects are only a few $\sigma$.
Moreover, it is only one experiment, {\it i.e.} LHCb, and an independent confirmation by Belle II would be highly desirable.  Also, if it is genuine LFUV it ought to show up irrespective of hadronic 
final states in $B$-decays. Thus one should see the corresponding $b \to s$ FCNC decays materializing into baryonic and other final states, such as $\Lambda_b\to \Lambda \ell^+\ell^-$.
It also should not depend on the spectator quark. Thus charged and neutral $B$ and also $B_s$ decays ought to exhibit
similar signs of LFUV. In particular, LHCb already seems to have indications  that the observed
rate for $B_s \to \phi \mu^+ \mu^-$ is seemingly below ``SM'' expectations~\cite{Bsphimumu} but the absolute rate calculations may suffer from some long-distance (non-local) contaminations, so a direct test of $\mu-e$ universality via a measurement of $B_s \to \phi e^+ e^-$ 
would be very valuable.

Let us briefly add that we are primarily focusing on the LFUV anomalies as they are theoretically cleaner 
and for now
we are choosing not to include some other possible indications of deviations from the SM, 
such as angular observables
or absolute rate for $B \to K^{(*)} \mu^+ \mu^-$~\cite{Aebischer:2019mlg, Alguero:2019ptt, Alok:2019ufo, Ciuchini:2019usw, Kowalska:2019ley, Arbey:2019duh}) and also rate for $B_s \to \phi \mu^+ \mu^-$~\cite{Bsphimumu} as in these 
cases there can be non-perturbative contributions 
from non-local effects especially in the region of low $q^2$ that are not under full theoretical control.

Before closing this subsection, it is worth pointing out here that the hints of LFUV are only seen in the semileptonic $B$-decays. Analogous semileptonic decays of charmed mesons do not show any such deviations from the SM. For instance, BESIII has recently reported a measurement of the ratio of BRs in the $D^+$-decay~\cite{BESIII:2020dbj}, {\it viz.} 
\begin{align}
    \frac{{\rm BR}(D^+\to \omega\mu^+\nu_\mu)}{{\rm BR}(D^+\to \omega e^+\nu_e)} \ = \ 1.05\pm 0.14 \, ,
\end{align} 
which agrees with the SM prediction $(0.93-0.96)$~\cite{Soni:2018adu, Faustov:2019mqr} within uncertainties. This further justifies our approach of linking the $B$-anomalies to BSM physics treating the third family as special. 

\subsection{\texorpdfstring{Muon $g-2$}{Muon g-2}}

Another interesting observable that has since long time been hinting towards BSM physics is the anomalous magnetic moment of the muon.
The existing BNL experimental result~\cite{Bennett:2006fi} for the $(g-2)_\mu$ reads~\cite{Tanabashi:2018oca}
\begin{equation}
 a_\mu^\text{exp} \ = \ (11,659,209.1 \pm 5.4~{\rm (stat)} \pm 3.3~{\rm (sys)}) \times 10^{-10} ~.
\end{equation}
The $(g-2)$ experiment at Fermilab~\cite{Grange:2015fou} is expected to improve the experimental accuracy by a factor of about four in the next few years. 

The SM prediction for $a_\mu$ can be decomposed in contributions from QED, from the electro-weak interactions, from hadronic vacuum polarization and from hadronic light-by-light scattering:
\begin{equation}
 a_\mu^\text{SM} \ = \ a_\mu^\text{QED} + a_\mu^\text{EW} + a_\mu^\text{VP} + a_\mu^\text{LbL} ~,
\end{equation}
The QED and electro-weak contributions are known with high accuracy~\cite{Aoyama:2017uqe,Gnendiger:2013pva}
\begin{eqnarray}
 a_\mu^\text{QED} &\ = \ & (11,658,471.897 \pm 0.007) \times 10^{-10} ~, \\
 a_\mu^\text{EW} &\ = \ & (15.36 \pm 0.10) \times 10^{-10} ~.
\end{eqnarray}

The hadronic vacuum polarization contribution can be determined using $e^+e^- \to $~hadrons data and dispersion relations. A recent such analysis gives~\cite{Davier:2019can} (see also Ref.~\cite{Keshavarzi:2019abf})
\begin{equation}
 a_\mu^\text{VP} \ = \ [ (693.9 \pm 4.0) - (9.9 \pm 0.1) + (1.24 \pm 0.01) ]\times 10^{-10} ~,
\end{equation}
where the first, second and third terms correspond to the LO, NLO, and NNLO contributions, respectively. The value is in good agreement with the findings of a hybrid approach that uses the best part of lattice results along with the best part of the experimental data and continuum dispersion relation data~\cite{Blum:2018mom}, 
and tends to favor the BSM interpretation of the data. This is particularly significant since in the traditional $R$-ratio
dispersion analysis there is appreciable concern due to the discrepancy of $\approx 2 \sigma$ between the BaBar data and the KLOE data~\cite{AidaEK_INT_Sept2019}. Indeed the lattice hybrid approach does not use the somewhat 
conflicting input data from BaBar or KLOE.

A recent model estimate of the light-by-light contribution 
reads~\cite{Nyffeler:2016gnb,Colangelo:2014qya,Keshavarzi:2018mgv}
\begin{equation}
 a_\mu^\text{LbL} \ = \ (10.1 \pm 2.6) \times 10^{-10} ~,
\end{equation}
Important lattice results for the light-by-light contribution have recently become available~\cite{Blum:2019ugy}. 
%\textcolor{red}{(WA: need to add refs.)}
These are consistent with phenomenological estimates and reinforce the expectation that  they are quite small 
$\approx 10 \times 10^{-10}$ compared to the hadronic vacuum polarization contribution $\approx 700 \times 10^{-10}$~\cite{Blum:2018mom}.

Combining the results collected above leads to a discrepancy between experiment and SM prediction at $3.3\sigma$ CL~\cite{Davier:2019can}: 
\begin{equation}
 \Delta a_\mu \ = \ a_\mu^\text{exp} - a_\mu^\text{SM} \ = \ (26.1 \pm 7.9)\times 10^{-10}~.
\end{equation}

%More recently there has been a different estimate~\cite{Davier:2019can} yielding,
%\begin{equation}
% a_\mu^\text{VP} \ = \ (693.9 \pm 4.0) \times 10^{-10} ~,
%\end{equation}
%leading to about a $3.3 \sigma$ confidence level (CL) deviation from the experimental number.

For this  anomaly the next year is likely to be pivotal. The new Muon $(g-2)$ experiment at Fermilab~\cite{Grange:2015fou} already seems to have collected
about two times the data used by the BNL experiment; the analysis of that accumulated data is expected in the next few months.
How this new result compares with the previous BNL result would be crucial for the BSM interpretation. 

On the lattice front, about a factor of 3 reduction in the error is anticipated in the next few months by the RBC-UKQCD Collaboration~\cite{CL_PC} and this could also have a critical bearing 
on the BSM interpretation. Also phenomenological approaches are pursued both for the hadronic vacuum polarization and the light-by-light scattering contribution~\cite{Colangelo:2015ama,Colangelo:2017qdm,Hoferichter:2018dmo,Colangelo:2018mtw,Hoferichter:2019gzf}. 
At the moment, the so-called ``hybrid" method of RBC-UKQCD~\cite{Blum:2018mom} which uses part of the continuum dispersive calculation and in part the lattice calculation in regions which complement each other seems to
tentatively favor the BSM interpretation. But it would be much better if pure lattice techniques can further reduce their error by factor of 2 to 3 so it does not use any input from experiment especially since two of the best experimental results from KLOE and BaBar have $\approx 2 \sigma$  disagreement between them. Therefore  pure lattice calculations with reduced errors would be very welcome in providing input for the fate of the BSM interpretation in muon $g-2$. It appears we will need to wait for another year or so for this to happen.

The theory uncertainty on the hadronic vacuum polarization contribution can also be reduced by about a factor of 2 at the proposed MUonE experiment~\cite{Abbiendi:2016xup, Dorigo:2020cae} which will make a very high-precision measurement of elastic $\mu-e$ scattering at a QED-dominated momentum exchange of $q^2 = {\cal O}(100~{\rm MeV})^2$.  This measurement will be quite robust and insensitive to any BSM physics that could be responsible for the muon $(g-2)$ anomaly~\cite{Dev:2020drf, Masiero:2020vxk}.

\subsection{Anomalous ANITA Events}

The Antarctic Impulsive Transient Antenna (ANITA) experiment~\cite{prop} is primarily designed for the detection of the ultra-high energy (UHE) cosmogenic neutrino flux via the Askaryan effect in ice~\cite{Askaryan:1962hbi}. A recent anomalous observation in UHE cosmic ray (UHECR) air showers made by the ANITA collaboration has also hinted at some BSM physics. Two anomalous upward-going events with deposited shower energies of $0.6\pm 0.4$ EeV~\cite{Gorham:2016zah} and $0.56^{+0.3}_{-0.2}$ EeV~\cite{Gorham:2018ydl} (1 EeV $= 10^9$ GeV) have been reported. Both these events originate from well below the horizon, with large negative elevation angles of $(-27.4\pm 0.3)^\circ$ and $(-35.0\pm 0.3)^\circ$, respectively. They do not exhibit phase inversion (opposite polarity) due to Earth's geomagnetic effects, and hence, are unlikely to be downgoing UHECR air showers reflected off the Antarctic ice surface, although there is some uncertainty in modeling the roughness of the surface ice~\cite{Prohira:2018mmv, Dasgupta:2018dzp, Shoemaker:2019xlt}. Potential background events from anthropogenic radio signals that might mimic the UHECR characteristics, or unknown processes that might lead to a non-inverted polarity on reflection from the ice cap are estimated to be $\leq$ 0.015, resulting in a $\gtrsim 3\sigma$ evidence for direct upward-moving Earth-emergent UHECR-like air showers above the ice surface~\cite{Gorham:2018ydl}. This poses considerable difficulty for interpretation of such events within the SM framework due to the low survival rate ($\lesssim 10^{-6}$) of EeV-energy neutrinos over long chord lengths in Earth$\sim 5000$ km, even after accounting for the probability increase due to $\nu_\tau$ regeneration~\cite{Gorham:2016zah, Fox:2018syq, Romero-Wolf:2018zxt}. Moreover, as pointed out in earlier studies~\cite{Collins:2018jpg,Chauhan:2018lnq, Shoemaker:2019xlt, Safa:2019ege}, the strength of isotropic cosmogenic neutrino flux needed to account for the two events is in severe tension with the upper limits set by Pierre Auger~\cite{Aab:2015kma, Zas:2017xdj} and IceCube~\cite{Aartsen:2016ngq,  Aartsen:2020vir}. Therefore, a BSM explanation with an anisotropic astrophysical source with some exotic generation and propagation mechanism of upgoing events is desirable to solve the ANITA anomaly, provided it stands further scrutiny after more data release from future ANITA  flights. In what follows, we will provide an explanation of the ANITA anomaly, in conjunction with the $B$-anomalies and the $(g-2)_\mu$ anomaly discussed above, within our RPV3 framework.\footnote{For alternative BSM interpretations of the ANITA anomaly, see e.g. Refs.~\cite{Cherry:2018rxj,Huang:2018als,Anchordoqui:2018ucj,Dudas:2018npp,Connolly:2018ewv,Anchordoqui:2018ssd,Heurtier:2019git,Cline:2019snp,Esteban:2019hcm,Heurtier:2019rkz,Borah:2019ciw,Abdullah:2019ofw,PhysRevD.100.043019,Esmaili:2019pcy,PhysRevD.100.063011}.}
%AS suggest....the section below is here temporarily.
%%%%%%%%%%%%%%%%%%%%%%%%%%%%%%%%%%%%%%%%%%%%%%%%%
%%%%%%%%%%%%%%%%%%%%%%%%%%%%%%%%%%%%%%%%%%%%%%%%%
\section{RPV Explanation of the Anomalies} \label{sec:RPV}
%%%%%%%%%%%%%%%%%%%%%%%%%%%%%%%%%%%%%%%%%%%%%%%%%

%AS....Now a few words to motivate the RPV

As we suggested before~\cite{Altmannshofer:2017poe}, 
RPV SUSY is a particularly interesting theoretical framework to address the flavor anomalies.  For one thing, for the charged-current tree level indication of BSM physics, RPV is a natural candidate and if LFUV is involved then this is especially so.
Moreover, since members of the third family, namely, $b$ and $\tau$ are involved in $B \to D^{(*)} \tau \nu$, it may well be that this anomaly is a hint that it is related to the issue of the radiative stability of the Higgs mass which is an important persistent problem of the SM. Motivated by the naturalness arguments and to keep the RPV SUSY scenario minimal, for reasons of simplicity, we have suggested that it may well be that the third generation superpartners are the lightest. In that scenario proton stability issues are less relevant and for that reason too $R$-parity breaking is a viable option~\cite{Brust:2011tb}. Lastly, we have shown that even with such an economical setup
involving effectively only one generation of superpartners a very attractive feature of SUSY, namely unification, is retained. Finally we also want to remark that our objective is to use the latest experimental data with the current set of indications to constrain as best we can the parameters of this interesting theoretical  construction.

We start from the $LQD$ part of the RPV SUSY Lagrangian that contains the $\lambda'$ couplings which are relevant for an explanation of $R_{D^{(*)}}$~\cite{Altmannshofer:2017poe,Deshpande:2012rr,  Zhu:2016xdg,Deshpand:2016cpw,Trifinopoulos:2018rna, Hu:2018lmk,Trifinopoulos:2019lyo, Wang:2019trs} and $R_{K^{(*)}}$~\cite{Biswas:2014gga, Deshpand:2016cpw, Das:2017kfo,  Earl:2018snx,Trifinopoulos:2018rna,Trifinopoulos:2019lyo, Hu:2019ahp}: 
\begin{align} 
{\cal L}_{LQD} \ = \  & \lambda'_{ijk}\big[\widetilde{\nu}_{iL}\bar{d}_{kR}d_{jL}+\widetilde{d}_{jL}\bar{d}_{kR}\nu_{iL}+\widetilde{d}^*_{kR}\bar{\nu}^c_{iL}d_{jL}\nonumber \\
-& \widetilde{e}_{iL}\bar{d}_{kR}u_{jL}-\widetilde{u}_{jL}\bar{d}_{kR}e_{iL}-\widetilde{d}^*_{kR}\bar{e}^c_{iL}u_{jL}\big]+{\rm H.c.}
\label{Eq.lambda_prime}
\end{align}
As we will see below, for explanations of the $R_{K^{(*)}}$ anomaly and the $(g-2)_{\mu}$ anomaly it is useful to also include the effect of the $LLE$ part of the RPV SUSY Lagrangian which contains the $\lambda$ couplings \cite{Trifinopoulos:2018rna}:
\begin{align}
{\cal L}_{LLE} \ = \ & \frac{1}{2}\lambda_{ijk}\big[ \widetilde{\nu}_{iL} \bar{e}_{kR} e_{jL} +\widetilde{e}_{jL} \bar{e}_{kR}\nu_{iL} +\widetilde{e}_{kR}^{*} \bar{\nu}_{iL}^c e_{jL} \nonumber \\
& \qquad - (i\leftrightarrow j) \big]+{\rm H.c.}
\label{Eq.RPVLLE}
\end{align}
One thing to keep in mind is that the $\lambda$ couplings are anti-symmetric in the first two indices: $\lambda_{ijk}=-\lambda_{jik}$. Also note that the simultaneous presence of $\lambda$ and $\lambda'$ couplings is consistent with proton decay constraints, as long as we do not switch on the relevant  $\lambda^{\prime\prime}$ ($UDD$-type) couplings.\footnote{The current proton lifetime constraint $\tau_{p\to \pi^0 \ell^+}\gtrsim 10^{34}$ years~\cite{Miura:2016krn} (with $\ell=e,\mu$) leads to a stringent upper bound of $\lambda'_{i1k}\lambda''^*_{11k}\lesssim 10^{-27}(m_{\widetilde{d}_{kR}}/100~{\rm GeV})^2$ (with $i=1,2$) on the RPV couplings~\cite{Barbier:2004ez}.} 

Following Ref.~\cite{Altmannshofer:2017poe}, throughout this work we will assume, for minimality, that the third-generation squarks, sleptons and sneutrinos are considerably lighter than the first and second generation ones. Integrating out the heavier SUSY particles we therefore can neglect the first and second generation sfermions, as their effect is suppressed by a higher mass scale in the RPV3 scenario.
Out of the 27 independent RPV couplings $\lambda^\prime_{ijk}$ in Eq.~(\ref{Eq.lambda_prime}) and the 9 independent $\lambda_{ijk}$ in Eq.~(\ref{Eq.RPVLLE}), there are 19 $\lambda^\prime$-type and 7 $\lambda$-type couplings that involve light third generation sfermions, namely, the right-handed sbottom $\widetilde b_{R}$, left-handed stop $\widetilde t_L$, left-handed tau-sneutrino $\widetilde \nu_\tau$ and both left- and right-handed staus $\widetilde \tau_{L,R}$.
%Due to $SU(2)_L$ invariance, only four out of the six soft SUSY breaking masses of those particles are independent: $m_{\widetilde b_L} = m_{\widetilde t_L}$, $m_{\widetilde b_R}$, $m_{\widetilde \tau_L} = m_{\widetilde \nu_\tau}$, and $m_{\widetilde \tau_R}$. 
We will treat these five masses as free parameters in our numerical analysis in Section~\ref{sec:results}. In addition, we require a light long-lived bino ($\widetilde{\chi}_1^0$) for the ANITA anomaly. 

As for the choice of couplings, we first analyze each of the experimental anomalies discussed above in the RPV-SUSY context and show the dependence of the observables on the relevant couplings. Then in the following Section~\ref{sec:results}, we present three different scenarios for our parameter set-up and the corresponding fit results. 

%\red{(these seem to be not exactly correct?)}
%{\bf (WA: I think it is a correct statement at the level of soft SUSY breaking masses. Actual particle masses are corrected by terms of relative size $v^2/m^2_{\widetilde f}$)}

%Working in the mass eigenbasis for the quarks and assuming that also sfermions are in their mass eigenstates, we obtain the following four-fermion operators at the tree-level
%
%\begin{align}
%{\cal L}_{\rm eff} \supset \ & \frac{\lambda'_{ij3}\lambda^{\prime *}_{mn3}}{2m^2_{\widetilde{b}_{R}}} \bigg[\bar\nu_{mL}\gamma^\mu \nu_{iL}\bar{d}_{nL}\gamma_\mu d_{jL}\nonumber \\
%& +\bar{e}_{mL}\gamma^\mu e_{iL}\left(\bar u_L V_{\rm CKM}\right)_n \gamma_\mu \left(V_{\rm CKM}^\dag u_L\right)_j \nonumber \\
%& -\nu_{mL}\gamma^\mu e_{iL}\bar d_{nL}\gamma_\mu \left(V_{\rm CKM}^\dag u_L\right)_j ~+~\text{h.c.} \bigg] \nonumber \\
%& -\frac{\lambda'_{ijk}\lambda^{\prime *}_{mjn}}{2m^2_{\widetilde{t}_{L}}}
%\bar e_{mL}\gamma^\mu e_{iL}\bar{d}_{kR}\gamma_\mu d_{nR} ~,
%\label{Leff}
%\end{align}
%where we only show the terms relevant for the following discussion.

%%%%%%%%%%%%%%%%%%%%%%%%%%%%%%%%%%%%%%%%%%%%%%%%%%
\subsection{Explanation of \texorpdfstring{$R_D$ and $R_{D^*}$}{}} \label{sec:IIIA}
%%%%%%%%%%%%%%%%%%%%%%%%%%%%%%%%%%%%%%%%%%%%%%%%%%
%%%%%%%%%%%diagram for rdrd
\begin{figure}[t!]
	\begin{subfigure}{.23\textwidth}
		\centering
		\includegraphics[width=1\linewidth]{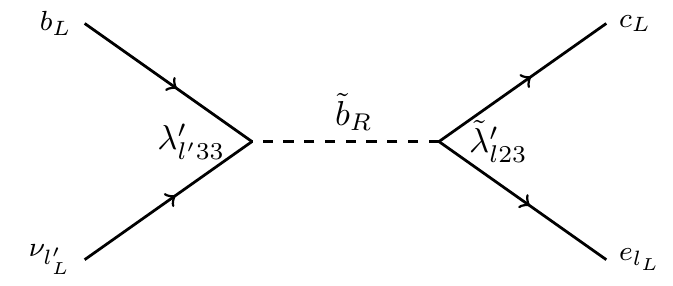}  
    	\caption{}
	    \label{fig:gm2bRD}
	\end{subfigure}
	\begin{subfigure}{.23\textwidth}
		\centering
		\includegraphics[width=1\linewidth]{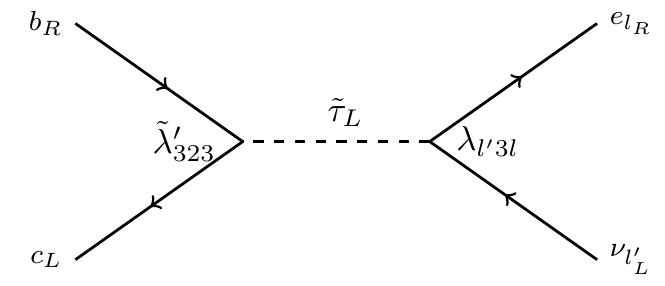}  
    	\caption{}
	    \label{fig:gm2aRD}
	\end{subfigure}
	\caption{Contributions to the $R_D$ and $R_{D^*}$ from $\lambda'$ and $\lambda$ in RPV SUSY: (a) with $LQD$ couplings only; (b) with both $LLE$ and $LQD$ couplings. Here $\widetilde{\lambda'}_{ijk}$ is defined as $\lambda'_{ilk}V_{jl}$ (with $V_{jl}$ being the CKM matrix elements). }
	\label{fig:RDdiagram}
\end{figure}
In Ref.~\cite{Altmannshofer:2017poe} we had identified BSM contributions to $b \to c \tau \nu$ transitions in the RPV setup, which can arise at the tree level from sbottom exchange [cf.~Fig.~\ref{fig:RDdiagram}(a)].
The sbottom exchange leads to contributions to the decay amplitude that have the same chirality structure as the SM contribution and thus modify $R_D$ and $R_{D^*}$ in a  universal way. Here we note that in the presence of the $LLE$ couplings, also diagrams with light sleptons, in particular a light left-handed stau, can contribute to the decays [cf.~Fig.~\ref{fig:RDdiagram}(b)]. However, in the scenarios we will consider below, the left-handed stau will be fairly heavy (specifically, we set $m_{\tilde \tau_L} = 10$ TeV in the benchmark scenarios of Section \ref{sec:results}) and the corresponding contributions will be negligible. We will therefore focus only on the sbottom contribution from the diagram in Fig.~\ref{fig:RDdiagram}(a).

It is important to note that $R_D$ and $R_{D^*}$ measured by BaBar and Belle correspond to ratios of the tauonic decay modes to an average of the muonic and electronic modes, while the LHCb measurements  are ratios of tauonic to muonic modes. 
Using the notation from Ref.~\cite{Trifinopoulos:2018rna}, 
we find in our setup 
\begin{equation} \label{eq:RDLHCb}
 \frac{R_D^\text{LHCb}}{R_D^\text{SM}} \ = \  \frac{R_{D^*}^\text{LHCb}}{R_{D^*}^\text{SM}} = \frac{|\Delta^c_{31}|^2+|\Delta^c_{32}|^2+|1+\Delta^c_{33}|^2}
{|\Delta^c_{21}|^2+|1+\Delta^c_{22}|^2+|\Delta^c_{23}|^2} ~,
\end{equation}
where %
\begin{align}\label{eq:delta}
\Delta_{ll'}^c &\ = \ \frac{v^2}{4 m^2_{\widetilde b_R}} \lambda_{l'33}^\prime \left( \lambda_{l33}^\prime + \lambda_{l23}^\prime \frac{V_{cs}}{V_{cb}} + \lambda_{l13}^\prime \frac{V_{cd}}{V_{cb}} \right)  %\nonumber \\
%+& \frac{v^2}{\blue{4} m^2_{\widetilde {\tau}_L}} \lambda_{l'3l} \left( \lambda_{333}^\prime + \lambda_{323}^\prime \frac{V_{cs}}{V_{cb}} + \lambda_{313}^\prime \frac{V_{cd}}{V_{cb}} \right)
~,
\end{align}
$v = 246$~GeV is the Higgs VEV and $V_{ij}$ are the CKM matrix elements. 

In the case of the $B$-factories, we instead have
\begin{widetext}
\begin{equation} \label{eq:RDBfac}
 \frac{R_D^\text{B-fac.}}{R_D^\text{SM}} \ = \  \frac{R_{D^*}^\text{B-fac.}}{R_{D^*}^\text{SM}} \ = \  \frac{|\Delta^c_{31}|^2+|\Delta^c_{32}|^2+|1+\Delta^c_{33}|^2}
{\xi_e (|1+\Delta^c_{11}|^2+|\Delta^c_{12}|^2+|\Delta^c_{13}|^2)+(1-\xi_e)(|\Delta^c_{21}|^2+|1+\Delta^c_{22}|^2+|\Delta^c_{23}|^2)} ~,
\end{equation}
\end{widetext}
where $\xi_e$ parameterizes the relative weight of the electronic and muonic decay modes in the $R_{D^{(*)}}$ measurements at the $B$-factories. We note that $\xi_e$ can in principle be different for each experimental analysis but we expect $\xi_e \sim 50\%$ (see e.g.~\cite{Dungel:2010uk}). We explicitly checked that varying $\xi_e$ has no significant impact on our results. This is due to the fact that $\mu-e$ universality in $b \to c \ell \nu$ decays is observed with high accuracy. Translating the results from Ref.~\cite{Jung:2018lfu} into our RPV scenario, we have  
\begin{equation}
 \frac{|1+\Delta^c_{11}|^2+|\Delta^c_{12}|^2+|\Delta^c_{13}|^2}{|\Delta^c_{21}|^2+|1+\Delta^c_{22}|^2+|\Delta^c_{23}|^2} \ = \ 1.022 \pm 0.024 ~.
\end{equation}
Therefore, it is an excellent approximation to combine the LHCb and $B$-factory results as done in Section~\ref{sec:2A1}. In that case we find\footnote{The parameter space explaining the $R_{D^{(*)}}$ data  automatically explains the  $R_{J/\psi}$ data, because the underlying transition is the same $b\to c\ell \nu$. Therefore, we do not discuss the $R_{J/\psi}$ fits separately.} 
\begin{equation}
 \frac{R_D}{R_D^\text{SM}} \ = \  \frac{R_{D^*}}{R_{D^*}^\text{SM}} \ = \ 1.15 \pm 0.04 ~,
\end{equation}
both for the LHCb and the $B$-factory expressions [cf.~Eqs.~(\ref{eq:RDLHCb}) and (\ref{eq:RDBfac})].
\subsubsection{\texorpdfstring{Implications of the observed $q^2$ distribution and of the $D^*$ polarization}{}}

Recently, Ref.~\cite{Murgui:2019czp} in an interesting study have included $q^2$  (where $q$ is the 4-momentum carried by the leptonic pair) and also the longitudinal polarization of the $D^*$ in addition to the integrated rates in order to discriminate against models.
To analyze the data in a model independent manner they allow all possible current structures in the weak Hamiltonian subject only to the constraint that only left-handed neutrinos are involved in the interaction;  thus,
\begin{align}
{\cal H}_{\text{eff}}^{b\to c \ell \nu}  \ = \  & \frac{4G_F}{\sqrt{2}}V_{cb}\big [ \left(  1 + C_{V_L} \right){\cal O}_{V_L} +C_{V_R} {\cal O}_{V_R}\nonumber\\ 
&+C_{S_R} {\cal O}_{S_R} 
+C_{S_L} {\cal O}_{S_L}+C_{T} {\cal O}_T \big ]  + \text{H.c.} 
\label{eq:effH}
\end{align}
with the operators
\begin{align}
    {\cal O}_{V_{L,R}} \ = \ & \left( \bar{c}\, \gamma^{\mu} b_{L,R} \right)\left( \bar{\ell}_L \gamma_{\mu}  \nu_{\ell L} \right),\nonumber\\
    {\cal O}_{S_{L,R}} \ = \ & \left( \bar{c}b_{L,R} \right)\left( \bar{\ell}_R \nu_{\ell L} \right),\quad \nonumber\\
    {\cal O}_{T} \ = \ & \left( \bar{c}\, \sigma^{\mu \nu} b_L \right)\left( \bar{\ell}_R \sigma_{\mu \nu} \nu_{\ell L} \right)\, , \label{eq:effop}
\end{align}
and weighted by the corresponding Wilson coefficients $C_i$. 
In this representation, the operator ${\cal O}_{V_L}$ is of special significance as it encapsulates the SM interaction. In their study of the existing experimental data,  Ref.~\cite{Murgui:2019czp} find that the simplest solution to the charge-current anomaly is with a small non-vanishing value of $C_{V_L}\approx 0.08$, with all other $C$'s equal to zero.

This has the important consequence that the polarization of the $D^*$ or for that matter of the $\tau$   will not be different from the SM. Recently Belle collaboration reported, for the longitudinal polarization of the $D^*$\cite{Abdesselam:2019wbt}
\begin{align}
F_L({D^*}) \ = \  0.60 \pm 0.08 {\rm (stat)} \pm 0.04 {\rm (sys)} \, ,
\label{eq:D_pol}
\end{align}
which is in mild tension of about 1.6 $\sigma$ with the SM which predicts~\cite{Alok:2016qyh, Huang:2018nnq, Bhattacharya:2018kig}
\begin{align}
F_{L}(D^*)_{\rm SM} \ = \ 0.46 \pm 0.03  \, .
\label{eq:D_pol_SM}
\end{align}
In the past $\approx 2$ years, Belle collaboration has also attempted to measure the polarization of the $\tau$ and found~\cite{Hirose:2016wfn}
\begin{align}
P_{\tau}(D^*)=-0.38 \pm 0.51 {\rm (stat)} \pm ^{+ 0.21}_{ - 0.16} {\rm (syst)} 
\label{eq:tau_pol}
\end{align}
At this point this result on tau polarization  within its large errors  is consistent with the SM
expectations of~\cite{Tanaka:2010se}
\begin{align}
P_{\tau}(D^*)_{\rm SM} \ = \ -0.497 \pm 0.013 \, .
\end{align}

The fact that the experimentally observed $q^2$ distribution in the semileptonic $ B \to D^{(*)}$ decays supports a small non-vanishing value, $C_{V_L}\approx 0.08$ is also very significant for our RPV3 BSM scenario. One can see from
Eq.~\eqref{Eq.lambda_prime} that as long as only the $LQD$ interactions are relevant, in RPV3 the dimension-6 effective interaction for the semileptonic decays is essentially identical to the $(V - A)\times (V -A)$ structure of the SM effective Hamiltonian with the difference being just in the overall coefficient. Whereas in the SM the overall coefficient is $G_F \times V_{cb}/\sqrt2$, RPV3 has the overall coefficient $\lambda^{\prime}\times \lambda^{\prime}/m_{\widetilde b}^2$. Thus the coefficient, $C_{V_L} \approx 0.08$ is consistent with $m_{\widetilde b} \approx 2 \,\rm TeV$ for $\lambda'\lesssim 0.5$, as we will explicitly see below in the numerical fits. 

\subsubsection{Bino Contribution}

There is an additional contribution to the $B \to D^{(*)} \ell \nu$ decays that can arise in our RPV scenario.
If the bino, $\widetilde{\chi}_1^0$, is extremely light and has a very long lifetime (as motivated by an explanation of the ANITA anomaly, see Section~\ref{ANITAdetail} below), then the decays $B \to D^{(*)} \ell {\chi}$ can be open and mimic the $B \to D^{(*)} \ell \nu$ decays. In this case, we could have the $B \to D^{(*)} \ell {\chi}$ processes via either left-handed stau or right-handed sbottom exchange which effectively give contributions to operators of the form $(\bar c P_R b)(\bar \tau P_R \chi)$ and $(\bar c \sigma_{\mu\nu} P_R b)(\bar \tau \sigma^{\mu\nu} P_R \chi)$. Details are given in Appendix~\ref{app:bino}. Evaluating these contributions, we find that the effect is rather small: ${\rm BR}(B\to D \tau \chi)/{\rm BR}(B\to D \tau \nu)_{\rm SM} \lesssim 1\%$ and thus this extra channel cannot significantly affect $R_{D^{(*)}}$. Note that an adequate analysis of sizable contributions from operators beyond $(V - A)\times (V -A)$ might require more involved tools~\cite{Bernlochner:2020tfi}.

We also did a similar analysis with regard to the possible contribution from the extra bino-channel to the longitudinal polarization fraction $F_L({D^*})$ of $B \to D^{*} \ell \nu$.\footnote{There is no correction to $F_L({D^*})$ from the RPV contribution to $B \to D^{(*)} \ell \nu$ as shown in Fig.~\ref{fig:RDdiagram}(a) due to the fact that the corresponding BSM operator has the same structure as the SM operator.} 
We do expect a non-zero correction to $F_L({D^*})$ coming from the extra Bino channel because of the different operators that are involved. However, we find that the effect is tiny $\Delta F_L({D^*}) \lesssim 8\times 10^{-5}$ which is not significant given the large uncertainties in the current experimental value [cf.~Eq.~\eqref{eq:D_pol}] and  the SM value [cf.~Eq.~\eqref{eq:D_pol_SM}]. 

%%%%%%%%%%end of section on q^2 and polarization%%%%%%%%%%%%%%%%%%%%%%%%%%

%%%%%%%%%%%%%%%%%%%%%%%%%%%%%%%%%%%%%%%
\subsection{Explanation of \texorpdfstring{$R_K$ and $R_{K^*}$}{}}
%%%%%%%%%%%%%%%%%%%%%%%%%%%%%%%%%%%%%%%
%%%%%%%%%%%%%%%%%%%%%%%%%%%%%%%%%%%%%%%%%%%%%%%%%%%%%%%%%%%%%%
The BSM contributions to the rare decays $B\to K \mu^+\mu^-$ and $B\to K^* \mu^+\mu^-$ are conveniently described by shifts in the Wilson coefficients of semileptonic four-fermion operators in the effective Hamiltonian~\cite{Aebischer:2019mlg}
\begin{equation}
    \mathcal H_\text{eff} \ = \ - \frac{4 G_F}{\sqrt{2}} V_{ts}^* V_{tb} \frac{e^2}{16\pi^2} \sum_{i = 9,10}\Big[ (C_i)^\ell (Q_i)^\ell + (C_i^\prime)^\ell (Q_i^\prime)^\ell \Big]  
    \label{eq:Heff}
\end{equation}
with the operators
\begin{align}
 (Q_9)^\ell \ & = \ (\bar s \gamma_\alpha P_L b)(\bar \ell \gamma^\alpha \ell) \,, \label{eq:Q9}\\
 (Q_{10})^\ell \ & = \ (\bar s \gamma_\alpha P_L b)(\bar \ell \gamma^\alpha \gamma_5 \ell) \, , \label{eq:Q10} 
\end{align}
and $Q_{9,10}^\prime$ are obtained from $Q_{9,10}$ by replacing $P_L \to P_R$. Recall that in the SM, the Wilson coefficients are 
\begin{equation}
    (C_9)^\ell \ \simeq \ -(C_{10})^\ell \ \simeq \  4~,\qquad (C_9^\prime)^\ell \ \simeq \  (C_{10}^\prime)^\ell \ \simeq \ 0~,
\end{equation}
universally for all $\ell = e, \mu, \tau$.
Fits of $R_K$ and $R_{K^*}$ show that the observed pattern can be accommodated with BSM in the coefficients $(C_9)^e$, $(C_{10})^e$, $(C_9)^\mu$, $(C_{10})^\mu$, as long as BSM in the primed coefficients is subdominant, otherwise it leads to an anti-correlated effect in $R_K$ and $R_{K^*}$, contradicting the current data. 

Global fits of all relevant data on rare $B$ decays find a particular consistent BSM picture which is characterized by non-standard effects in muonic coefficients in the combination of Wilson coefficients $(C_9)^\mu = -(C_{10})^\mu$~\cite{Aebischer:2019mlg} (see also~\cite{Alguero:2019ptt, Alok:2019ufo, Ciuchini:2019usw, Kowalska:2019ley, Arbey:2019duh}).
As we will see below, our RPV SUSY scenario will generate contributions to both $(C_9)^\mu = -(C_{10})^\mu$ and $(C_9^\prime)^\mu = -(C_{10}^\prime)^\mu$. Such a scenario provides an excellent fit to the data for the following values~\cite{Aebischer:2019mlg}
\begin{eqnarray}
    && (C_9)^e \ \simeq \ (C_{10})^e \ \simeq \  (C_9^\prime)^e \ \simeq \ (C_{10}^\prime)^e \ \simeq \ 0 ~, \\
    && (C_9)^\mu \ = \ -(C_{10})^\mu \ \simeq \ -0.55 \pm 0.10 ~, \label{eq:unprimed} \\ 
    && (C_9^\prime)^\mu \ = \ -(C_{10}^\prime)^\mu \ \simeq \ 0.20 \pm 0.11 ~. \label{eq:prime}
\end{eqnarray}
Note that the combination $(C_9)^\mu \simeq -(C_{10})^\mu$ corresponds to BSM that mainly affects left-handed muons. All other coefficients are compatible with zero at the 2$\sigma$ level.
The correction to the SM values of the Wilson coefficients $C_9^\text{SM} \simeq - C_{10}^\text{SM} \simeq 4$ is at the level of $-15\%$ for the muon flavor, while for the electron flavor the corrections vanish. 
The above BSM values for the coefficients explain not only the observed values for $R_K$ and $R_{K^*}$, but also other (theoretically less clean) anomalies in rare $B$ decays, like the angular observable $P_5^\prime$ or the branching ratio of $B_s \to \phi \mu\mu$ (see Refs.~\cite{Aebischer:2019mlg, Alguero:2019ptt, Alok:2019ufo, Ciuchini:2019usw, Kowalska:2019ley, Arbey:2019duh}). 

Note that in our RPV setup the simultaneous presence of muon and electron couplings would likely lead to extremely stringent constraints from searches for $\mu \to e$ transitions, like the $\mu \to e \gamma$ decay, or $\mu \to e$ conversion in nuclei~\cite{KKS2014}. We therefore focus on muonic couplings only.

%%%%%%%%%%%%%
\begin{figure}[t!]
	\begin{subfigure}{.23\textwidth}
		\centering
		\includegraphics[width=1\linewidth]{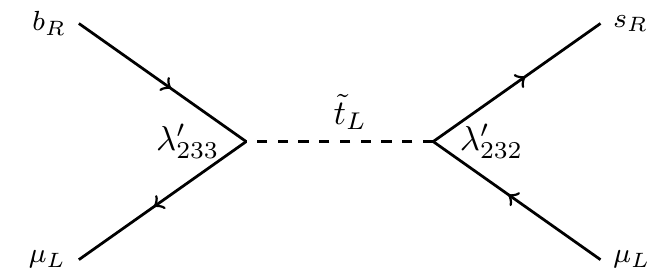}  
    	\caption{}
	    \label{fig:4-1}
	\end{subfigure}
	\begin{subfigure}{.23\textwidth}
		\centering
		\includegraphics[width=1\linewidth]{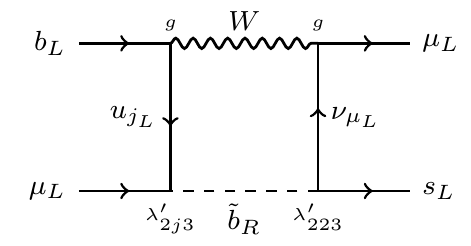}  
    	\caption{}
	    \label{fig:4-2}
	\end{subfigure}
		\begin{subfigure}{.23\textwidth}
		\centering
		\includegraphics[width=1\linewidth]{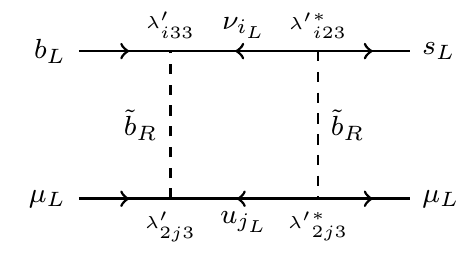}  
    	\caption{}
	    \label{fig:4-3}
	\end{subfigure}
		\begin{subfigure}{.23\textwidth}
		\centering
		\includegraphics[width=1\linewidth]{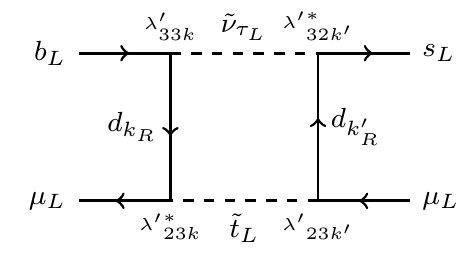}  
    	\caption{}
	    \label{fig:4-4}
	\end{subfigure}
		\begin{subfigure}{.23\textwidth}
		\centering
		\includegraphics[width=1\linewidth]{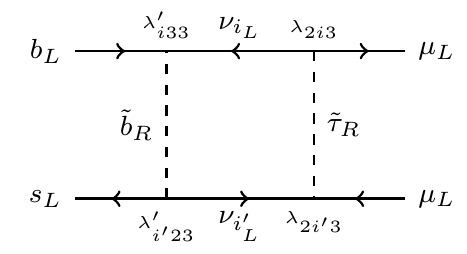}  
    	\caption{}
	    \label{fig:4-5}
	\end{subfigure}
	\begin{subfigure}{.23\textwidth}
		\centering
		\includegraphics[width=1\linewidth]{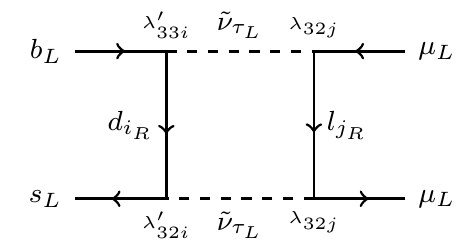}  
		\caption{}
		\label{fig:4-6}
	\end{subfigure}
	\caption{Different classes of contribution to the $b \to s \mu^+\mu^-$ transition in RPV SUSY: (a) tree level stop exchange; (b) sbottom-$W$ boson loop; (c) sbottom loop; (d) stop-sneutrino loop; (e) sbottom-stau loop; and (f) sneutrino loop. }
	\label{fig:C9C10diagrams}
\end{figure}
%%%%%%%%%%%%%%%%%%%%%%%%%%%%%%%%%%%%%%%%%%%%%%%%%%

In the considered RPV scenario, contributions to $b \to s \ell \ell$ transitions arise both at the tree level and the loop level.
Tree-level exchange of stops (see Fig.~\ref{fig:C9C10diagrams}a) gives contributions to the wrong chirality Wilson coefficients. In agreement with Ref.~\cite{Das:2017kfo} we find
\begin{equation}
 (C_9^\prime)^\mu \ = \  - (C_{10}^\prime)^\mu = - \frac{v^2}{2m_{\widetilde t_L}^2} \frac{\pi}{\alpha_\text{em}} \frac{\lambda^\prime_{233} \lambda^\prime_{232}}{V_{tb} V_{ts}^*} ~,
\end{equation}
where $\alpha_\text{em}$ is the fine structure constant. 
The above-discussed preferred ranges for these coefficients in Eq.~(\ref{eq:prime}) translate into the approximate bound 
\begin{equation}
\left| \lambda^\prime_{233} \lambda^\prime_{232} \right| \ \lesssim  \ 10^{-3} \times \left( \frac{m_{\widetilde t_L}}{1\,\text{TeV}} \right)^2 ~. \label{eq:primebound}
\end{equation}
In addition, there are various classes of 1-loop contributions to the $b \to s \mu \mu$ decays that we consider (see Fig.~\ref{fig:C9C10diagrams}b-f). There are loops with right-handed sbottoms and $W$ bosons (Fig.~\ref{fig:C9C10diagrams}b), with two right-handed sbottoms (Fig.~\ref{fig:C9C10diagrams}c), as well as with stops and sneutrinos (Fig.~\ref{fig:C9C10diagrams}d).\footnote{We neglect diagrams from loops involving winos that were discussed in Ref.~\cite{Earl:2018snx}, assuming that winos are sufficiently heavy in our RPV3 scenario. Note that this does not necessarily spoil the gauge coupling unification in RPV3~\cite{Altmannshofer:2017poe}, as the renormalization group (RG) running is logarithmic, and ${\cal O}(10~{\rm TeV})$ winos (and similar mass for the gluino to satisfy the stringent LHC constraints), along with light bino (and Higgsinos), are acceptable.} These contributions are all governed by the $\lambda^\prime$ RPV couplings. In the presence of the $\lambda$ RPV couplings there are additional 1-loop effects (as first pointed out by Ref.~\cite{Trifinopoulos:2018rna}). We take into account loops with right-handed sbottoms and staus (Fig.~\ref{fig:C9C10diagrams}e), as well as with left-handed sneutrinos (Fig.~\ref{fig:C9C10diagrams}f). All those diagrams give contributions to the left-handed Wilson coefficients and therefore can in principle explain the anomalies in $R_K$ and $R_{K^*}$. Summing up all these contributions we get~\cite{Das:2017kfo,Earl:2018snx,Trifinopoulos:2018rna}
%
%%%%%%%
%\blue{It seems from \{Das:2017kfo,Earl:2018snx\} that the third term and the second term should have same combination of lambda primes, and their sign should be different when written in this way (the red plus sign should be minus).}
\begin{eqnarray} \label{eq:C9C10} 
(C_9)^\mu & \ = \ & - (C_{10})^\mu \ = \  \frac{m_t^2}{m_{\widetilde b_R}^2} \frac{|\lambda'_{233}|^2}{16\pi\alpha_\text{em}} - \frac{v^2}{16 m_{\widetilde b_R}^2} \frac{X_{bs} X_{\mu\mu}}{e^2 V_{tb} V_{ts}^*}  \nonumber \\
&& -\frac{v^2}{16 ( m_{\widetilde t_L}^2-m_{\widetilde \nu_\tau}^2)} \log\left(\frac{m_{\widetilde t_L}^2}{m_{\widetilde \nu_\tau}^2}\right) \frac{X_{b\mu} X_{s\mu}}{e^2 V_{tb} V_{ts}^*} \nonumber\\
&& -\frac{v^2}{16 ( m_{\widetilde b_R}^2-m_{\widetilde \tau_R}^2)} \log\left(\frac{m_{\widetilde b_R}^2}{m_{\widetilde \tau_R}^2}\right) \frac{\widetilde X_{b\mu} \widetilde X_{s\mu}}{e^2 V_{tb} V_{ts}^*} \nonumber\\
&& -\frac{v^2}{16 m_{\widetilde \nu_\tau}^2} \frac{\widetilde X_{bs} \widetilde X_{\mu\mu}}{e^2 V_{tb} V_{ts}^*} ~,
\end{eqnarray}
where the $X$ and $\widetilde X$ factors are the following combinations of RPV couplings:
\begin{eqnarray}
X_{bs} & \ = \ & \lambda^\prime_{133} \lambda^\prime_{123} + \lambda^\prime_{233} \lambda^\prime_{223} + \lambda^\prime_{333} \lambda^\prime_{323} ~, \nonumber \\
\widetilde X_{bs} & \ = \ & \lambda^\prime_{331} \lambda^\prime_{321} +\lambda^\prime_{332} \lambda^\prime_{322} +\lambda^\prime_{333} \lambda^\prime_{323} ~, \nonumber \\
X_{\mu\mu} & \ = \ & |\lambda^\prime_{213}|^2 + |\lambda^\prime_{223}|^2 + |\lambda^\prime_{233}|^2  ~, \nonumber \\
\widetilde X_{\mu\mu} & \ = \ & |\lambda_{231}|^2 + |\lambda_{232}|^2 + |\lambda_{233}|^2  ~, \nonumber \\
X_{b\mu} & \ = \ & \lambda^\prime_{331} \lambda^\prime_{231} + \lambda^\prime_{332} \lambda^\prime_{232} + \lambda^\prime_{333} \lambda^\prime_{233} ~, \nonumber \\
X_{s\mu} & \ = \ & \lambda^\prime_{321} \lambda^\prime_{231} +\lambda^\prime_{322} \lambda^\prime_{232} +\lambda^\prime_{323} \lambda^\prime_{233}~, \nonumber \\
\widetilde X_{b\mu} & \ = \ & \lambda^\prime_{133} \lambda_{123} + \lambda^\prime_{333} \lambda_{323} ~, \nonumber \\
\widetilde X_{s\mu} & \ = \ & \lambda^\prime_{123} \lambda_{123} +\lambda^\prime_{323} \lambda_{323}~. 
\end{eqnarray}
It is intriguing that the RPV setup produces BSM contributions that follow the $(C_9)^\mu = -(C_{10})^\mu$ pattern that is preferred by the data. Note that the first term in~(\ref{eq:C9C10}) arises from the sbottom-$W$ boxes and has the wrong sign, {\it i.e.} it always worsens the agreement with data. The coupling combinations that enter in the other terms are constrained for example by $B_s$ mixing and $B \to K \nu\bar\nu$. The last two terms in  (\ref{eq:C9C10}) involve both the $\lambda^\prime$ and $\lambda$ couplings (the last one was not included in Ref.~\cite{Trifinopoulos:2018rna}).
These additional terms provide more freedom to explain the $R_{K^{(*)}}$ anomalies in the context of RPV SUSY. An explanation of the anomalies requires negative $C_9$. Given that $V_{ts} \simeq -0.04$, this in turn requires some of the $\lambda^\prime$ or $\lambda$ couplings to be negative.

Finally, let us also mention that in our RPV setup there are contributions to the related $b \to s \gamma$ decay. The constraints from $b \to s \gamma$ are discussed in Section \ref{sec:bsgamma}, where we show that they only lead to weak bounds on the RPV3 parameter space considered here.

\begin{figure}[t!]
	\begin{subfigure}{.23\textwidth}
		\centering
		\includegraphics[width=1\linewidth]{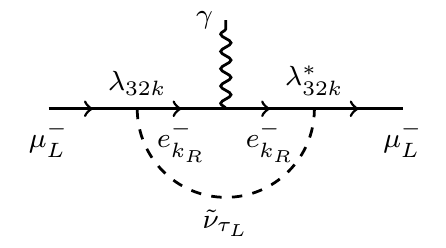}  
    	\caption{}
	    \label{fig:gm2a}
	\end{subfigure}
	\begin{subfigure}{.23\textwidth}
		\centering
		\includegraphics[width=1\linewidth]{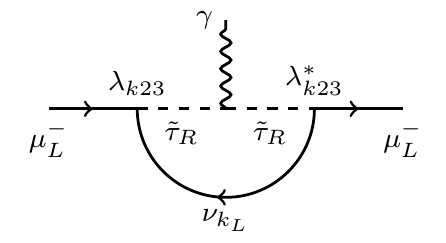}  
    	\caption{}
	    \label{fig:gm2b}
	\end{subfigure}
		\begin{subfigure}{.23\textwidth}
		\centering
		\includegraphics[width=1\linewidth]{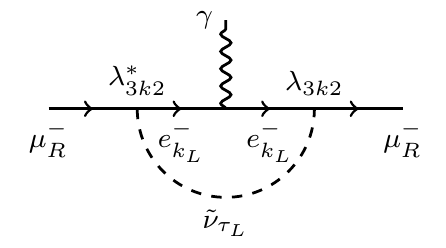}  
    	\caption{}
	    \label{fig:gm2c}
	\end{subfigure}
		\begin{subfigure}{.23\textwidth}
		\centering
		\includegraphics[width=1\linewidth]{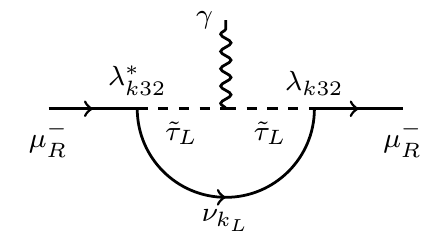}  
    	\caption{}
	    \label{fig:gm2d}
	\end{subfigure}
	\begin{subfigure}{.23\textwidth}
		\centering
		\includegraphics[width=1\linewidth]{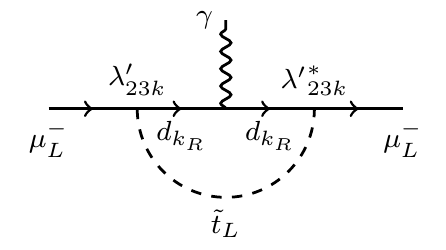}  
    	\caption{}
	    \label{fig:gm2alp}
	\end{subfigure}
	\begin{subfigure}{.23\textwidth}
		\centering
		\includegraphics[width=1\linewidth]{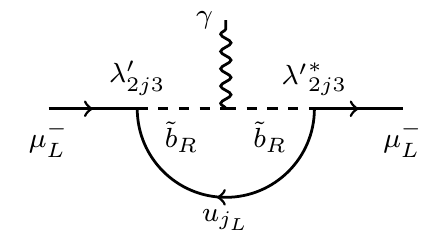}  
    	\caption{}
	    \label{fig:gm2blp}
	\end{subfigure}
		\begin{subfigure}{.23\textwidth}
		\centering
		\includegraphics[width=1\linewidth]{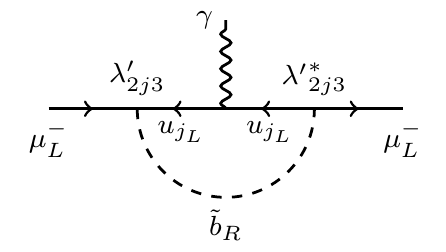}  
    	\caption{}
	    \label{fig:gm2clp}
	\end{subfigure}
		\begin{subfigure}{.23\textwidth}
		\centering
		\includegraphics[width=1\linewidth]{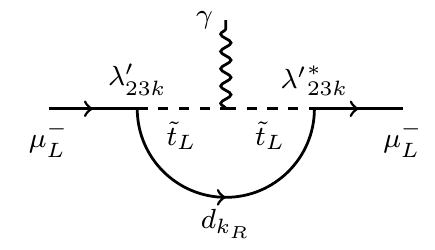}  
    	\caption{}
	    \label{fig:gm2dlp}
	\end{subfigure}
	\caption{Contribution to the $(g-2)_{\mu}$ from $\lambda$ (subfigures a--d) and $\lambda'$ (subfigures e--h) couplings in RPV SUSY.
	}
	\label{fig:muongm2}
\end{figure}

%%%%%%%%%%%%%%%%%%%%%%%%%%%%%%%%%%%%%%%
\subsection{Explanation of \texorpdfstring{$(g-2)_\mu$}{}}
%%%%%%%%%%%%%%%%%%%%%%%%%%%%%%%%%%%%%%%

The contributions to $(g-2)_{\mu}$ can arise in RPV SUSY both from the $\lambda$ and $\lambda^\prime$ couplings. The diagrams involving $\lambda$ are shown in Figs.~\ref{fig:muongm2}a-d and those involving $\lambda'$ (with sleptons and leptons in the loop switched to squarks and quarks) are shown in Figs.~\ref{fig:muongm2}e-h. In our RPV3 setup, these contributions can be summarized as~\cite{Kim:2001se} 
\begin{align}
    \Delta a_{\mu} & \ = \ \frac{m_{\mu}^2}{96\pi^2}\sum_{k=1}^3 \left(\frac{2(|\lambda_{32k}|^2+|\lambda_{3k2}|^2)}{m^2_{\widetilde{\nu}_{\tau}}}\right.\nonumber \\
    & \qquad \left.-\frac{|\lambda_{3k2}|^2}{m^2_{\widetilde{\tau}_{L}}}-\frac{|\lambda_{k23}|^2}{m^2_{\widetilde{\tau}_{R}}}+\frac{3|\lambda'_{2k3}|^2}{m_{\widetilde{b}_R}^2} \right)\label{eq:gm2l} \, .
%    \Delta a_{\mu}^{\lambda'}& \ = \ \frac{m_{\mu}^2}{32\pi^2}\sum_{k=1}^3 \frac{|\lambda'_{2k3}|^2}{m_{\widetilde{b}_R}^2} ~.\label{eq:gm2lp}
\end{align}
We find that the net contribution from the $\lambda$-dependent terms is typically dominant, as the relevant $\lambda$ couplings tend to be less constrained than the $\lambda^\prime$ couplings (cf.~Table~\ref{tab:constraintSum}).

It is worth noting here that the electron $g-2$ also has a $\sim2.4\sigma$ discrepancy between the experimental measurement~\cite{Hanneke:2010au} and SM prediction~\cite{Aoyama:2014sxa}, due to a new measurement of the fine structure constant~\cite{Parker:2018vye}: 
\begin{align}
    \Delta a_e \ = \ (-8.7\pm 3.6)\times 10^{-13} \, .
\end{align}
It is difficult to explain the opposite sign with respect to $\Delta a_\mu$ using RPV couplings only. However, within the minimal supersymmetric SM (MSSM), it is possible to explain $\Delta a_e$ by either introducing explicit lepton flavor violation~\cite{Dutta:2018fge} or using threshold corrections to the lepton Yukawa couplings~\cite{Endo:2019bcj} or arranging the bino-slepton and chargino-sneutrino contributions differently between the electron and muon sectors~\cite{Badziak:2019gaf}. Since this is independent of the RPV sector, we do not include the electron $(g-2)$ in our subsequent discussion. 

%%%%%%%%%%%%%%%%%%%%%%%%%%%%%%%%%%%%%%%%%%%%%%%%%%%%%%%%%%%%%%%%%%%%%%%%%%%%%%%%%
\subsection{Explanation of ANITA Upgoing Events}\label{ANITAdetail}
\begin{figure}[t!]
	\begin{subfigure}{.23\textwidth}
		\centering
		\includegraphics[width=1\linewidth]{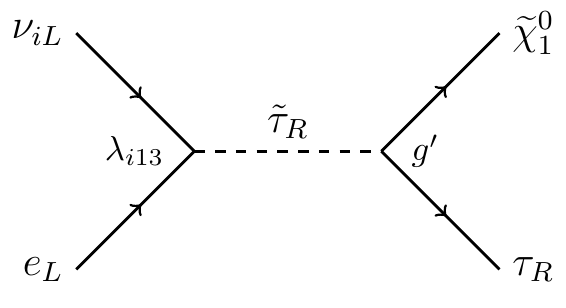}  
    	\caption{}
	    
	\end{subfigure}
	\begin{subfigure}{.23\textwidth}
		\centering
		\includegraphics[width=1\linewidth]{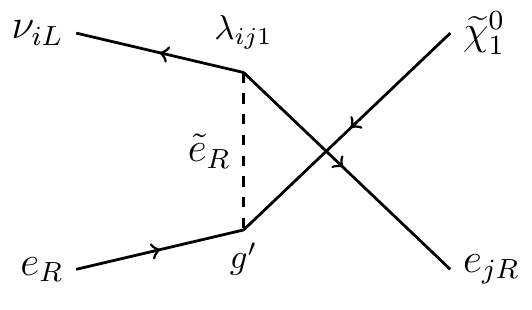}  
    	\caption{}
	    
	\end{subfigure}
	\begin{subfigure}{.23\textwidth}
		\centering
		\includegraphics[width=1\linewidth]{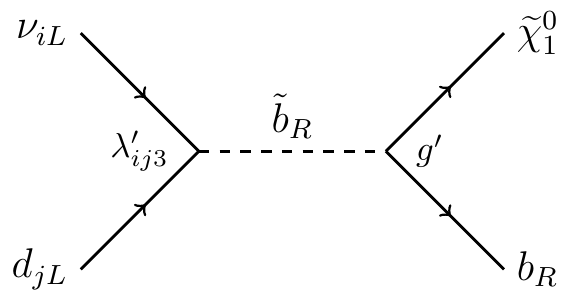}  
    	\caption{}
	    
	\end{subfigure}
	\begin{subfigure}{.23\textwidth}
		\centering
		\includegraphics[width=1\linewidth]{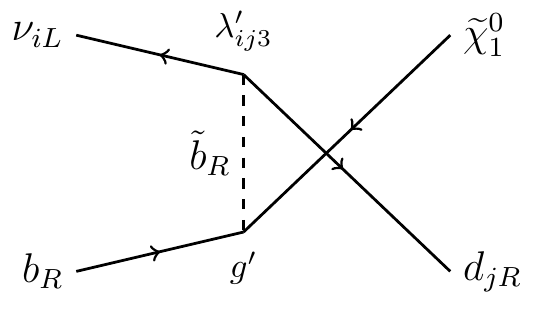}  
    	\caption{}
	    
	\end{subfigure}
	\caption{Feynman diagrams for production of bino from UHE neutrino interaction with quarks. (a) and (b) involve the $\lambda$ couplings, while (c) and (d) involve the $\lambda'$ couplings. The $s$-channel processes (a) and (c) give the dominant contribution at resonance energies. The decay of bino can be described by reversing the diagrams with the same interactions. We can ignore the process (b) with selectron propagator in our RPV3 framework.}
	\label{fig:ANITA_diagram}
\end{figure}

We interpret the ANITA upgoing anomalous events \cite{Gorham:2016zah, Gorham:2018ydl} as signals from the decay of long-lived bino in RPV SUSY, produced by interactions between UHE neutrinos and nucleons/electrons inside Earth matter via exchange of a TeV-scale sparticle mediator. As first discussed in Ref.~\cite{Collins:2018jpg}, the whole process could be divided into four sub-processes, namely, the generation of the bino on the far-side of Earth, its propagation through Earth matter, followed by its decay in the atmosphere and signal detection at ANITA. The generation and the decay of bino could both be described by Fig.~\ref{fig:ANITA_diagram} with one of the vertices coming from either $\lambda$ or $\lambda'$ sector, while the other being $U(1)_Y$ gauge coupling $g'$. The contribution from the $\lambda$ sector involving the $\nu-e$ interactions turns out to be sub-dominant in our case due to the choice of small $\lambda_{i13}$ and the lower probability to have an $s$-channel resonance for $\nu-e$ interactions as compared to $\nu-q$ interactions, since all three down-type quark PDFs are sizable at EeV energies~\cite{Collins:2018jpg}.
%%%%%%%%%%%%%%%%%%%%%%%%%%%%%%%%%%%%%%%%%%%%%%%%%%%%%%%%%%%%%%%%
% \begin{figure}[t!]
% 	\centering
% 	\includegraphics[width=1\linewidth]{ANITA_bino_diagram_v2.pdf}  
% 	\caption{Feynman diagrams for production of bino from UHE neutrino interaction with quarks. (a) and (b) involve the $\lambda$ couplings, while (c) and (d) involve the $\lambda'$ couplings. The $s$-channel processes (a) and (c) give the dominant contribution at resonance energies. The decay of bino can be described by reversing the diagrams with the same interactions. We can ignore the process (b) with selectron propagator in our RPV3 framework.}
% 	\label{fig:ANITA_diagram}
% \end{figure}
%%%%%%%%%%%%%%%%%%%%%%%%%%%%%%%%%%%%%%%%%%%%%%%%%%%%%%%%%%%%%

After the bino is generated, it is required to have a long lifetime to travel through a chord length of $\sim 5000$ km, as inferred from the ANITA events. The decay width of bino is parameterized by its mass $m_{\widetilde{\chi}_1^0}$, the mediator sbottom or stau mass and the $\lambda$, $\lambda'$, $g'$ couplings  as:
\begin{equation}
    \Gamma(\widetilde{\chi}_1^0) \ \simeq \ \frac{g'^2 m_{\widetilde{\chi}_1^0}^5}{512\pi^3}\sum_{ij}\left[\frac{3|\lambda'_{ij3}|^2}{m_{\widetilde{b}_{R}}^4}+\frac{|\lambda_{ij3}|^2}{m_{\widetilde{\tau}_{R}}^4}\right] \, .
    \label{eq:binodecay}
\end{equation}
As mentioned above, the $\lambda$-contribution is subdominant and we will only keep the $\lambda'$ terms in Eq.~\eqref{eq:binodecay}. 
The longevity of the bino in our model comes from a combination of two effects: (i) It is electrically neutral and interacts with the nucleons in earth matter very weakly: $\sigma(\widetilde{\chi}_1^0\ q \to {\rm anything}) \lesssim 10^{-36} \rm cm^2$ at EeV energies. (ii) It is produced with a very high Lorentz boost factor of $\gamma\sim 10^6$. So as long as the bino has a mean lifetime of $\sim 10$ ns in its rest-frame, which translates to a lifetime $\sim 0.01$ s in the lab frame, it can safely propagate through a chord length of $\sim 5000$ km without losing much energy. From Eq.~\eqref{eq:binodecay}, we find that this happens for a relatively light bino with $m_{\widetilde{\chi}_1^0}\sim$ a few GeV. See Appendix~\ref{app:anita} for the variation of bino mean free path with energy. After propagating the chord length of a few thousand km, as it reaches near the surface of Earth, it undergoes a 3-body decay back to quarks (or leptons) and neutrinos, followed by hadronization of the quarks, producing an extensive air shower due to the Askaryan effect~\cite{Askaryan:1962hbi}. The radio signal from the air showers is then detected by the ANITA balloon detector.

The expected number of events can be estimated as follows~\cite{Collins:2018jpg}: 
\begin{equation}\label{eventNest}
N  \ = \ \int d E_{\nu} \ \langle A_{\rm eff}\cdot\Delta\Omega \rangle \cdot T \cdot \Phi_{\nu} \, ,
\end{equation}
where we have taken $T = 53$ days for the total effective exposure time, $\Phi_{\nu}(E_\nu) = 2\times10^{-20} \rm{(GeV\cdot cm^2 \cdot s\cdot sr)^{-1}}$ for the cosmic neutrino flux,\footnote{This is consistent with the recent upper bound for transient sources, based on a joint analysis of ANITA detection and IceCube non-detection results~\cite{Aartsen:2020vir}. To be more specific, our transient anisotropic flux value $\Phi_{\nu}$ integrated over the small solid angle $\Delta \Omega$ corresponding to the uncertainty in the observed elevation angles for the ANITA events is $\Phi_{\rm int}= 4.9\times10^{-24} \rm{(GeV\cdot cm^2 \cdot s)^{-1}}$ at $0.5$ EeV, to be compared with the upper bound on $\Phi_{\rm int} \leq 8\times10^{-24} \rm{(GeV\cdot cm^2 \cdot s)^{-1}}$ for the steady analysis~\cite{Aartsen:2020vir}.} and $\langle A_{\rm eff}\cdot\Delta\Omega \rangle$ is the effective area integrated over the relevant solid angle, averaged over the probability for interaction and decay to happen over the specified geometry. The effective area contains all the information of the geometry, decay width of bino and the cross section for the bino generation process; see Ref.~\cite{Collins:2018jpg} for the explicit expression. From Eq.~\eqref{eventNest}, we know that the overall event number $N$ is a function of $m_{\widetilde{\chi}_1^0}$, $m_{\widetilde{b}_R}$ and $\lambda'_{ij3}$ for our RPV3 scenario. Therefore, comparing the simulated event numbers with the ANITA observation of two anomalous events gives us the best-fit parameter region at a given CL.

\section{Numerical Results} \label{sec:results}
%%%%%%%%%%%%%%%%%%%%%%%%%%%%%%%%

%%%%%%%%%%%%%%%%%%%%%%%%%%%%%%%%
\begin{table*}[t!]
\centering
\begin{tabular}{|c|c|c|}\hline\hline
		Observable & Parameter dependence & Relevant terms \\ \hline\hline
		$R_{D^{(*)}}$ & $\lambda'_{i33}, \lambda'_{3j3}, \lambda'_{2j3}, m_{\widetilde b_R}, $& $\frac{\lambda'_{i33}\cdot \lambda'_{3j3}\nonumber}{m_{\widetilde b_R}^2},\ 
		-\frac{\lambda'_{i33}\cdot \lambda'_{2j3}\nonumber}{m_{\widetilde b_R}^2}$, \\
		& $\lambda_{i33}$, $\lambda_{i32}$, $m_{\widetilde \tau_L}$ & $\frac{\lambda_{i33}\cdot \lambda'_{3j3}\nonumber}{m_{\widetilde \tau_L}^2},\ 
		\frac{\lambda_{i32}\cdot \lambda'_{3j3}\nonumber}{m_{\widetilde \tau_L}^2}
		$
		\\\hline \hline
		%%%%%%%%%%%%%%%%%%%%%%%%%%%%%%%%%%%%%%%%%%%%%%%%%%%%
		%%%%%%%%%%%%%%%%%%%%%%%%%%%%%%%%%%%%%%%%%%%%%%%%%%%
		 &   &
		$ \frac{|\lambda'_{233}|^2}{m_{\widetilde b_R}^2}$, \\
		%%%%%%%%%%%
		& $\lambda'_{331},\lambda'_{332},\lambda'_{321},\lambda'_{322},\lambda'_{231},\lambda'_{232},$ & $\frac{(\lambda'_{i33}\cdot \lambda'_{i23})|\lambda'_{2j3}|^2}{m_{\widetilde b_R}^2}$,\\
		%%%%%%%%%%%
		$R_{K^{(*)}}$ & $\lambda'_{i33},\lambda'_{i23},\lambda'_{213},\lambda'_{312},\lambda_{32k}, \lambda_{3j2}, $ & $\frac{\log\left({m_{\widetilde t_L}^2}/{m_{\widetilde \nu_\tau}^2}\right)}{ ( m_{\widetilde t_L}^2-m_{\widetilde \nu_\tau}^2)} (\lambda'_{33i}\cdot\lambda'_{32i})|\lambda'_{23i}|^2$,  \\
		%%%%%%%%%%%
		& $m_{\widetilde b_R}$, $m_{\widetilde t_L}, m_{\widetilde \tau_R} $ & $\frac{\log\left({m_{\widetilde b_R}^2}/{m_{\widetilde \tau_R}^2}\right)}{ ( m_{\widetilde b_R}^2-m_{\widetilde \tau_R}^2)} \lambda'_{i33}\lambda'_{i'23}\lambda_{2i3}\lambda_{2i'3}$, \\
		%%%%%%%%%%%
		& & $\frac{1}{m_{\widetilde \nu_{\tau}}^2} \lambda'_{33i}\lambda'_{3i2}\lambda_{32j}\lambda_{3j2} \nonumber$   \\\hline\hline
		%%%%%%%%%%%%%%%%%%%%%%%%%%%%%%%%%%%%%%%%%%%%%%%%%%%%%%%%%%%%%%%%%%%%%%%%%%%%%%%%%%%%%%%%%%%%%%%%%%%%%%%%%%%%%%%%%%
		 &   & 
		$|\lambda_{32k}|^2\frac{2}{m^2_{\widetilde{\nu}_{\tau}}}$, \\
		%%%%%%%%%%%%%%%%%
		& $\lambda_{32k}, \lambda_{3k2}, \lambda_{k23}$  &$|\lambda_{3k2}|^2\left(\frac{2}{m^2_{\widetilde{\nu}_{\tau}}}-\frac{1}{m^2_{\widetilde{\tau}_{L}}}\right)$,  \\
		$(g-2)_{\mu}$ & $\lambda'_{233}, \lambda'_{223}, \lambda'_{213}, $ &$-|\lambda_{k23}|^2\frac{1}{m^2_{\widetilde{\tau}_{R}}}$,  \\
		& $m_{\widetilde{b}_R}, m_{\widetilde{\tau}_{R}}, m_{\widetilde{\tau}_{L}}, m_{\widetilde{\nu}_{\tau}} $ &$\frac{|\lambda'_{233}|^2}{m_{\widetilde{b}_R}^2-m_t^2}$,  \\
		& &$\frac{1}{m_{\widetilde{b}_R}^2}(|\lambda'_{213}|^2+|\lambda'_{223}|^2)$ \\ \hline\hline
		%%%%%%%%%%%%%%%%%%%%%%%%%%%%%%%%%%%%%%%%%%%%%%%%
		%%%%%%%%%%%%%%%%%%%%%%%%%%%%%%%%%%%%%%%%%%%%%%%%%%%
		ANITA  & $\lambda'_{123},\lambda'_{223},\lambda'_{233},\lambda'_{323},\lambda'_{333},m_{\widetilde{b}_R}, m_{\widetilde{\chi}_1^0}$ & $\frac{|\lambda'_{ij3}|^2m^5_{\widetilde{\chi}_1^0}}{m_{\widetilde{b}_R}^4}$ \\
		\hline \hline
	\end{tabular}
	\caption{\label{tab:parameterSum} The parameter dependence and dominant terms in the expressions for the  $R_{D^{(*)}}$, $R_{K^{(*)}}$, $(g-2)_{\mu}$ and ANITA anomalies in our RPV3 scenario. }
\end{table*}

%\subsubsection{Parameter Set up}

After examining Eqs.~\eqref{eq:delta}, \eqref{eq:C9C10}, \eqref{eq:gm2l} and \eqref{eq:binodecay}, all the relevant parameters contributing to the anomalies discussed above in our RPV3 scenario are summarized in Table~\ref{tab:parameterSum}. For  convenience, we also collect the dominant terms in the expressions for anomalies in Table \ref{tab:parameterSum}. The same is done for the relevant experimental constraints in Table~\ref{tab:constraintSum} which we discuss in detail in the following  Section~\ref{sec:constraints}.

As mentioned before, in our RPV3 setup, there are six free mass parameters relevant for the anomalies, namely, 
\begin{align}
\{m_{\widetilde{b}_{R}},m_{\widetilde{t}_{L}}, m_{\widetilde{\tau}_{L}},m_{\widetilde{\tau}_{R}}, m_{\widetilde{\nu}_{\tau}}, m_{\widetilde{\chi}_1^0}\} \, .
\end{align}
As for the choice of RPV couplings shown in Table~\ref{tab:parameterSum}, we apply certain symmetry rules to reduce the number of parameters. We consider the following three different cases and present our numerical fit results in each case.\footnote{Other example structures of the RPV couplings using flavor symmetry can be found in Ref.~\cite{Barbier:2004ez}.} 

\subsection{Case 1: CKM-like Structure} \label{sec:CKM}
This symmetry is inspired by the observed hierarchy in the CKM mixing matrix in the quark sector. This is brought out most clearly in the Wolfenstein parameterization of the CKM-matrix~\cite{wolf83}, where the first generation plays the central role. The coupling of first-to-first generation quarks are of order one, whereas the coupling of the first to the second carries a suppression factor of $\lambda\simeq \sin \theta_{C} \approx 0.23$. Similarly, the coupling of second generation to the third carries a suppression of $\lambda^2$, and the coupling of first generation to the third carries a suppression factor of $\lambda^3$. Inspired by this structure, in our RPV scenario which is third-generation-centric, 
%role of the third generation in RPV and and the first generation in the SM gets interchanged. Thus coupling of third to second %generation carries an unknown suppression factor which we assume is $\lambda$. Coupling of third generation to first is taken to %go as $\lambda^3$, coupling of 2nd generation to first generation now goes as $\lambda^2$. 
we postulate the $\lambda'$-couplings to be of the form 
\begin{align}
\lambda'_{ijk} \ = \ \lambda'_{333}\, \epsilon^{(3-i)+(3-j)+(3-k)} \, ,
\end{align}
with $\lambda'_{333}\sim {\cal O}(1)$ and each time any of the three indices $\{i,j,k\}$ differs from 3, we pay an appropriate factor of $\epsilon$, which is a tunable small parameter in the model. A similar rule is applied to the $\lambda$ sector, where we choose for the nonzero $\lambda$'s:\footnote{Note that $\lambda_{ijk}$ vanishes for $i=j$ [cf.~Eq.~\eqref{Eq.RPVLLE}].} 
\begin{align}
\lambda_{ijk} \ = \ \lambda_{233}\, \epsilon^{(2-i)+(3-j)+(3-k)} \, ,
\end{align}
where $i<j$ and $\lambda_{233}\sim {\cal O}(1)$. This setup reduces the number of couplings from 27 ($\lambda'_{ijk}$)+9 ($\lambda_{ijk}$)=36 to only 3, namely, 
\begin{align}\label{eq:variableSetUp1}
    \{\lambda'_{333},\lambda_{233},\epsilon
\} \, .
\end{align}

%case1%%%%%%%%%%%%%%%%%%%%%%%%%%%%%%%
\begin{figure*}[th!]
	\centering
	\includegraphics[width=0.85\textwidth]{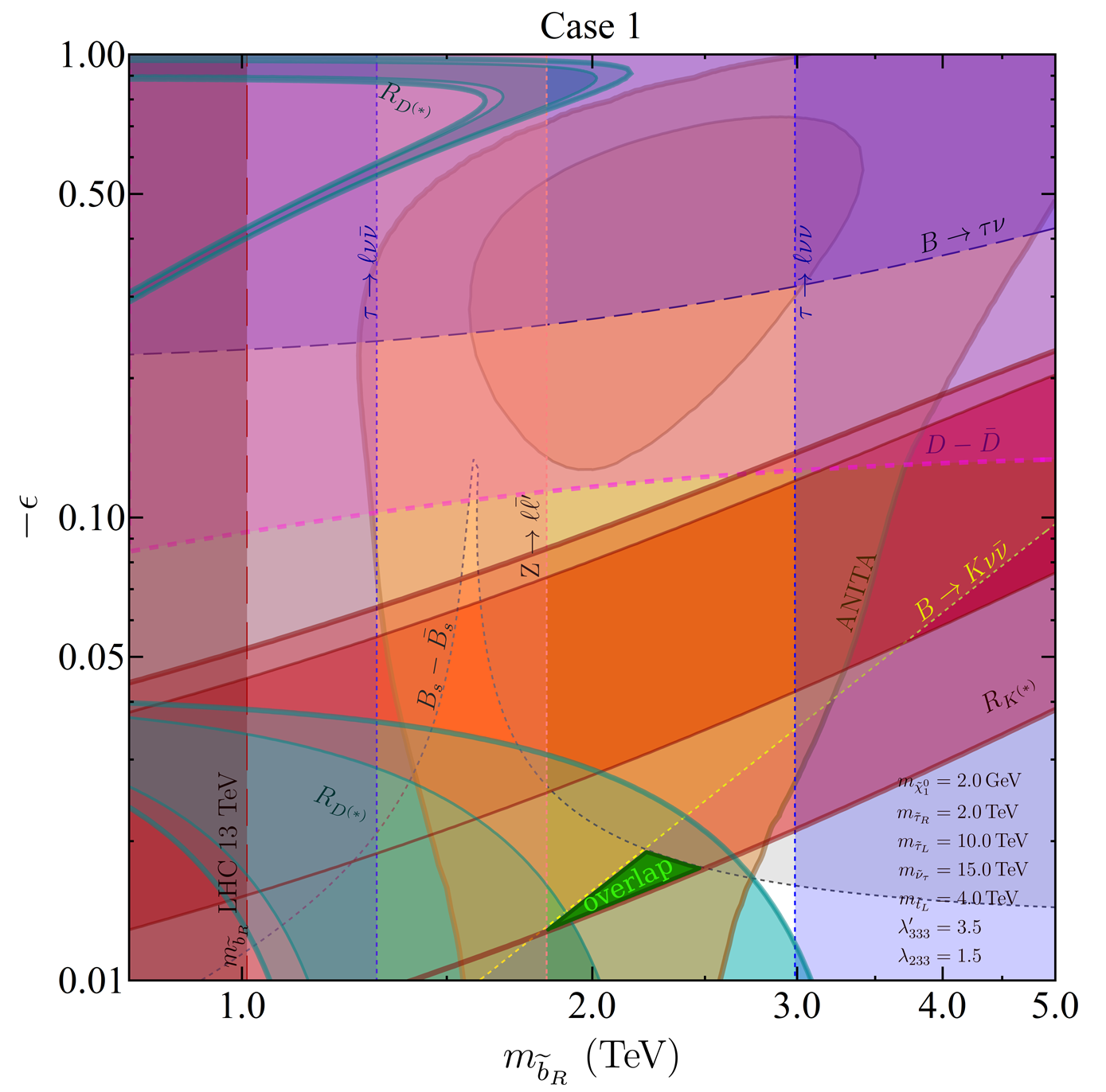}
	\caption{Benchmark scenario for Case 1 (with CKM-like symmetry) in the two-dimensional parameter plane $(m_{\widetilde b}, -\epsilon)$, while keeping other free parameters fixed as shown in the figure. The shaded regions with solid boundaries are the $2\sigma$ (thin) and $3\sigma$ (thick) favored regions to explain the $R_{D^{(*)}}$ (cyan), $R_{K^{(*)}}$ (red) and ANITA (orange) anomalies. The $(g-2)_\mu$ anomaly requires  $-\epsilon\sim {\cal O}(10)$, and therefore, not shown here. The shaded regions with dashed/dotted boundaries are the current experimental bounds on the parameter space from $B\to K\nu \bar\nu$ (yellow),  
	$B_s-\overline{B}_s$ mixing (grey),  
	%$D\to\mu\mu$,
	$D-\overline{D}$ mixing (magenta),
	$B\to\tau \nu$ (dark blue),
	$Z\to \ell \ell'$ (pink), and 
	$\tau\to \ell \nu \bar\nu$ (blue). The overlap region simultaneously explaining the $R_{D^{(*)}}$ and $R_{K^{(*)}}$ anomaly is shown by the green shaded region, and the region also explaining the ANITA anomaly along with $R_{D^{(*)}}$ and $R_{K^{(*)}}$ is shown by the green shaded region with thick boundaries.} \label{fig:ANITARdRkgm2CKM}
\end{figure*}
%%%%%%%%%%%%%%%%%%%%%%%%%%%%%%%%

In Fig.~\ref{fig:ANITARdRkgm2CKM}, we show a benchmark scenario for Case 1 in the $(m_{\widetilde{b}_{R}},\epsilon)$ plane, while fixing the other free parameters as follows:
\begin{align}
 &   \lambda'_{333} \ = \ 3.5\, , \quad \lambda_{233} \ = \ 1.5 \, , \nonumber \\
 &  m_{\widetilde{\tau}_{R}} \ = \ 2.0\ {\rm TeV}\, , \quad  m_{\widetilde{\tau}_{L}} \ = \ 10.0\ {\rm TeV} \, ,  \nonumber \\
& m_{\widetilde{\nu}_{\tau}} \ = \ 15.0\ {\rm TeV} \, , \quad 
m_{\widetilde{t}_{L}} \ = \ 4.0\ {\rm TeV} \, , \nonumber \\
& m_{\widetilde{\chi}_1^0}\ = \ 2.0\ {\rm GeV} \, . 
\label{eq:bp1}
\end{align}
The two coupling values are mainly chosen to simultaneously maximize the overlap region where the anomalies can be explained, as well as to evade the current existing bounds. A particularly stringent constraint comes from  $\tau\to\ell\nu\bar\nu$ (see Section~\ref{sec:taudecay}) which involves both $\lambda'_{333}$ and $\lambda_{233}$ couplings, and the masses of right-handed stau $m_{{\widetilde \tau}_R}$ and right-handed bottom,  $m_{{\widetilde b}_R}$. Thus we need to change $\lambda'_{333}$ and $\lambda_{233}$ together so that their overall effect mostly cancels to give a narrow allowed window from $\tau\to\ell\nu\bar\nu$. These two couplings are set as large as possible so that the cancellation takes place, and meanwhile gives a maximized overlap region as long as the other constraints do not become too strong.  
The masses chosen here are consistent with the 13 TeV LHC constraints~\cite{Tanabashi:2018oca}. The stau mass is chosen close to the experimental limit of 900 GeV to obtain the maximally allowed parameter space, while satisfying the bound from $\tau\to\ell\nu\bar\nu$, {\it i.e.} choosing a larger stau mass will shrink the available parameter space shown in Fig.~\ref{fig:ANITARdRkgm2CKM}, while a smaller stau mass will shrink the window of the allowed region from $\tau\to\ell\nu\bar\nu$. As for the choice of the sneutrino mass, from Table~\ref{tab:constraintSum} we could see that the term involving $m_{\widetilde{\nu}_{\tau}}$ contributes dominantly to the $B_s-\overline B_s$ bound and thus to alleviate this bound, we set $m_{\widetilde{\nu}_{\tau}}$ at a relatively larger value of 15 TeV. We choose $m_{\widetilde{\tau}_{L}}$ to be 10 TeV to suppress the possible contribution to $R_{D^{(*)}}$ from $LLE$ couplings. Also, $m_{\widetilde{t}_{L}}$ is set at 4 TeV to suppress the tree-level contribution to $b\to s\ell\ell$ as mentioned in Eq~\eqref{eq:primebound}.  

The favored regions for explaining the $R_{D^{(*)}}$, $R_{K^{(*)}}$ and ANITA anomalies are shown in Fig.~\ref{fig:ANITARdRkgm2CKM} by cyan, red and orange-shaded regions, with the $2\sigma$ and $3\sigma$ regions depicted by thin and thick solid contours respectively. The $\epsilon$ parameter is required to take negative values in order to find overlap between $R_{D^{(*)}}$ and $R_{K^{(*)}}$ regions. This is due to the fact that we need $C_9<0$ to fit the data [cf.~Eq.~\eqref{eq:unprimed}] and since $C_9$ is composed of odd powers of $\epsilon$ with positive definite factors [cf.~Eq.~\eqref{eq:C9C10}],  this inevitably sets $\epsilon$ negative. On the other hand, the $R_{D^{(*)}}$-favored regions are divided into two different branches due to the polynomial dependence of $\lambda'_{ijk}$ and $\lambda_{ijk}$ upon $\epsilon$ [cf.~Eq.~\eqref{eq:delta}]. As for the ANITA-favored region, it is mostly governed by the bino mass which is set at 2.0 GeV, apart from the sbottom mass and $\lambda'$ couplings. 

Other shaded regions in Fig.~\ref{fig:ANITARdRkgm2CKM} with dashed/dotted boundaries are the relevant experimental constraints; see Section~\ref{sec:constraints} and Table~\ref{tab:constraintSum} for details. The main constraints come from $B_s-\overline{B}_s$ mixing~\cite{Amhis:2019ckw} and $B\to K\nu\bar{\nu}$~\cite{Lees:2013kla,Buttazzo:2017ixm,Bordone:2018hqs} measurements. Note that the $B_s$-meson mixing bound has a branch-cut feature which is due to the cancellation between the terms in Eq.~\eqref{eq:bbmixing}. Somewhat less constraining bounds come from $B\to\tau\nu$~\cite{Amhis:2019ckw}, $D - \overline D$ mixing~\cite{Peng:2014oda}, $\tau\to\ell\nu\bar\nu$~\cite{Tanabashi:2018oca},  and $Z\to\ell\bar\ell'$ data~\cite{Tanabashi:2018oca}. Finally, the vertical shaded region below $m_{\widetilde{b}_{R}}<1.0$ TeV is excluded from direct sbottom searches at the LHC~\cite{Tanabashi:2018oca}. 

The overlap region between $R_{D^{(*)}}$, $R_{K^{(*)}}$ and ANITA is highlighted by the green shaded region in Fig.~\ref{fig:ANITARdRkgm2CKM}  around $(m_{\widetilde{b}_{R}},\epsilon)=(2.2\,{\rm TeV},\,-0.015)$. This is remarkable, given how simple the coupling choice is, even though it occurs only at the $3\sigma$ CL. However, a major drawback of this scenario is that the $(g-2)_{\mu}$-favored region lies around $-\epsilon \sim O(10)$, which is far away from our CKM-like assumption that $|\epsilon|\ll 1$; therefore, it is not shown in Fig.~\ref{fig:ANITARdRkgm2CKM}.

\subsection{Case 2: Flavor Symmetry} \label{sec:ST}
The second benchmark point we study is inspired by a $U(2)_q\times U(2)_\ell$ flavor symmetry proposed in Ref.~\cite{Trifinopoulos:2018rna}. In this case, the values of  $\lambda'_{ijk}$ and $\lambda_{ijk}$ couplings are decided by the specific flavon VEVs in the model. They have the generic structure $\lambda'_{ijk}\sim c'_{ijk}\epsilon'$ and $\lambda_{ijk}\sim c_{ijk}\epsilon$, where the $\epsilon'$ and $\epsilon$ values may differ for each coupling, while $c'_{ijk}$ and $c_{ijk}$ are $O(1)$ free parameters. Here we choose a simplified version of this model and assume that $c'_{ijk}$ and $c_{ijk}$ are strictly equal to the overall scales of $\lambda'$ and $\lambda$ respectively, {\it i.e.} $\lambda'_{ijk} \sim \lambda' \epsilon'$ and $\lambda_{ijk} \sim \lambda \epsilon$ with $\epsilon'$ and  $\epsilon$ fixed by the flavor structure parameters as indicated in Ref.~\cite{Trifinopoulos:2018rna}. Moreover, to accommodate $ R_{K^{(*)}}$, we choose $\lambda'_{333}$ to be negative and set it as a free parameter to be fit numerically. All other $\lambda'_{ijk}$ values are fixed by the overall scale $\lambda'$, {\it i.e.}
\begin{align}
   & \lambda'_{1jk} \ = \  \lambda'_{211} \ = \  \lambda'_{231} \ = \  \lambda'_{213} \nonumber \\
    & \qquad \ = \  \lambda'_{311} \ = \  \lambda'_{331} \ = \  \lambda'_{313} \ \simeq \ 0\, , \nonumber \\
    & \lambda'_{221} \ = \ \lambda'_{212} \ \simeq \ \lambda'\epsilon_\ell \epsilon'_q \, , \nonumber \\
     & \lambda'_{321} \ = \ \lambda'_{312} \ \simeq \  \lambda'\epsilon'_q \, , \nonumber \\
      & \lambda'_{222} \ = \ \lambda'_{223} \ = \ \lambda'_{232}\ \simeq \ \lambda'\epsilon_\ell \epsilon_q \, , \nonumber \\
      & \lambda'_{322} \ = \ \lambda'_{323} \ = \ \lambda'_{332}\ \simeq \ \lambda' \epsilon_q \, , \nonumber \\
     & \lambda'_{233} \ \simeq \ \lambda'\epsilon_\ell \, ,
     \label{eq:STlambp}
\end{align}
where $\epsilon_q\approx m_s/m_b\simeq 0.025$, $\epsilon'_q\approx \epsilon_q\sqrt{m_d/m_s}\simeq 0.005$ and $\epsilon_\ell\simeq 1$~\cite{Trifinopoulos:2018rna}. Similarly, all $\lambda_{ijk}$ values are fixed by the overall scale $\lambda$, {\it i.e.}
\begin{align}
& \lambda_{121} \ = \ \lambda_{131} \ = \ \lambda_{133} \ \simeq \ 0 \, , \nonumber \\
& \lambda_{123} \ = \ \lambda_{132} \ = \ \lambda_{231} \ \simeq \ \lambda \epsilon'_\ell \, , \nonumber \\
& \lambda_{232} \ \simeq \ \lambda \epsilon_{\ell S}\, , \quad 
\lambda_{122} \ \simeq \ \lambda \epsilon_\ell \epsilon'_\ell \, , \nonumber \\
& \lambda_{233} \ \simeq \ \lambda \epsilon_\ell \, , 
\end{align}
where $\epsilon'_\ell\simeq 0.004$ and $\epsilon_{\ell S}\simeq 0.06$~\cite{Trifinopoulos:2018rna}. Therefore, this choice is equivalent to taking 3 free parameters for the couplings, {\it i.e.} 
\begin{align}
\{\lambda'_{333}, \lambda', \lambda\} \, ,
\label{eq:variableSetUp2}
\end{align}
which is the same number of parameters as in Case 1 [cf.~Eq.~\eqref{eq:variableSetUp1}]. 
%%%%%%%%%%%%%%%%%%%%%%%%%%%%%%%%%%%%%%%%%%%%%%%%%%%%%%%%%%%

\begin{figure*}[tbh!]
	\centering
	\includegraphics[width=0.85\textwidth]{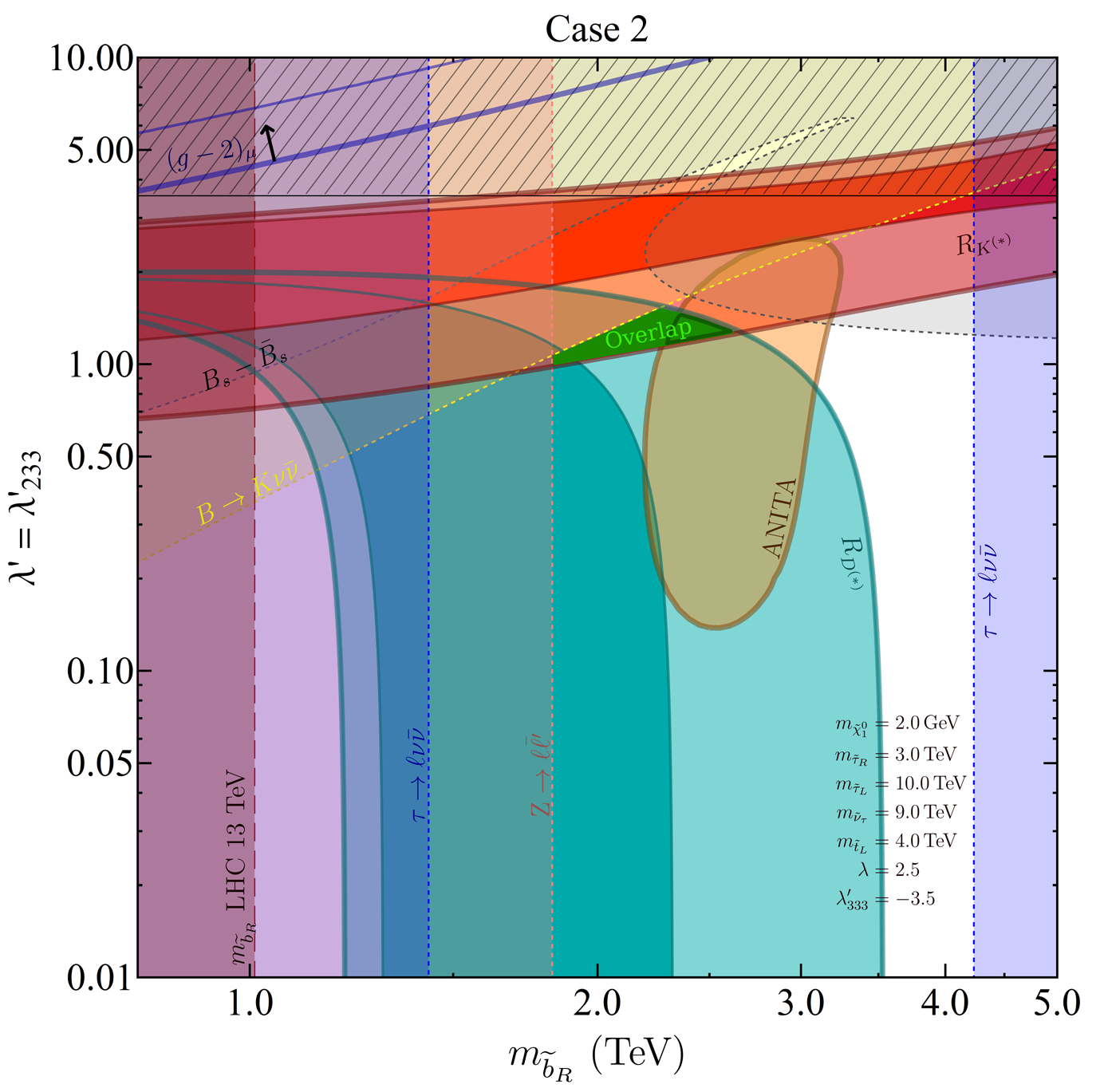}
	\caption{Benchmark scenario for Case 2 (with flavor symmetry) in the two-dimensional parameter plane $(m_{\widetilde b}, \lambda')$, while keeping other free parameters fixed as shown in the figure. The labels for the shaded regions are the same as in Fig.~\ref{fig:ANITARdRkgm2CKM}. The horizontal hatched region is theoretically disfavored from perturbativity constraint on $\lambda'\leq \sqrt{4\pi}$. The thin (thick) blue line at the upper left corner shows the $2\sigma$ ($3\sigma$) region favored by the $(g-2)_\mu$ anomaly.}
 \label{fig:ANITARdRkgm2ST}
\end{figure*}

In Fig.~\ref{fig:ANITARdRkgm2ST}, we show a benchmark scenario for Case 2 in the $(m_{{\widetilde b}_R}, \lambda')$ plane, while keeping the $m_{\widetilde{t}_{L}}$ and $m_{\widetilde{\chi}_1^0}$ fixed at the same values as in Case 1 [cf.~Eq.~\eqref{eq:bp1}, and the other five coupling parameters fixed at 
\begin{align}
&\lambda'_{333} \ = \ -3.5 \, , \quad 
\lambda = 2.5 \, \nonumber \\
&  m_{\widetilde{\tau}_{R}} \ = \ 3.0\ {\rm TeV} \, , \quad  m_{\widetilde{\tau}_{L}} \ = \ 10.0\ {\rm TeV} \, ,  \nonumber \\
& m_{\widetilde{\nu}_{\tau}} \ = \ 9.0\ {\rm TeV}  \, .
\label{eq:bp2}
\end{align}
The choice of the combination of $\lambda$, $\lambda'_{333}$, $m_{\widetilde{\tau}_{R}}$ and $m_{\widetilde{\tau}_{L}}$  is mainly due to the consideration of enlarging the overlapping region and avoiding current constraints. Larger magnitude of $\lambda$ and $\lambda'_{333}$ will push $R_{K^{(*)}}$ region downwards and $R_{D^{(*)}}$ upwards giving a larger overlap. However, both $B_s-\overline{B}_s$ mixing, and the $B\to K\nu\bar{\nu}$ and $\tau\to\ell\nu\bar\nu$ decays are sensitive to the choice of these four parameters (see Table~\ref{tab:constraintSum}) and most of them become stronger as we increase the couplings. The more complicated relation comes from $\tau\to\ell\nu\bar\nu$ which involve $\lambda$, $\lambda'_{333}$, $m_{\widetilde{\tau}_{R}}$ and $m_{\widetilde{b}_{R}}$. As described in Eq.~\eqref{eq:tautolnunu}, the two dominant terms of $\tau\to\ell\nu\bar\nu$, involving $\lambda$, $m_{\widetilde{\tau}_{R}}$ and  $\lambda'_{333}$, $m_{\widetilde{b}_{R}}$ respectively cancel each other. Thus we choose $m_{\widetilde{\tau}_{R}}=3.0$ TeV to maintain a window in the right range of $m_{\widetilde{b}_{R}}\sim 2.5$ TeV where $R_{K^{(*)}}$, $R_{D^{(*)}}$ and ANITA overlap. A smaller $m_{\widetilde{\tau}_{R}}$ will shrink the window and move it to the left, but choosing $m_{\widetilde{\tau}_{R}}$ to be larger will cause the $R_{K^{(*)}}$ region to shrink, due to nonlinear dependence on $m_{{\widetilde \tau}_R}$. Meanwhile, we increase $\lambda$, $\lambda'_{333}$ simultaneously so that their effects on $\tau\to\ell\nu\bar\nu$ window mostly cancel. To avoid RG running problems ({\it i.e.} hitting the Landau pole too close to the TeV-scale), $\lambda'_{333}$ is set at its largest possible magnitude of $-3.5$. This large coupling results in severe $B_s-\overline{B}_s$ mixing bound and to alleviate this, we choose $m_{\widetilde{\nu}_{\tau}}$ to be 9 TeV. $m_{\widetilde{\tau}_{L}}$ is chosen, different from $m_{\widetilde{\tau}_{R}}$, at 10 TeV, as mentioned in the previous case to suppress the possible contribution to $R_{D^{(*)}}$ from $LLE$ couplings.
%since a larger $m_{\widetilde{\tau}_{L}}$ result in $R_D$ band region moving downwards reducing overlap region due to the counter balance between $m_{\widetilde{\tau}_{L}}$ and $\lambda'$ in $R_D$.
The color scheme for the shaded regions is the same as in Fig.~\ref{fig:ANITARdRkgm2CKM}. Now we also show the $2\sigma$ ($3\sigma$) preferred region for $(g-2)_{\mu}$ at the upper left corner of Fig.~\ref{fig:ANITARdRkgm2ST} by the thin (thick) blue line with the arrow pointing into the allowed region. The horizontal hatched region is theoretically disfavored from perturbativity constraint on $\lambda'\leq \sqrt{4\pi}$. 

The location and shape of the favored regions for $R_{D^{(*)}}$ and $R_{K^{(*)}}$ anomalies are different from Case 1 mainly due to the fact that the parameter planes are different. In Fig.~\ref{fig:ANITARdRkgm2CKM}, the $y$-axis shows the $\epsilon$ parameter which plays the role as the relative scale between two $\lambda'$ or two $\lambda$ couplings, while in Fig.~\ref{fig:ANITARdRkgm2ST} the $y$-axis shows the overall scale for the $\lambda'$-couplings. Generally speaking, the overall scale could be larger but the relative scale should be heavily suppressed due to the polynomial dependence.  Therefore, the overlap region in Fig.~\ref{fig:ANITARdRkgm2ST} has $\lambda'\sim 0.8$, as compared to that in Fig.~\ref{fig:ANITARdRkgm2CKM} which has $\epsilon\sim -0.01$. 

Also note that in Case 2, the $3\sigma$ allowed region for ANITA shrinks dramatically, in both $m_{\widetilde{b}_R}$ and $\lambda'_{233}$ directions, which is mainly due to the structure of the $\lambda'$ couplings in Eq.~\eqref{eq:STlambp}.
The favored region shrinks in the $m_{\widetilde{b}_R}$ direction because there are larger $\lambda'$ couplings and thus the simulated number of events for ANITA gets more sensitive to change of $m_{\widetilde{b}_R}$. Shrinking in the $\lambda'$ direction is a combined effect of the structural change of the $\lambda'$s and the change of $y$-axis from relative scale ($\epsilon$ in Case 1) to overall scale ($\lambda'_{233}$ in Case 2).  

% To be more specific, only one $\lambda'$ is of $O(1)$ in case 1 and three other $\lambda'$ are of $O(0.02)$ with all others of $10^{-4}$ while for case 2, two $\lambda'$ is of $O(1)$ and at least six other $\lambda'$s are of $O(0.06)$ with other $\lambda'$s of $O(0.01)$. So $\lambda'$ is generally larger in case 2 compared to case 1 and that  

% Similarly ANITA flavor region is much narrower in case 2 compared with case 1 because there are more $O(1)$ couplings contributing to ANITA compared with case 1. The main contribution for ANITA comes from $\lambda'_{ij3}$, which has only one $O(1)$ term ($\lambda'_{333}$) in case 1 while having at least two $O(1)$ terms ($\lambda'_{333}$, $\lambda'_{233}$) for case 2. We know from \cite{} that to fit with experiment data, the decay length $l_{\rm decay}$ of the bino particle $\widetilde{\chi}_1^0$ need to in the similar range as the chord length $10^3 \,{\rm km}$.  
The overlap region of $R_{D^{(*)}}$, $R_{K^{(*)}}$ and ANITA anomalies is marked by the green block around $(m_{\widetilde b_R}, \lambda') \sim (2.5\, {\rm TeV},\, 1.1)$. No overlap could be achieved with $(g-2)_{\mu}$ region in this parameter setup. We find that $(g-2)_{\mu}$ is most sensitive to $m_{\widetilde \nu_{\tau}}$ and we have tried an extreme case of setting $m_{\widetilde \nu_{\tau}}$ at the current LHC lower bound of 900 GeV~\cite{Tanabashi:2018oca}, which does expand the $(g-2)_{\mu}$ region downward but not enough to have an overlap while in the meantime $B_s$ meson mixing bound becomes much severe and rules out the whole parameter region. Thus in this case $(g-2)_{\mu}$ cannot be accounted for. 

The bounds also appear differently in Case 2 than in Case 1 due to the change of $y$-axis. The most stringent bounds in this case are $\tau\to\ell \nu\bar{\nu}$ \cite{Tanabashi:2018oca} and $B_s$ meson mixing processes \cite{Amhis:2019ckw}. Similar to Fig.~\ref{fig:ANITARdRkgm2CKM}, the branch-cut feature in the $B_s$-meson mixing bound is due to the cancellation between the terms in Eq.~\eqref{eq:bbmixing}.

\subsection{Case 3: No Symmetry} \label{sec:ours}
In this final benchmark scenario, we do not invoke any symmetries. Instead, we adopt a pragmatic approach to choose our parameters so that we maintain the necessary freedom to explain all the anomalies while satisfying all experimental constraints. At the same time, we want to keep the total number of free parameters the same as in the other two cases, {\it i.e.} six mass parameters and three couplings. Thus, we try to equalize the non-zero parameters as much as possible. We end up with the following 3 free coupling parameters,
\begin{align}\label{eq:variableSetUp3}
    \{ & \lambda'_{223}\, , \quad 
    \lambda' \ \equiv \ \lambda'_{123} \ = \ \lambda'_{233} \ = \ \lambda'_{323}\, , \nonumber \\ & \qquad 
    \lambda \ \equiv \ \lambda_{132}\ = \ \lambda_{231}\ = \ \lambda_{232}
\},
\end{align}
with all the other $\lambda'$ and $\lambda$ couplings are set to be very small (essentially zero in practice).

\begin{figure*}[tbh!]
	\centering
	\includegraphics[width=0.85\textwidth]{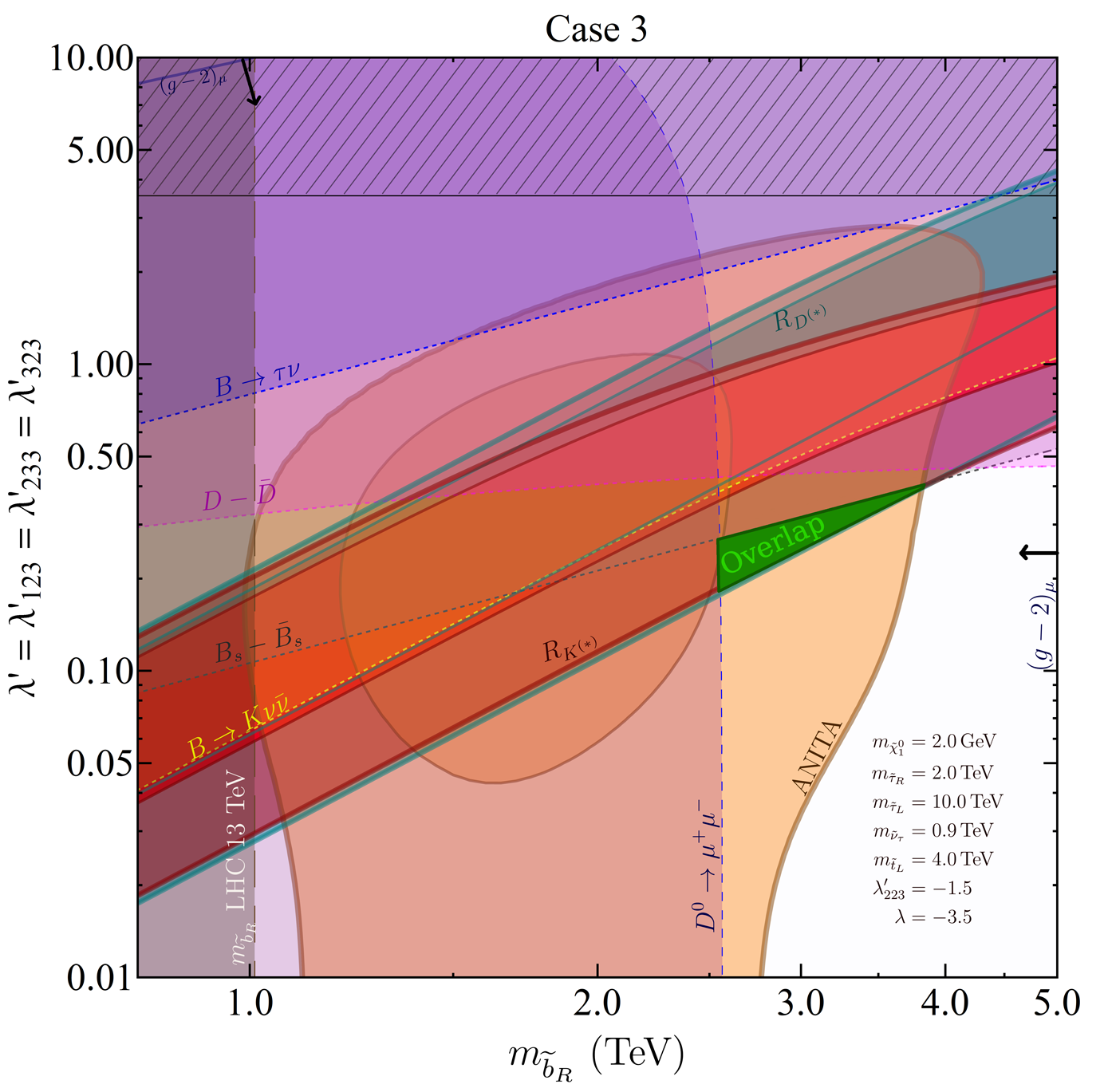}
	\caption{Benchmark scenario for Case 3 (with no symmetry) in the two-dimensional parameter plane $(m_{\widetilde b}, \lambda')$, while keeping other free parameters fixed as shown in the figure. The labels for the shaded regions are the same as in Fig.~\ref{fig:ANITARdRkgm2ST}. In addition, the $D^0\to \mu^+\mu^-$ constraint is shown by the blue shaded region (marked by the dashed blue boundary). The $2\sigma$ $(g-2)_\mu$ region covers almost the entire shown parameter space, so the $3\sigma$ region is 
	not shown. Also, as in Fig.~\ref{fig:ANITARdRkgm2ST}, the horizontal hatched region is theoretically disfavored from perturbativity constraint on $\lambda'\leq \sqrt{4\pi}$.
 }
	\label{fig:ANITARdRkgm2}
\end{figure*}

As shown in Fig.~\ref{fig:ANITARdRkgm2}, our benchmark point in this scenario is set as
\begin{align}
& \lambda'_{223} \ = \ -1.5 \, , \quad \lambda \ = \ -3.5 \, , \quad m_{\widetilde{\chi}_1^0} \ = \ 2.0~{\rm GeV}
 \, , \nonumber \\
& m_{\widetilde{\tau}_R} \ = \ 2.0~{\rm TeV} \, , \quad m_{\widetilde{\tau}_L} \ = \ 10.0~{\rm TeV} \, , \quad  \nonumber \\
& m_{\widetilde{\nu}_\tau} \ = \ 0.9~{\rm TeV} \, , \quad 
m_{\widetilde{t}_L} \ = \ 4.0~{\rm TeV} \, ,
\label{eq:bp3}
\end{align}
while we vary the remaining two parameters $\lambda'$ and $m_{\widetilde{b}_{R}}$ to find the common overlap region for $R_{D^{(*)}}$, $R_{K^{(*)}}$, $(g-2)_\mu$ and ANITA. We are able to do so around $(m_{\widetilde{b}_{R}},\lambda')=(3.0{\rm\, TeV}, 0.3)$. The overlap region is highlighted as the green block in Fig.~\ref{fig:ANITARdRkgm2}. In this parameters setup, $R_{D^{(*)}}$ and $R_{K^{(*)}}$ are brought together mainly by setting a large negative $\lambda'_{223}=-1.5$. When combined with setting $\lambda'_{333}=0$, this setup results in $R_{D^{(*)}}$ being dominated by $-X_c^{\mu}\sim -\lambda'_{223}\lambda'/m_{\widetilde{b}_R}$, which gives a positive contribution as we want. Meanwhile, for $R_{K^{(*)}}$, the dominant term is the second term from Eq.~\eqref{eq:C9C10} $\sim \lambda'^3_{223}\lambda'/m_{\widetilde{b}_R}^2$, which gives a negative contribution as required. The $(g-2)_{\mu}$-favored region in this case is vastly expanded compared to Case 2, and covers pretty much the entire parameter space shown in Fig.~\ref{fig:ANITARdRkgm2}. This is mainly due to the choice of small  $m_{\widetilde{\nu}_{\tau}}$ and the multiple $O(1)$ $\lambda$s, where we choose $\lambda$ to be $-3.5$, which give larger overlap compared to the positive value due to the dominant $\lambda$ term contribute to the denominator of $R_{D^{(*)}}$. This setting guarantees the dominant contribution to be the $\lambda$ terms in Eq.~\eqref{eq:gm2l} and thus the subdominant $\lambda'$ terms could have a much larger range. 
In this case, the effect of $m_{{\widetilde \tau}_R}$ on $(g-2)_\mu$ and $R_{K^{(*)}}$ is gone due to the vanishing couplings $\lambda_{k23}$. So the only influence of $m_{{\widetilde \tau}_R}$ is on $D - \overline D$ mixing bound, which inversely depends on $m_{{\widetilde \tau}_R}^2$ (see Section~\ref{sec:DDmixing}). Therefore, we simply set $m_{{\widetilde \tau}_R} = $2 TeV, same as in Case 1. On the other hand, from the same consideration of reducing the effect of $LLE$ coupling on $R_{D^{(*)}}$ like previous two cases, we set $m_{{\widetilde \tau}_L}=$10 TeV. 
%decrease of $m_{{\widetilde \tau}_L}$ will result in $R_D$ region to bend to the left reducing the overlap region because the increase in magnitude of the term $-\lambda\,\lambda'/m_{{\widetilde \tau}_L}^2$ (this is positive since $\lambda<0$) requires the increase of magnitude of the counter term $-\lambda_{223}\,\lambda'/m_{{\widetilde b}_R}^2$ by decreasing $m_{{\widetilde b}_R}$ to balance the total value of $R_D$.

The relevant bounds, including $B\to\tau \nu$~\cite{Amhis:2019ckw}, $D-\overline{D}$ mixing~\cite{Peng:2014oda}, $B_s-\overline{B}_s$ mixing~\cite{Amhis:2019ckw}, $B\to K\nu \bar\nu$~\cite{Lees:2013kla,Buttazzo:2017ixm,Bordone:2018hqs} and $D^0\to\mu^+\mu^-$~\cite{Aaij:2013cza}, are also shown in Fig.~\ref{fig:ANITARdRkgm2} by dark blue, magenta, gray, yellow and blue shaded regions respectively, while the LHC bound on sbottom mass is shown by the vertical brown-shaded region. In this case, the most stringent constraints come from $B_s-\overline{B}_s$ mixing and $D\to\mu\mu$ which shrink the overlap region substantially. The $B_s-\overline{B}_s$ mixing, as mentioned earlier in Case 1 and Case 2, is a typical bound for our RPV3 model trying to explain the $B$-anomalies since the relevant couplings $\lambda'_{i33}$, $\lambda'_{i23}$ and $\lambda'_{i32}$ all contribute to $B$-meson mixing. The branch-cut feature of the $B_s-\overline{B}_s$ mixing bound seen in Figs.~\ref{fig:ANITARdRkgm2CKM} and \ref{fig:ANITARdRkgm2ST} is absent in Fig.~\ref{fig:ANITARdRkgm2} because in this case there is no cancellation in Eq.~\eqref{eq:bbmixing}, as the third term dominates due to the choice of small sneutrino mass. On the other hand, the $D\to\mu\mu$ bound is crucial mainly due to the important role of $\lambda'_{223}$ in this particular Case 3. Note that in this case the $\tau\to\ell\nu\bar\nu$ bound is not relevant due to vanishing couplings $\lambda'_{333}=\lambda_{233}=0$; see Section~\ref{sec:taudecay} for more details.

\section{Constraints} \label{sec:constraints}

For the record let us briefly mention that just before the advent of the two asymmetric $B$-factories, the general perception was that RPV had so many parameters and that it was so completely unconstrained that it can accommodate just about anything; see {\it e.g.} p.921, Table 13.6 in Ref.~\cite{BABAR_PHYS_BK97}. On the contrary, what we will show 
here is that the situation 
now has dramatically improved, thanks to the enormous experimental and theoretical progress in the past two decades. In fact, despite the many parameters our RPV3 scenario is remarkably well-constrained as we discuss below so much so that
more accurate measurements of say $R_{D^{(*)}}$ preserving the central value could have appreciable adverse consequences at least for the version of RPV that we are now finding to be favorable. 

In this section, we discuss all relevant constraints on our RPV3 scenario shown in Figs.~\ref{fig:ANITARdRkgm2CKM}, \ref{fig:ANITARdRkgm2ST} and \ref{fig:ANITARdRkgm2}, with the  parameter dependence and dominant terms in the corresponding expressions summarized in Table~\ref{tab:constraintSum}.  

\begin{table*}[tbh!]
	\begin{tabular}{|c|c|c|}\hline\hline
		Constraint & Parameter dependence & Relevant terms\\ \hline \hline
		$B\to \tau\nu$ & $\lambda'_{\ell'33}$, $\lambda'_{3j3}$ , $m_{\widetilde b_R}$ &  $\frac{\lambda'_{\ell'33}\cdot \lambda'_{3j3}\nonumber}{m_{\widetilde b_R}^2}$
		\\ \hline 
		%%%%%%%%%%%%%%%%%%%%%%%%%%%%%%%%%%%%%%%%%%%%%%%%%%%%
		%%%%%%%%%%%%%%%%%%%%%%%%%%%%%%%%%%%%%%%%%%%%%%%%%%%
		$B\to K^{(*)}\nu \bar\nu$ & $\lambda'_{\ell'33}$, $\lambda'_{\ell 23}$ , $m_{\widetilde b_R}$  & $\frac{\lambda'_{\ell'33}\cdot \lambda'_{\ell 23}\nonumber}{m_{\widetilde b_R}^2}, \, \frac{\lambda'_{\ell'33}\cdot \lambda'_{\ell 32}\nonumber}{m_{\widetilde b_L}^2}$
		 
		 \\ \hline
		 
		 %%%%%%%%%%%%%%%%%%%%%%%%%%%%%%%%%%%%%%%%%%%%%%%%%%%%
		 %%%%%%%%%%%%%%%%%%%%%%%%%%%%%%%%%%%%%%%%%%%%%%%%%%%
		 $B\to \pi/\rho\,\nu \bar\nu$ & $\lambda'_{\ell'33}$, $\lambda'_{\ell 13}$ , $m_{\widetilde b_R}$  & $\frac{\lambda'_{\ell'33}\cdot \lambda'_{\ell 13}\nonumber}{m_{\widetilde b_R}^2}$
		 \\ \hline
		 
		 %%%%%%%%%%%%%%%%%%%%%%%%%%%%%%%%%%%%%%%%%%%%%%%%%%%%
		 %%%%%%%%%%%%%%%%%%%%%%%%%%%%%%%%%%%%%%%%%%%%%%%%%%%
		& $\lambda'_{i33}$, $\lambda'_{i23}$, $\lambda'_{i32}$ ,   & $\frac{\lambda'_{i23}\lambda'_{i33}\lambda'_{j23}\lambda'_{j33}}{m_{\widetilde b_R}^2}$, \\
		  $B_s-\overline{B}_s$ mixing  & $m_{\widetilde b_R}$, $m_{\widetilde \nu}$ &  $\frac{\lambda'_{i23}\lambda'_{i32}\lambda'_{j33}\lambda'_{j33}}{m_{\widetilde b_R}^2}$  ,\\
		  & &  $\frac{\lambda'_{332}\lambda'_{323}}{m_{\widetilde \nu}^2}$  \\ \hline
		  $D-\overline{D}$ mixing & $\lambda'_{323}$, $m_{\widetilde b_R}$, $m_{\widetilde \tau_R}$ & $\frac{\lambda'^4_{323}}{m_{\widetilde b_R}^2}$, $\frac{\lambda'^4_{323}}{m_{\widetilde \tau_R}^2}$
		 \\ \hline
		 $D^0\to\mu^+\mu^-$ & $\lambda'_{2j3}$, $m_{\widetilde b_R}$ & $\frac{\lambda'_{2j3}\lambda'_{2j'3}}{m_{\widetilde b_R}^2}$\\\hline 
		 
		%%%%%%%%%%%%%%%%%%%%%%%%%%%%%%%%%%%%%%%%%%%%%%%%
		%%%%%%%%%%%%%%%%%%%%%%%%%%%%%%%%%%%%%%%%%%%%%%%%%%%
		$\tau\to\ell \nu \bar\nu$ & $\lambda_{323}$, $\lambda'_{333}$, $m_{\widetilde \tau_R}$, $m_{\widetilde b_R}$ & $\frac{{\lambda^2_{323}}}{m_{\widetilde \tau_R}^2} $, $\frac{{{\lambda'}^2_{333}}}{m_{\widetilde b_R}^2} $\\\hline
		$Z\to \ell \bar \ell'$& $\lambda'_{333}$, $m_{\widetilde b_R}$ &$\frac{{\lambda'}^2_{333}}{m_{\widetilde b_R}^2}$ \\
		\hline
		\hline
	\end{tabular}
	\caption{The parameter dependence and the dominant terms in the expressions for the relevant constraints in our RPV3 scenario. }\label{tab:constraintSum}
\end{table*}

\subsection{\texorpdfstring{$B \to \tau \nu$ and $B_c \to \tau \nu$}{}}
%B to tau nu

%%%%%%%%%%%%%%%%%%%%%%%%%%%%%%%%%%%%%%%%%%%%%%%%%%
%%%%%%%%%%%diagram for B to tau nu
\begin{figure}[tbh!]
	\begin{subfigure}{.23\textwidth}
		\centering
		\includegraphics[width=1\linewidth]{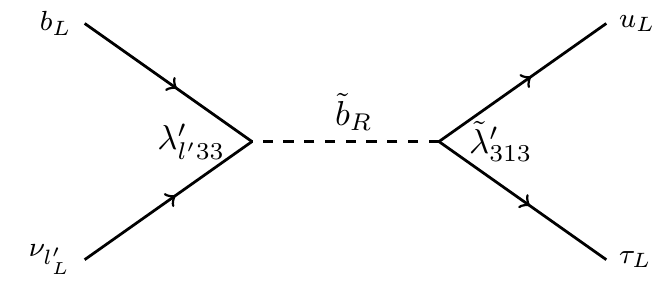}  
    	\caption{}
	    \label{fig:b2taunua}
	\end{subfigure}
	\begin{subfigure}{.23\textwidth}
		\centering
		\includegraphics[width=1\linewidth]{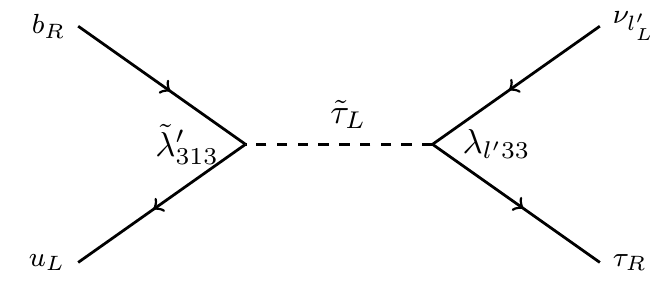}  
    	\caption{}
	    \label{fig:b2taunub}
	\end{subfigure}
	\caption{Contributions to the $B\to\tau\nu$ decay in RPV3: (a) with $LQD$ couplings only, and (b) with both $LLE$ and $LQD$ couplings.}
	\label{fig:b2taunu}
\end{figure}
In the notation of Ref.~\cite{Trifinopoulos:2018rna}, for $B^\pm\to \tau \nu$, we have
\begin{align}
 &\frac{{\rm BR}(B \to \tau\nu)}{{\rm BR}(B \to \tau\nu)_\text{SM}} \ = \  \sum_{l'=1}^3\left|\delta_{3l'}+\Delta_{3l'}^\mu
 \right|^2 \, , \label{eq:Btaunu}
 %\nonumber\\
 %&\Delta_{ll'}^\mu = \sum_{j=2}^{3} \frac{\sqrt{2}}{4 G_f} \frac{\lambda'_{l'33}\lambda'_{lj3}}{2 m_{\widetilde{b}_R}^2}\frac{V_{uj}}{V_{ub}} \label{eq:bctaunv1}
% \left| 1 + \frac{v^2}{2 m^2_{\widetilde b_R}} \text{Re}(X_u)  \right| ^2~,\\
% X_u &=& \lambda_{333}^\prime \left( \lambda_{333}^\prime + \lambda_{323}^\prime \frac{V_{us}}{V_{ub}} + \lambda_{313}^\prime \frac{V_{ud}}{V_{ub}} \right) ~, \nonumber
\end{align}
where the sum over $l'$ is for all flavors of neutrinos in final state, and 
\begin{align}
    \Delta_{ll'}^\mu \ = \ & \frac{\sqrt{2}}{4 G_F} \sum_{j=2}^{3}  \left(\frac{\lambda'_{l'33}\lambda'_{lj3}}{2 m_{\widetilde{b}_R}^2} \right. \nonumber\\
    & \qquad \left. + \frac{m_{B}^2}{(m_b+m_u) m_\tau} \frac{\lambda_{l'33}\lambda'_{lj3}}{2 m_{\widetilde{\tau}_L}^2} \right)\frac{V_{uj}}{V_{ub}} \, ,
    \label{eq:bctaunv2}
\end{align}
which includes processes involving both $LLE$ and $LQD$ vertices; see Fig.~\ref{fig:b2taunu}. Notice that the extra factor in front of the second term is due to the difference between vector and pseudoscalar current. The $B\to\tau\nu$ channel has been experimentally measured and the most updated results is reported in Ref.~\cite{Amhis:2019ckw}: 
\begin{equation}
    {\rm BR}(B\to \tau \bar{\nu})_{\rm exp} \ = \  (1.06\pm0.19)\times10^{-4}\, ,
\end{equation}
with a SM prediction of ~\cite{Altmannshofer:2017poe}:
\begin{equation}
    {\rm BR}(B\to \tau \bar{\nu})_{\rm SM} \ = \  (0.947\pm0.182)\times10^{-4}\, .
\end{equation} 
Comparing these numbers for the experimental measurement and SM calculation, a constraint could be imposed on the combination of RPV couplings and masses of sparticles in Eq.~\eqref{eq:Btaunu}. In Figs.~\ref{fig:ANITARdRkgm2CKM} and \ref{fig:ANITARdRkgm2}, this constraint has been shown by the blue shaded region with dashed dark blue boundary. The constraint turns out to be not relevant for the parameter choice in  Fig.~\ref{fig:ANITARdRkgm2ST}.

Similarly, the decay $B_c\to\tau\nu$ also gets a contribution from  Eq.~(\ref{eq:bctaunv2}) with $V_{uj}/V_{ub}$ replaced by $V_{cj}/V_{cb}$. This channel has not been measured and may not be measured in the near future. Previously, constraints have been imposed  using the life time of $B_c$, $\tau_{B_c} = 0.51\times 10^{-12} \rm s$  \cite{Tanabashi:2018oca}) and a 10\%-40\% estimate on the maximal allowed ${\rm BR}(B_c\to\tau\nu)$ ~\cite{Alonso:2016oyd, Celis:2016azn, Akeroyd:2017mhr}. We do not use this channel as a constraint, since we find that in our scenarios $B\to\tau\nu$ gives always stronger bounds. 
For completeness, we provide the predictions for ${\rm BR}(B_c\to\tau\nu)$ for our benchmark points: 
25.6\% (Case 1), 0.9\% (Case~2), and 2.0\% (Case~3). 
The corresponding ratio of the ${\rm BR}(B_c\to\nu\tau)$ between the RPV3 scenario and SM is found to be  
$\frac{{\rm BR}(B_c\to\tau\nu)_{\rm RPV3}}{{\rm BR}(B_c\to\tau\nu)_{\rm SM}}   =  34.2$ (Case 1), 1.2 (Case 2), and 2.7 (Case 3).

\subsection{\texorpdfstring{$B \to K^{(*)} \nu \bar{\nu}$ and $B \to \pi \nu \bar{\nu}$}{}}
%%%%%%%%%%%%%%%%%%%%%%%%%%%%%%%%%%%%%%%%%%%%%%%%%%
%%%%%%%%%%%diagram for B to K nu nu
\begin{figure}[tbh!]
	\begin{subfigure}{.23\textwidth}
		\centering
		\includegraphics[width=1\linewidth]{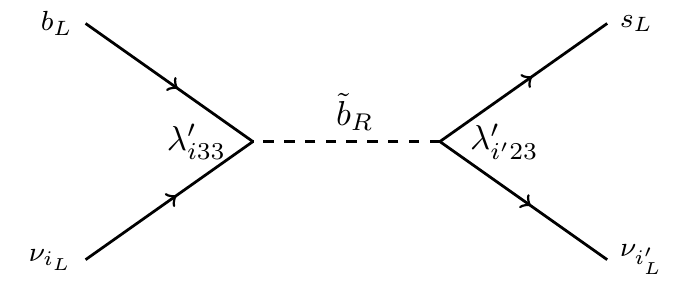}  
    	\caption{}
	    \label{fig:B2Knunua}
	\end{subfigure}
	\begin{subfigure}{.23\textwidth}
		\centering
		\includegraphics[width=1\linewidth]{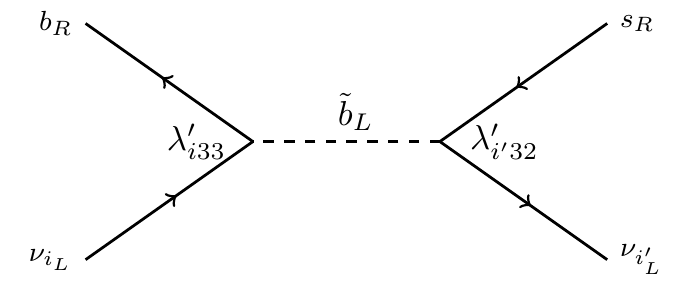}  
    	\caption{}
	    \label{fig:B2Knunub}
	\end{subfigure}
	\caption{Contributions to $B\to K^{(*)}\nu \bar\nu$ via $\lambda'$ interactions in RPV3.}
	\label{fig:B2Knunu}
\end{figure}
Tree-level exchange of sbottoms contributes to the decays $B \to K \nu \bar\nu$ and $B \to K^{*} \nu \bar\nu$; see Fig.~\ref{fig:B2Knunu}. Taking into account decay modes into different neutrino flavor combinations we get for the branching ratios:
\begin{eqnarray}
 && R_{B\to K^{(*)}\nu \bar\nu} \ \equiv \  \frac{{\rm BR}(B \to K^{(*)} \nu \bar\nu)}{{\rm BR}(B \to K^{(*)} \nu \bar\nu)_\text{SM}} \nonumber \\
  && \ = \ \frac{1}{3}\left| \delta_{ii'} + \frac{v^2 \pi s_w^2}{2 \alpha_\text{em}} \frac{\lambda^\prime_{i33} }{V_{tb}V_{ts}^*} \left(\frac{\lambda^\prime_{i'23}}{m_{\widetilde b_{R}}^2}+\frac{\lambda^\prime_{i'32}}{m_{\widetilde b_{L}}^2}\right) \frac{1}{X_t}  \right|^2.
%  && ~~ = \frac{1}{3} + \frac{1}{3}\left| 1 + \frac{v^2}{2 m_{\widetilde b_R}^2} \frac{\pi s_w^2}{\alpha_\text{em}} \frac{\lambda^\prime_{333} \lambda^\prime_{323}}{V_{tb}V_{ts}^*} \frac{1}{X_t}  \right|^2  \nonumber \\
%  && ~~ + \frac{1}{3}\left| 1 + \frac{v^2}{2 m_{\widetilde b_R}^2} \frac{\pi s_w^2}{\alpha_\text{em}} \frac{\lambda^\prime_{233} \lambda^\prime_{223}}{V_{tb}V_{ts}^*} \frac{1}{X_t}  \right|^2  \nonumber \\
%  && ~~ + \frac{1}{3}\frac{v^4}{4 m_{\widetilde b_R}^4} \frac{\pi^2 s_w^4}{\alpha_\text{em}^2} \frac{|\lambda^\prime_{233} \lambda^\prime_{323}|^2 + | \lambda^\prime_{333}\lambda^\prime_{223}|^2}{|V_{tb}V_{ts}^*|^2} \frac{1}{X^2_t} \nonumber\\
%  && ~~+ \frac{1}{3}\frac{v^4}{4 m_{\widetilde b_R}^4} \frac{\pi^2 s_w^4}{\alpha_\text{em}^2} \frac{|\lambda^\prime_{233} \lambda^\prime_{123}|^2 +  |\lambda^\prime_{333} \lambda^\prime_{123}|^2}{|V_{tb}V_{ts}^*|^2} \frac{1}{X^2_t} 
\end{eqnarray}
with the top loop function $X_t = 1.469\pm0.017$ \cite{Brod:2010hi} and $s_w$ being the weak mixing angle. Note that we consider both $\widetilde b_L$ and $\widetilde b_R$ exchanges, a feature only valid for final state with two neutrinos. Depending on the chosen benchmark, this equation simplifies into different forms and we use $m_{\widetilde{b}_L}=m_{\widetilde{b}_R}$ for numerical purposes. A bound for this ratio has been given by \cite{Buttazzo:2017ixm,Bordone:2018hqs} $R_{B\to K^{(*)}\nu \bar\nu} < 5.2$ at 95\% CL, which is adopted for our parameter setting and indicated in Figs.~\ref{fig:ANITARdRkgm2CKM},~\ref{fig:ANITARdRkgm2ST} and~\ref{fig:ANITARdRkgm2} as the yellow-shaded regions with dashed yellow boundary. 

An analogous expression holds for the decays $B \to \pi \nu \bar\nu$ and $B \to \rho \nu \bar\nu$:
\begin{eqnarray}
 && \frac{{\rm BR}(B \to \pi \nu \bar\nu)}{{\rm BR}(B \to \pi \nu \bar\nu)_\text{SM}} \ = \ \frac{{\rm BR}(B \to \rho \nu \bar\nu)}{{\rm BR}(B \to \rho \nu \bar\nu)_\text{SM}} \nonumber \\
  && \ = \ \frac{1}{3}\left| \delta_{ii'} + \frac{v^2 \pi s_w^2}{2 \alpha_\text{em}} \frac{\lambda^\prime_{i33} }{V_{tb}V_{td}^*} \left(\frac{\lambda^\prime_{i'13}}{m_{\widetilde b_{R}}^2}+\frac{\lambda^\prime_{i'31}}{m_{\widetilde b_{L}}^2}\right) \frac{1}{X_t}  \right|^2.
%  && ~~ = \frac{1}{3} + \frac{1}{3}\left| 1 + \frac{v^2}{2 m_{\widetilde b_R}^2} \frac{\pi s_w^2}{\alpha_\text{em}} \frac{\lambda^\prime_{333} \lambda^\prime_{313}}{V_{tb}V_{td}^*} \frac{1}{X_t}  \right|^2  \nonumber \\
%  && ~~ + \frac{1}{3}\left| 1 + \frac{v^2}{2 m_{\widetilde b_R}^2} \frac{\pi s_w^2}{\alpha_\text{em}} \frac{\lambda^\prime_{233} \lambda^\prime_{213}}{V_{tb}V_{td}^*} \frac{1}{X_t}  \right|^2 \nonumber \\
%  && ~~ + \frac{1}{3}\frac{v^4}{4 m_{\widetilde b_R}^4} \frac{\pi^2 s_w^4}{\alpha_\text{em}^2} \frac{|\lambda^\prime_{233} \lambda^\prime_{313}|^2 + |\lambda^\prime_{213} \lambda^\prime_{333}|^2}{|V_{tb}V_{td}^*|^2} \frac{1}{X^2_t} ~.
\end{eqnarray}
% Assuming $\lambda_{313}^\prime =  \lambda_{213}^\prime = 0$, this bound is always satisfied.
However, the experimental bounds on those decays are much weaker than the $B\to K^{(*)}\nu\bar\nu$ bounds and are always satisfied for the parameter choice we have, and hence, are not shown in Figs.~\ref{fig:ANITARdRkgm2CKM},~\ref{fig:ANITARdRkgm2ST} and~\ref{fig:ANITARdRkgm2}.

\subsection{\texorpdfstring{$B_s-\overline{B}_s$ Mixing}{Bs Bs-bar mixing}}
%%%%%%%%%%%%%%%%%%%%%%%%%%%%%%%%%%%%%%%%%%%%%%%%%%
%%%%%%%%%%%diagram for BB mixing
\begin{figure}[tbh!]
	\begin{subfigure}{.23\textwidth}
		\centering
		\includegraphics[width=1\linewidth]{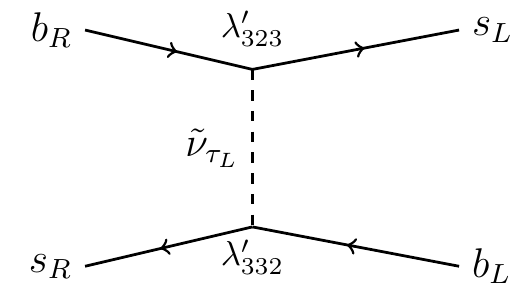}  
    	\caption{}
	    \label{fig:BsBsbara}
	\end{subfigure}
	\begin{subfigure}{.23\textwidth}
		\centering
		\includegraphics[width=1\linewidth]{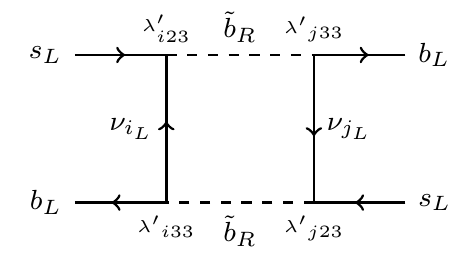}  
    	\caption{}
	    \label{fig:BsBsbarb}
	\end{subfigure}
	\begin{subfigure}{.23\textwidth}
		\centering
		\includegraphics[width=1\linewidth]{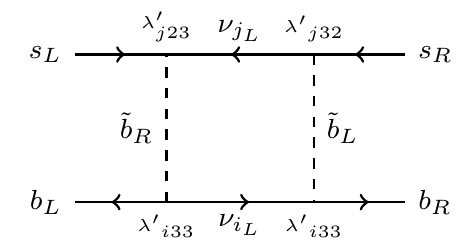}  
    	\caption{}
	    \label{fig:BsBsbarc}
	\end{subfigure}
	\caption{Dominant contributions to $B_s-\overline{B}_s$ mixing via $\lambda'$ couplings in RPV3.}
	\label{fig:BsBsbar}
\end{figure}

Here, RPV contributions can arise at the tree level from sneutrino exchange, as well at the one-loop level from box diagrams with sbottoms, sneutrinos, or stops; see Fig.~\ref{fig:BsBsbar}. Based on the derivation from Ref.~\cite{Trifinopoulos:2018rna}, we have:
\begin{align}\label{eq:bbmixing}
    &\Delta M_{B_s}^{\rm RPV} \ = \ \frac{2}{3} m_{B_s}
     f_{B_s}^2\left|P^{VLL}_1 \frac{\lambda'_{i23}\lambda'_{j33}\lambda'_{j23}\lambda'_{i33}}{128\pi^2 m_{\widetilde b_R}^2} \right. \nonumber\\ &\quad \left. +P^{LR}_1 \frac{\lambda'_{i23}\lambda'_{j33}\lambda'_{i32}\lambda'_{j33}}{128\pi^2 m_{\widetilde b_R}^2} + P^{LR}_2 \frac{\lambda'_{332}\lambda'_{323}}{2 m_{\widetilde \nu}^2}
    \right| \, ,
\end{align}
where we update the hadronic $P$ factors from Ref.~\cite{Buras:2001ra} with the latest lattice input from Ref.~\cite{Bazavov:2016nty} giving:
\begin{equation}
    P^{VLL}_1 \ = \ 0.80\, , \, 
    P^{LR}_1 \ = \ -2.52\, \, {\rm and}\, \, P^{LR}_2 \ = \ 3.08~.
\end{equation}

The mass difference $\Delta M_{B_s}$ in neutral $B_s$ meson mixing is measured with excellent precision, $\Delta M_{B_s} = (17.757 \pm 0.021)$\,ps$^{-1}$~\cite{Amhis:2019ckw}. The SM prediction, on the other hand, has sizable uncertainties stemming mainly from the hadronic matrix elements and the CKM matrix element $V_{cb}$~\cite{DiLuzio:2019jyq}. For the SM prediction, we use the latest lattice average of hadronic matrix elements from Ref.~\cite{Aoki:2019cca} (see also Refs.~\cite{Bazavov:2016nty,Boyle:2018knm,Dowdall:2019bea}), $ f_{B_s} \sqrt{\hat B_{B_s}} = ( 274 \pm 8 )$\,MeV, where $ f_{B_s}$ is the $B_s$ decay constant, and $\hat B_{B_s}$ a so-called bag parameter. For the CKM matrix element we use $|V_{cb}| = (41.0 \pm 1.4) \times 10^{-3}$, which is the conservative PDG average of recent inclusive and exclusive determinations~\cite{Tanabashi:2018oca}. We find $\Delta M_{B_s}^\text{SM} = (19.3 \pm 1.7)$\,ps$^{-1}$. This is in good agreement with the experimental value and a recent SM prediction based on light cone sum rule calculations~\cite{King:2019lal}. Combining our SM prediction with the experimental result we obtain the following bound at 95\% C.L.
\begin{equation}
 0.78 \ < \ \left| \frac{\Delta M_{B_s}}{\Delta M_{B_s}^\text{SM}} \right| \ < \ 1.12 ~.
\end{equation}

%This mass difference has been well measured experimentally and the most recent result is given in \cite{Amhis:2019ckw}
%\begin{equation}
%    \Delta M_{B_s}^{\rm exp}=(1.1684\pm0.0014)\times10^{-11}\, {\rm GeV}.
%\end{equation}
%The most recent SM calculation value for $\Delta M_{B_s}$ is given by Wolfgang company in \cite{} as:
%\begin{equation}
%    \Delta M_{B_s}^{\rm SM}=(1.2699\pm0.1119)\times10^{-11}\, {\rm GeV}.
%\end{equation}
The combination of the experimental result and the SM prediction puts a bound on the parameter space by confining the possible contribution to $\Delta M_{B_s}$ from RPV. This bound is indicated as the grey-shaded region in  Figs.~\ref{fig:ANITARdRkgm2CKM},~\ref{fig:ANITARdRkgm2ST} and~\ref{fig:ANITARdRkgm2}.

\subsection{\texorpdfstring{$D-\overline{D}$ Mixing}{DD-bar mixing}}\label{sec:DDmixing}
%%%%%%%%%%%%%%%%%%%%%%%%%%%%%%%%%%%%%%%%%%%%%%%%%%
%%%%%%%%%%%diagram for DD mixing
\begin{figure}[tbh!]
	\begin{subfigure}{.23\textwidth}
		\centering
		\includegraphics[width=1\linewidth]{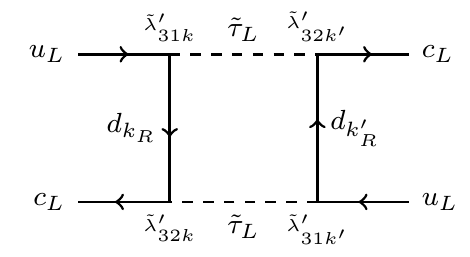}  
    	\caption{}
	    \label{fig:Dmixinga}
	\end{subfigure}
	\begin{subfigure}{.23\textwidth}
		\centering
		\includegraphics[width=1\linewidth]{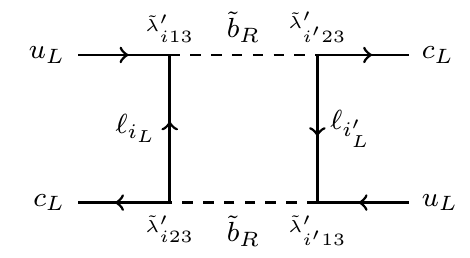}  
    	\caption{}
	    \label{fig:Dmixingb}
	\end{subfigure}
	\caption{Contribution to $D-\overline{D}$ Mixing from $\lambda'$ in RPV3 at 1-loop level.}
	\label{fig:Dmixing}
\end{figure}
Here the dominant contributions come from stau or sbottom loops, as shown in Fig.~\ref{fig:Dmixing}, which arise from the  $\widetilde{\lambda}'_{ijk}\equiv \lambda'_{ilk}V_{jl}$ couplings. The effective Hamiltonian for the loop-level contributions to $D-\overline{D}$ mixing from RPV is described as~\cite{Golowich:2007ka}
\begin{align}
    \mathcal{H}_{\rm RPV} \ = \ &\frac{1}{128\pi^2}\left(\frac{(\widetilde{\lambda}'_{32k}\widetilde{\lambda}'_{31k})^2}{m_{\widetilde{\tau}_{L}}^2}+\frac{(\widetilde{\lambda}'_{i23}\widetilde{\lambda}'_{i13})^2}{m_{\widetilde{b}_{R}}^2}\right)\nonumber \\
    & \times ({\bar u}_L \gamma_\mu c_L)({\bar u}_L\gamma^\mu c_L) \, .
\end{align}
Using this we can derive a bound on the RPV parameters and relate them to $x_D\equiv \Delta M_D/\Gamma_D$ (where $\Gamma_D$ is the mean decay width of $D$ meson):
\begin{align}
& \frac{1}{8}\left[\left(\frac{1\ {\rm TeV}}{m_{\widetilde{b}_{R}}}\right)^2+\left(\frac{1\ {\rm TeV}}{m_{\widetilde{\tau}_{R}}}\right)^2\right]{\lambda'^4_{323}} (V_{cs}V_{cd})^2\nonumber\\
& \qquad \leqslant (0.085)^2 x^{\rm expt}_{D} \, .
\end{align}
Combining this with the experiment result: $x_{D}^{\rm expt}= (5.8\pm 1.9)\times10^{-3}$~\cite{Peng:2014oda}, we get the bound for ($m_{\widetilde{b}_{R}}$, $\lambda'$), which is denoted by the pink-shaded region in Fig.~\ref{fig:ANITARdRkgm2}.

\subsection{\texorpdfstring{$D^0 \to \mu^+\mu^-$}{D0 -> mu+mu-}} \label{Dtomumu}
%%%%%%%%%%%%%%%%%%%%%%%%%%%%%%%%%%%%%%%%%%%%%%%%%%
%%%%%%%%%%%diagram for D to mu mu 
\begin{figure}[tbh!]
		\centering
		\includegraphics[width=0.25\textwidth]{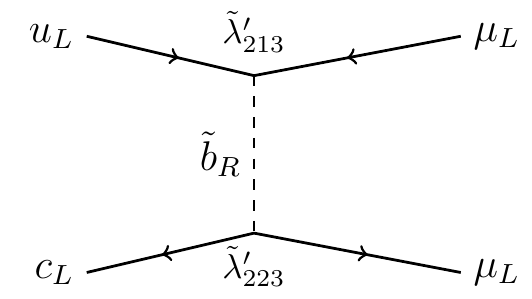}
	\caption{Contribution to $D^0 \to \mu^+\mu^-$ from $\lambda'$ in RPV3 at tree level.}
	\label{fig:Dmumu}
\end{figure}
As shown in Fig.~\ref{fig:Dmumu}, tree-level contributions from sbottom exchange to this rare $D^0$ decay width can be expressed as \cite{Deshpand:2016cpw}:
\begin{align}
    \Gamma(D^0 \to \mu^+ \mu^-) & \ = \  \frac{1}{128\pi}\left|\frac{\lambda'_{2j3}\lambda'_{2j'3} V_{uj'} V_{cj}}{m_{\widetilde b_R}^2}\right|^2 f_D^2 \nonumber\\
   &\times m_D m_{\mu}^2\sqrt{1-4m_{\mu}^2/m_D^2} \, , 
\end{align}
In case 3, this bound become most important and the expression reduces to a function of $\lambda'$, $\lambda'_{223}$ and $m_{\widetilde b_R}$.  An updated upper bound on this branching ratio \cite{Aaij:2013cza} is set at $7.6\times 10^{-9}$ at 95\% CL and the corresponding bound is shown as the light purple area in Fig.~\ref{fig:ANITARdRkgm2}. In other two cases, this bound is subdominant and is not shown in Fig \ref{fig:ANITARdRkgm2CKM} and Fig \ref{fig:ANITARdRkgm2ST}. 
% \subsection{\texorpdfstring{$b \to s \gamma$}{}}

% There are sbottom loops, sneutrino loops, and stop loops.

% \subsection{\texorpdfstring{$\tau \to \mu \gamma$ and $\tau \to 3 \mu$}{}}

% \red{this part already covered in Professor Soni's discussion}
% sbottom loops and stop loops.

% \subsection{\texorpdfstring{$\tau \to \mu K$}{}}

% tree level contributions from stop exchange.

\subsection{\texorpdfstring{$Z\to \ell \bar{\ell}'$}{Z -> l l'-bar}}
%%%%%%%%%%%%%%%%%%%%%%%%%%%%%%%%%%%%%%%%%%%%%%%%%%
%%%%%%%%%%%diagram for Z to ll
\begin{figure}[tbh!]
		\centering
		\includegraphics[width=0.26\textwidth]{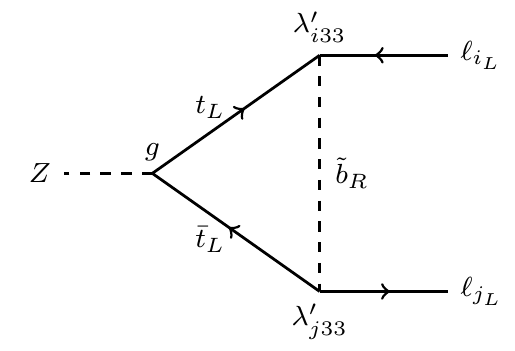}
	\caption{Contribution to $Z\to \ell \bar{\ell}'$ from $\lambda'$ in RPV3 at loop level.}
	\label{fig:Zll}
\end{figure}
This process gets modified by top-sbottom loops, as shown in Fig.~\ref{fig:Zll}. 
%More specifically, $Z$ will decay to an off-shell $t\bar t$ pair both of which through $LQD$ give a lepton and a sbottom. The sbottom line closes to form a loop, while the product of the process becomes a $l\bar l'$ final state. 
Due to different $i$ index in $\lambda_{i33}$, we may have different flavor final states ($Z \to \tau\tau$ or $Z \to \mu\mu$) or even flavor-violating final states such as $Z \to \tau\mu$.
A change in the $Z$ decay process from the SM prediction will affect the ratio of the vector and axial-vector couplings of the $Z$ boson with different lepton flavors. Experimental measurements on these couplings are given in Ref.~\cite{Tanabashi:2018oca} as:
\begin{align}
    \left(\frac{g_V^\tau}{g_V^e}\right)_{\rm exp} & \ = \  0.9588 \pm 0.02997 \, , \label{eq:zll1}\\
    \left(\frac{g_A^\tau}{g_A^e}\right)_{\rm exp} & \ = \  1.0019 \pm 0.00145 \, . \label{eq:zll2}
\end{align}
The contributions to these ratios from RPV model are given by~\cite{Trifinopoulos:2018rna} (see also \cite{Arnan:2019olv})
\begin{align}
    \left(\frac{g_V^\tau}{g_V^e}\right)_{\rm SM+RPV} & \ = \ 1 - \frac{2\,\delta g_{\ell_3 \ell_3}}{1-4\,s_w^2} \, , \nonumber\\
    \left(\frac{g_A^\tau}{g_A^e}\right)_{\rm SM+RPV} & \ = \ 1 - 2\,\delta g_{\ell_3 \ell_3} \, ,\nonumber
\end{align}
where $\delta g_{\ell_i \ell_j}$ is a simplification of Eq.~(30) in Ref.~\cite{Feruglio:2017rjo} where we only keep the top Yukawa-related terms. It is denoted as follows:
\begin{align}
    \delta g_{\ell_i \ell_j} & \ \simeq \ \frac{3 y_t^2}{32\sqrt{2}G_F \pi^2} \frac{\lambda'_{i33} \lambda'_{j33}}{m_{\widetilde b_R}^2}\left[\log\left(\frac{m_{\widetilde b_R}}{m_Z}\right)-0.612\right] \, .
\end{align}
Taking $i$, $j$ both equal to 3 and using Eqs.~\eqref{eq:zll1} and \eqref{eq:zll2}, we derive a bound on the parameter space, as shown by the vertical pink-shaded region in Figs.~\ref{fig:ANITARdRkgm2CKM} and \ref{fig:ANITARdRkgm2ST}. This bound is not shown in Fig.~\ref{fig:ANITARdRkgm2} because the choice $\lambda'_{333} \sim 0$ makes it irrelevant for Case 3.

In principle, bounds could also be put on $\lambda'_{333}\lambda'_{233}$ by evaluating the experimental bounds on the LFV branching ratio of $Z\to \tau\mu$ process. However, the current experimental bound for ${\rm BR}(Z\to\tau\mu)$ is of order of $10^{-5}$ while the contribution to this branching from RPV is typically $<10^{-7}$ \cite{Feruglio:2017rjo}. Therefore, no substantial bound can be put from the flavor violating $Z$ coupling. Also worth noting is that the $W$ couplings could also be altered by RPV loop processes. However, such bounds from the $W$ coupling variations are not shown here since they are not as strong as the bound from $\tau\to \ell \nu\bar{\nu}$ process~\cite{Feruglio:2017rjo}, which is described in the next subsection.

\subsection{\texorpdfstring{$\tau\to \ell \nu\bar{\nu}$}{tau -> l nu nu-bar}} \label{sec:taudecay}
%%%%%%%%%%%%%%%%%%%%%%%%%%%%%%%%%%%%%%%%%%%%%%%%%%
%%%%%%%%%%%diagram for tau to l nu nu
\begin{figure}[tbh!]
		\centering
		\includegraphics[width=.33\textwidth]{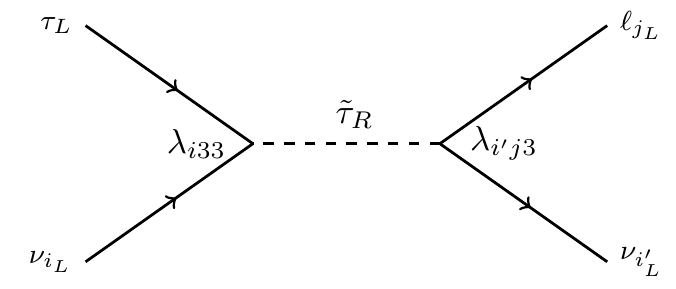}
	\caption{Contribution to $\tau\to \ell \nu\bar{\nu}$ from $\lambda'$ in RPV3 at tree level.}
	\label{fig:taudecay}
\end{figure}
The $LLE$ coupling will result in the change of the decay rate of $\tau\to e\nu\bar{\nu}$ and of $\tau\to \mu\nu\bar{\nu}$ via the exchange of $\widetilde{\tau}_R$, as shown in Fig.~\ref{fig:taudecay}. This effect could be tested by the ratio:
\begin{equation}
R_{\tau}^{\tau/\ell} \ = \ \frac
{{\rm BR}(\tau\to \ell\nu\bar{\nu})_{\rm exp}
	/
	{\rm BR}(\tau\to \ell\nu\bar{\nu})_{\rm SM}
	}
{{\rm BR}(\mu\to e\nu\bar{\nu})_{\rm exp}
	/
	{\rm BR}(\mu\to e\nu\bar{\nu})_{\rm SM}
} \, .
\end{equation}
Based on the derivation from Ref.~\cite{Trifinopoulos:2018rna}, in the SM+RPV case, we have:
\begin{equation}
R_{\tau}^{\tau/\ell} \ \simeq \ 1+\frac{\sqrt{2}}{4 G_F}\frac{\lambda_{323}^2}{m_{\widetilde{\tau}_R}^2}-\frac{3y_t^2}{16\sqrt{2}G_F \pi^2} \frac{\lambda'^2_{333}}{m_{\widetilde{b}_R}^2}\log\left(\frac{m_{\widetilde{b}_R}}{m_Z}\right) \, .\label{eq:tautolnunu}
\end{equation}
This can be used to put constraints on the parameter space when combined with the experimental values \cite{Pich:2013lsa}:
\begin{align}
(R_{\tau}^{\tau/\mu})_{\rm exp}& \ = \ 1.0022\pm0.0030 \, , \\
(R_{\tau}^{\tau/e})_{\rm exp}& \ = \ 1.0060\pm0.0030 \, .
\end{align}
The corresponding bound is displayed in Figs.~\ref{fig:ANITARdRkgm2CKM} and \ref{fig:ANITARdRkgm2ST} as dark blue region, while it is not shown in Fig.~\ref{fig:ANITARdRkgm2} because in Case~3 this bound becomes irrelevant due to $\lambda'_{333}\sim \lambda_{323}\sim 0$.

%choosing $\lambda=-3.3$, this put a lower bound on the ${\widetilde{\tau}_R}$ mass: $m_{\widetilde{\tau}_R}>5.4\ {\rm TeV}$.
\subsection{\texorpdfstring{$b \to s \gamma$}{b -> s gamma}} \label{sec:bsgamma}
%%%%%%%%%%%%%%%%%%%%%%%%%%%%%%%%%%%%%%%%%%%%%%%%%%
%%%%%%%%%%%diagram for b to s gamma
\begin{figure}[tbh!]
		\centering
		\includegraphics[width=.33\textwidth]{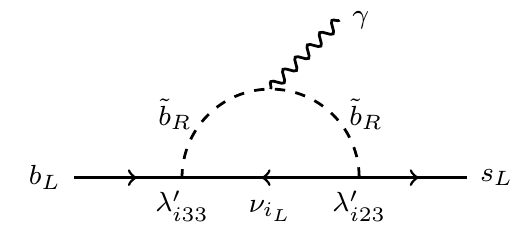}
	\caption{Contribution to $b \to s \gamma$ from $\lambda'$ in RPV3 at one-loop level.}
	\label{fig:bsgamma}
\end{figure}
The branching ratio of the rare decay $b\to s\gamma$ has been measured~\cite{Amhis:2019ckw} as:
\begin{equation}
    {\rm BR}(b\to s\gamma)_{\rm exp} \ =  \ (3.43 \pm 0.21 \pm 0.07) \times 10^{-4}\, ,
\end{equation}
which is consistent with the SM result~\cite{Misiak:2015xwa}
\begin{align}
  {\rm BR}(b\to s\gamma)_{\rm SM} \ = \ (3.36 \pm 0.23)\times 10^{-4}  \, .
\end{align}
However, as pointed out by Refs.~\cite{deCarlos:1996yh, Kong:2004cp, Besmer:2000rj, Dreiner:2013jta}, BSM effects from both $R$-parity conserving and violating terms could contribute to this channel either directly via one-loop diagrams (like in Fig.~\ref{fig:bsgamma}) or indirectly via RG running. Considering the direct RPV contribution only, we take the bound in Ref.~\cite{deCarlos:1996yh} adopting it to the updated measurement~\cite{Amhis:2019ckw}, which gives:
\begin{align}
    |\lambda'_{323}\lambda'_{333}|& \ \lesssim \ 0.025 \left| 2\left(\frac{100\ {\rm GeV}}{m_{\widetilde \nu_\tau}}\right)^2 -\left(\frac{100\ {\rm GeV}}{m_{\widetilde b_{R}}}\right)^2 \right|^{-1} , \label{eq:bsg1} \\
    |\lambda'_{332}\lambda'_{333}|& \ \lesssim \ 0.01 \left| \left(\frac{100 \ {\rm GeV}}{m_{\widetilde \tau_{L}}}\right)^2 -\left(\frac{100 \ {\rm GeV}}{m_{\widetilde b_{L}}}\right)^2 \right|^{-1}. \label{eq:bsg2}
\end{align}
Substituting the benchmark mass values for our three cases we find that the constraints are $\lambda'_{332}\lambda'_{333}\lesssim 1.64, 1.15, 1.00$ and $\lambda'_{323}\lambda'_{333} \lesssim 7.14, 25.94, 1.01$ respectively for Case 1, 2 and 3, while the actual values of these coupling products we have for all these cases are $O(0.01)$. Thus the $b \to s \gamma$ constraint is always satisfied for all our benchmark points. The weakness of this bound could be understood both from the partial cancellation between the two terms in Eqs.~\eqref{eq:bsg1} and \eqref{eq:bsg2}, and from the dependence of the upper bounds on the sparticle  masses.

\subsection{Neutrino Mass}
%%%%%%%%%%%%%%%%%%%%%%%%%%%%%%%%%%%%%%%%%%%%%%%%%%
%%%%%%%%%%%diagram for neutrino mass
\begin{figure}[tbh!]
	\begin{subfigure}{.23\textwidth}
		\centering
		\includegraphics[width=1\linewidth]{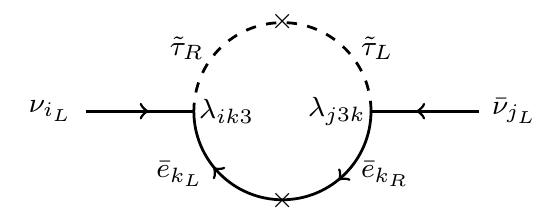}  
    	\caption{}
	    \label{fig:numassa}
	\end{subfigure}
	\begin{subfigure}{.23\textwidth}
		\centering
		\includegraphics[width=1\linewidth]{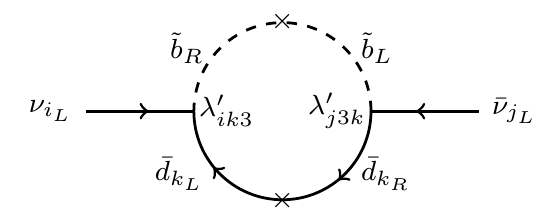}  
    	\caption{}
	    \label{fig:numassb}
	\end{subfigure}
	\caption{Contribution to neutrino mass from RPV3 at one-loop level.}
	\label{fig:numass}
\end{figure}
The trilinear RPV couplings in Eqs.~\eqref{Eq.lambda_prime} and \eqref{Eq.RPVLLE} contribute to neutrino masses at one-loop level through the lepton-slepton and quark-squark loops, as shown in Fig.~\ref{fig:numassa} and \ref{fig:numassb} respectively. Using the general expression~\cite{Hall:1983id, Babu:1989px, Barbier:2004ez} and dropping the terms involving the first two generation sfermions, we obtain: 
\begin{align}
    M^\nu_{ij} \ \simeq \ &  \frac{3}{16\pi^2}\sum_k\lambda'_{ik3}\lambda'_{j3k}m_{d_k}
    \frac{\left(\widetilde{m}^d_{LR}\right)^2_{33}}{m^2_{\widetilde b_{R}}-m^2_{\widetilde b_{L}}}\ln\left(\frac{m^2_{\widetilde b_{R}}}{m^2_{\widetilde b_{L}}}\right) \nonumber \\
    & + 
    \frac{1}{16\pi^2}\sum_k\lambda_{ik3}\lambda_{j3k}m_{e_k}
    \frac{\left(\widetilde{m}^e_{LR}\right)^2_{33}}{m^2_{\widetilde \tau_{R}}-m^2_{\widetilde \tau_{L}}}\ln\left(\frac{m^2_{\widetilde \tau_{R}}}{m^2_{\widetilde \tau_{L}}}\right)\nonumber \\
    & +(i\leftrightarrow j) \, , 
    \label{eq:numass}
\end{align}
where $(\widetilde{m}^d_{LR})^2$ and $(\widetilde{m}^e_{LR})^2$ are the left-right squark and slepton mixing matrices respectively, given by 
\begin{align}
    (\widetilde{m}^{d}_{LR})^2_{ij} \ = \ & \frac{v_d}{\sqrt 2}(A^{d}_{ij}-\mu\tan\beta y^{d}_{ij}) \, , 
    \label{eq:mixsquark}
\end{align}
(and similarly for $(\widetilde{m}^e_{LR})^2$ in terms of $A^e$ and $y^e$), where $A^{d,e}$ are the soft trilinear terms, $y^{d,e}$ are the Yukawa couplings, and $\tan\beta=v_u/v_d$ is the ratio of the VEVs of the two Higgs doublets in the MSSM. 

In the basis in which the charged lepton masses and the down quark masses are diagonal, it is customary to assume that the $A$-terms are proportional to the Yukawa couplings, {\it i.e.} $A^d_{33}=A^b y^b$ and $A^e_{33}=A^\tau y^\tau$. With this substitution, Eq.~\eqref{eq:numass} simplifies to  
\begin{align}
     M^\nu_{ij} \ \simeq \ & \frac{3}{8\pi^2}\left(\frac{A^b-\mu\tan\beta}{\overline{m}^2_{\widetilde b}}\right)\sum_k \lambda'_{ik3}\lambda'_{j3k}m_{d_k}m_b\nonumber \\
     & + \frac{1}{8\pi^2}\left(\frac{A^\tau-\mu\tan\beta}{\overline{m}^2_{\widetilde \tau}}\right)\sum_k \lambda_{ik3}\lambda_{j3k}m_{e_k}m_\tau \, ,
     \label{eq:Mnu1}
\end{align}
where $\overline{m}_{\widetilde b}$ and $\overline{m}_{\widetilde \tau}$ are the average sbottom and stau masses. We must ensure that the trace of the $M^\nu$ matrix in Eq.~\eqref{eq:Mnu1} ({\it i.e.} the sum of its eigenvalues $m_{\nu_i}$) should satisfy the cosmological bound on the sum $\sum_i m_{\nu_i}\lesssim 0.1$ eV~\cite{Aghanim:2018eyx}. For the three cases discussed earlier, we find that this requires  $(A^{b,\tau}-\mu\tan\beta)\lesssim {\cal O}(0.5~{\rm MeV})$ for Cases 1 and 2, while for Case 3, the upper bound is relaxed to about a GeV. With this choice, the neutrino mass constraint can be readily satisfied, and therefore, we do not include it in Figs.~\ref{fig:ANITARdRkgm2CKM}, \ref{fig:ANITARdRkgm2ST} and \ref{fig:ANITARdRkgm2}.   

\subsection{Neutrinoless Double Beta Decay}
\begin{figure}[tbh!]
		\centering
		\includegraphics[width=0.33\textwidth]{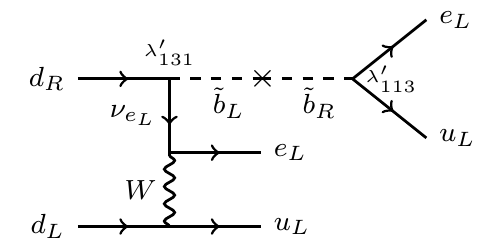}  
	\caption{Contribution to $0\nu\beta\beta$ from RPV3 at tree level. 
	%\blue{Add figure 6.22 from Ref.~\cite{Barbier:2004ez}.
	}
	\label{fig:0nubb}
\end{figure}

The same $\lambda'$ couplings responsible for nonzero Majorana neutrino mass could also induce a sizable rate for the rare neutrinoless double beta decay ($0\nu\beta\beta$) process.  There are several contributions, via processes involving the sequential $t$-channel exchange of two sfermions and a gaugino, where the sfermion may be a slepton or a squark, and the gaugino may be a neutralino or a gluino~\cite{Barbier:2004ez}. But all these contributions depend only on $\lambda'_{111}$, and are therefore, hugely suppressed or  vanish altogether in our RPV3 setup. 

There is another contribution~\cite{Babu:1995vh}, based on the $t$-channel scalar-vector type exchange of a sfermion and a $W$ boson linked together through an intermediate internal neutrino exchange, as shown in Fig.~\ref{fig:0nubb}. The amplitude for this process depends on the left-right down-type squark mixing given by Eq.~\eqref{eq:mixsquark}.  Using the latest lower limits on the $0\nu\beta\beta$ lifetime~\cite{KamLAND-Zen:2016pfg, Agostini:2018tnm}, we obtain a bound on the combination~\cite{Pas:1998nn} 
\begin{align}
|\lambda'_{131}\lambda'_{113}|\frac{(A^b-\mu\tan\beta)}{m_{\widetilde b_{R}}^4} \  \lesssim \ 10^{-14}~{\rm GeV}^{-3}.
\end{align}
We checked that this condition is easily satisfied in all three benchmark cases considered here, again due to the choice of the $\lambda'$ couplings in RPV3, and also due to the requirement of small $(A^b-\mu\tan\beta)$ for the neutrino mass.

% \subsection{\texorpdfstring{$W$ and $Z$ Couplings}{}}

%  Need to check the size.

\section{LFV Predictions} \label{sec:LFV}

% \subsection{\texorpdfstring{$B \to K^{(*)} \tau \tau$ and $B \to K^{(*)} \tau \mu$}{}}

% both tree level contributions from stop exchange and various 1-loop contributions.

%AS om tau LFV
%\subsection{}$\tau$ lepton flavor violating decays %arise  quite 
%naturally at tree and loop level, see equation 7 %and 8 in ADS.
%There are many interesting examples: $\tau \to \mu %\phi (\rho, \omega, \pi^0 \eta, \eta^{\prime}$, %$K^+ K^-$, $\pi^+ \pi^-$ etc.
%PDG gives current bounds on many of these modes of %around $10^{-8}$.
%In the next few years, Belle-II and possibly other %experiments should be able to improve on these by %1-2 orders of magnitudes.

%In our RPV setup these lepton FV decays of the %$\tau$ are accompanied by electrically neutral %hadronic current and also similarly a neutral lep

In this section, we make predictions for LFV decay modes of the $\tau$-lepton and rare decays of the $B$-mesons for our three benchmark cases, anticipating that future experiments like Belle II~\cite{Kou:2018nap} or upgraded LHCb~\cite{Bediaga:2018lhg} might be able to test some of these predictions.  

%%%%%%%%%%%%%%%%%%%%%%%%%%%%%%%%%%%%%%%%%%%%%%%%%%%%%%%%%%%%%
%table with sample of interesting modes of lfv
\begin{table*}[tbh!]
\begin{tabular}{|c|c|c|c|c|c|}\hline\hline
Flavor-violating  & $\lambda$,$\lambda'$  & \multicolumn{3}{c|}{RPV3 Prediction} & Current experimental \\ \cline{3-5}
decay mode & dependence &  
Case 1
 & Case 2 & Case 3 & bound/measurement  \\ 
           
\hline\hline

$\tau \to \mu \phi$  & $\lambda'_{332} \lambda'_{232}$, $\lambda_{323} \lambda'_{322}$ & 
%$2 \times 10^{-10}$ 

$1.9\times 10^{-15}
$ 

& $
3.8\times 10^{-10}
$
& $2.6\times10^{-12}$
& $<8.4 \times 10^{-8}$~\cite{Miyazaki:2011xe} \\

$\tau \to \mu K K$ & $\lambda'_{332} \lambda'_{232}$, $\lambda_{323} \lambda'_{322}$ & 
%$3 \times 10^{-11}$ 
$
1.2\times10^{-17}
$

& $
2.4\times 10^{-12}
$
& $2.9\times 10^{-13}$
& $<4.4 \times 10^{-8}$ ~\cite{Miyazaki:2012mx}\\

$\tau \to \mu K_s^0$ & $\lambda'_{332} \lambda'_{231}$, $\lambda'_{312} \lambda_{323}$  & 
%$6 \times 10^{-11}$ 
$4.5 \times 10^{-19}$ 
& $8.7 \times 10^{-12}$
& $3.1\times 10^{-13}$
& $<2.3 \times 10^{-8}$ ~\cite{Miyazaki:2010qb}\\

$\tau \to \mu \gamma$ & $\lambda'_{333}\lambda'_{233}$, $\lambda_{133} \lambda_{123}$ 
&  $1.3 \times 10^{-10}$
%$1.1 \times 10^{-11}$ 

& $1.3 \times 10^{-8}$
& $2.4 \times 10^{-10} $
& $<4.4 \times 10^{-8}$ ~\cite{Aubert:2009ag}\\

%$\tau \to \mu \ell^+ l^-$ & $\lambda'_{333}\times \lambda'_{233}$, $\lambda_{133}\times \lambda_{123}$ 
%&$4.2 \times 10^{-12}$
%& $2.4 \times 10^{-10}$
%& $\sim0$
%& $2 \times 10^{-8}$ %~\cite{Hayasaka:2010np}\\

$\tau \to \mu\mu\mu$ &  $\lambda_{323} \lambda_{322}$ & 
%$1.5 \times 10^{-10}$ 
$1.7 \times 10^{-11}$ 

& $1.2\times10^{-9}$
& $1.2\times 10^{-11}$
& $<2.1 \times 10^{-8}$ ~\cite{Hayasaka:2010np}\\ \hline

$B_{(s)} \to K^{(*)}(\phi) \mu \tau$ & 
$\lambda'_{333} \lambda'_{232}$, $\lambda'_{233} \lambda'_{332}$, $\lambda'_{332} \lambda_{323}$
%, $\lambda'_{332}\times \lambda_{323}$
& 
%\red{$7 \times 10^{-7}$ }
$
4.1\times 10^{-9}
$

& $
1.2\times10^{-7}
$
& $2.2\times10^{-10}$
&  $<2.8\times 10^{-5}$~\cite{Lees:2012zz} \\

$B_s \to \tau \mu$ & $\lambda'_{333} \lambda'_{232}$,
$\lambda'_{233} \lambda'_{332}$,  $\lambda'_{332} \lambda_{323}$ & 
%\red{$1.3 \times 10^{-8}$}
$
4.4\times 10^{-10}
$

& $
1.3\times10^{-8}
$
& $2.3\times10^{-11}$
& $<3.4\times 10^{-5}$~\cite{Aaij:2019okb} \\

$b \to s \tau \tau$ & $\lambda'_{333} \lambda'_{332}$ &  
{$3.4 \times 10^{-7}$ }
%\red{$7 \times 10^{-7}$ }

& {$2.8 \times 10^{-8}$ }
& $1.3\times10^{-13}$
& $N/A$ \\

$B \to K^{(*)} \tau \tau$ & $\lambda'_{333} \lambda'_{332}$ &  
{$3.7 \times 10^{-6}$ }
%\red{$7 \times 10^{-7}$ }

& {$4.2 \times 10^{-8}$ }
& $9.6\times10^{-12}$
& $<2.2 \times 10^{-3}$~\cite{TheBaBar:2016xwe} \\ %~\cite{Lees:2015uun}\\

$B_s \to \tau \tau$ & $\lambda'_{333} \lambda'_{332}$ & 
{$3.7 \times 10^{-8}$}
%\red{$1.3 \times 10^{-8}$}

& $3.0 \times 10^{-9}$
& $1.4\times10^{-14}$
& $<6.8 \times 10^{-3} $ ~\cite{Aaij:2017xqt}\\

$b \to s \mu \mu$ & $\lambda'_{233} \lambda'_{232}$, $\lambda'_{332} \lambda_{232}$ & 
{$5.9 \times 10^{-9}$ }
%\red{$7 \times 10^{-7}$ }

& $3.2 \times 10^{-8}$
& $8.8 \times 10^{-9}$
& $4.4 \times 10^{-6}$~\cite{Lees:2013nxa}\\

$B_s \to \mu \mu$ & $\lambda'_{233}\lambda'_{232}$, $\lambda'_{332} \lambda_{232}$ & 
{$4.1 \times 10^{-11}$}
%\red{$1.3 \times 10^{-8}$}

& $6.5 \times 10^{-11}$
& $1.8\times10^{-11}$
& $3.0 \times 10^{-9}$~\cite{Aaij:2017vad}\\ \hline \hline
\end{tabular}
\caption{RPV3 contributions to the branching ratios of the flavor-violating decay modes of $\tau$ and of $B$-mesons in the three benchmark cases considered here. Also shown are the current experimental bounds at 90\% CL for each channel. There is no existing bound on $b\to s\tau\tau$, so that entry is labeled as $N/A$. For the last two decay modes, namely, the inclusive $B\to X_s\mu^+\mu^-$ and exclusive $B_s\to \mu^+\mu^-$, we show the central values of the experimental measurements. The values for Case 1 are calculated with the parameter set in Eq.~\eqref{eq:bp1} along with $-\epsilon = 0.02$ and $m_{\widetilde{b}_R} = 2.0$ TeV from the overlap region in Fig.~\ref{fig:ANITARdRkgm2CKM}. For case 2, the parameters are set in Eq.~\eqref{eq:bp2}, along with $\lambda' = 0.8$ and $m_{\widetilde{b}_R} = 2.0$ TeV from the overlap region in Fig.~\ref{fig:ANITARdRkgm2ST}. For case 3, the parameters are set in Eq.~\eqref{eq:bp3} with $\lambda'=0.2$ and $m_{\widetilde{b}_R} = 3.0$ TeV from the overlap region in Fig.~\ref{fig:ANITARdRkgm2}. %Notice that for entries of $B_{(s)} \to K^{(*)}(\phi) \mu \tau$ and $B_s \to \tau \mu$, $\lambda'_{(2)333} \lambda'_{(3)232}$ means $\lambda'_{333} \lambda'_{232}$ or $\lambda'_{233} \lambda'_{332}$.  
}
\label{tab.LFVtauB}
\end{table*}
%%%%%%%%%%%%%%%%%%%%%%%%%%%%%%%%%%%%%%%%%%%%%%%%%%

\subsection{Tree-level LFV \texorpdfstring{$\tau$}{} Decays}
In our RPV3 setup $\tau$-LFV decays arise  quite 
naturally at tree and loop level, see Ref.~\cite{Altmannshofer:2017poe}.
There are many interesting channels at tree level: $\tau \to \ell Y$  (where $\ell=e,\mu$ and $Y$ stands for $\phi,\rho, \omega, \pi^0, \eta, \eta^{\prime}$, $K^+ K^-$, $\pi^+ \pi^-$ etc.). The PDG~\cite{Tanabashi:2018oca} gives current bounds on the branching ratios of many of these modes at around $10^{-8}$ level.
In the next few years, Belle-II and possibly other experiments like LHCb should be able to improve on these by 1-2 orders of magnitudes. Since the branching ratios scale as $(m_W/M)^{4}$, where $M$ is the mediator mass, it is important to understand that these existing stringent bounds of $10^{-8}$ do not necessarily mean that the masses of the LFV interactions are 100 times heavier than $m_W$ since we also expect rotations in flavor space to carry suppression factors, in complete analogy with what we see in weak interactions of the SM. In fact in the SM, the magnitude of the observed CP asymmetries are an even better illustration of the effect of rotations in flavor space. Due to mixing angles in flavor space we witness ${\cal O}(1)$ CP asymmetries in some decays involving the $b$-quark whereas they  become ${\cal O}(10^{-3})$ or even smaller in strange and charm decays. 

For illustrative purposes, let us first consider the simple Case 1 with CKM-like coupling structure. Concretely, we plan to implement the third-generation centric rotations due to RPV interactions in complete analogy with the SM.
We just have to bear in mind that in RPV3 we interchange the role of the first and third generations compared to the SM. Moreover, as in the SM, the order parameter, $\lambda \simeq 0.23$ in the Wolfenstein representation~\cite{Wolfenstein:1983yz} can be used for flavor rotations in our RPV3 set up. In particular, when RPV interactions $\tau \to$ $u$ and $\mu \to$ $u$ are involved, in a similar fashion, these can be accompanied by suppression factors,  
say, $\epsilon_{31} \epsilon_{21}$, where $\epsilon_{31}\approx \lambda^2$ and $\epsilon_{21}\approx \lambda^3$. In line with our thinking that superpartners of third generation quarks are the lightest, these rotations may be analogous to $V_{ub}$ and $V_{cb}$ respectively with the product causing a suppression in the rate
of order $\lambda^{10} \approx 4\times 10^{-7}$. Thus,  with a mediator mass of $M\simeq 1.6$ TeV (20 times heavier than $W$), this can result in a branching ratio  of ${\cal O}(10^{-12})$ and be completely consistent with the current bounds.

\begin{figure}[t!]
		\includegraphics[width=0.33\textwidth]{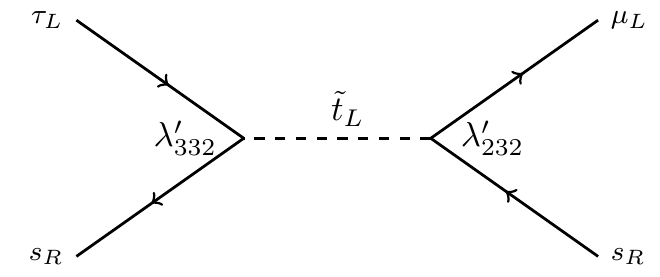}
	\caption{Contribution to $\tau \to \mu s \bar s$ from $\lambda'$ in RPV3 at tree level.}
	\label{fig:tautomuqq}
\end{figure}

So clearly there is significant model dependence involved at this stage and  we will just need to dig the appropriate effects of these rotations in flavor space from the experimental data. In this third-generation centric RPV3 model of ours, it would seem that 
$\tau \to \mu \bar{s} s$ final states may be less suppressed than those with $uu$, $dd$ and $sd$. The $\tau \to \mu \bar s s$ process, shown in Fig.~\ref{fig:tautomuqq}, gives rise to distinctive final states such as $\tau \to \mu \phi ~[K^+ K^-]$.
%%%%%%%%%%%%%%%%%%%%%%%%%%%%%%%%%%%%%%%%%%%%%%%%%%
%%%%%%%%%%%diagram for tau to mu  q q
Making the ad hoc assumption that these couplings go as $\epsilon_{32} \approx \lambda \approx 0.23$, a mediator mass of 1.6 TeV can lead to 
\begin{align}
   {\rm BR}(\tau \to \mu \phi) & \ \simeq \ 
\frac{\lambda^6}{|V_{us}|^2}\left(\frac{\lambda'_{333}}{g}\right)^4 \left(\frac{m_W}{m_{\tilde t}}\right)^4  {\rm BR} (\tau \to \nu K^{*}) \nonumber \\
& \ \approx \ 1.2\times 10^{-9} \, ,
\label{eq:taumuphi}
\end{align}
where we have used BR($\tau \to \nu K^*$) $\approx 1.2 \%$ and $\lambda'_{333} \sim 3.5$, which is taken as the value from case 1 with $g\sim0.66$ being the weak coupling constant. The prediction in Eq.~\eqref{eq:taumuphi} is 
consistent with current bounds and perhaps within reach of LHC experiments as well as Belle~II. 

Similarly we can estimate BR($\tau \to \mu K K$) $\approx 8.0 \times 10^{-10}$
by normalizing to the SM mode BR($\tau \to \nu K K$) $\approx 1.5 \times 10^{-3}$.

Yet another simple mode where we can make a statement about the branching ratio  is
%28	
$\tau \to \mu K^0$. This can be normalized conveniently to the SM mode $\tau \to \nu K^+$ which has a branching ratio  of about $ 7 \times 10^{-3}$. Note that
%29	
as above the $\tau \to s$ RPV vertex will carry a suppression of $\lambda$. The $\mu \to d$ vertex couples second generation to first; thus this is analogous to $V_{cb}$ in the SM and the rate goes as $(\lambda^3/|V_{cb}|)^2\approx \lambda^2$.
%30	
Putting all the factors together, one finds BR($\tau \to \mu K^0$) $\approx 5 \times 10^{-10}$. 

%%%%%%%%%%%%%%%%%%%%%%%%%%%%%%%%%%%%%%%

Another interesting example is $\tau \to \mu\mu\mu$. This arises at tree level via
use of $LLE$ couplings of RPV [cf.~Eq~\eqref{Eq.RPVLLE}]. We again assume a suppression 
of $\epsilon_{32} \simeq \lambda \approx 0.23$. Then again for a mediator mass of 1.6 TeV,
we can get 
\begin{align}
 {\rm BR}(\tau \to 3 \mu) \ \simeq \ \lambda^2 \left(\frac{m_W}{m_{\widetilde \nu_\tau}}\right)^4
\left(\frac{\lambda_{323}}{g}\right)^4 {\rm BR}(\tau \to \mu \nu \bar{\nu}) \, ,
\end{align}
where $\lambda_{323}\sim 1.5$ is taken as the value from case 1 with $g\sim0.66$ being the weak coupling constant.
% the index $k$ in $\lambda_{32k}$ is either 2 or 3. 
In this calculation we have assumed that when the third-generation sneutrino couples to two muons which are from second generation, there is a suppression of ${\cal O}(\lambda^2)$ in the vertex. Using the SM $\tau$ branching ratio  for leptonic decays of $\approx 16 \%$, we
get BR($\tau \to 3\mu$) $\approx 7.5\times 10^{-9}$ %1.5 \times 10^{-10}$
whereas the current bound is $2 \times 10^{-8}$. 

In Table~\ref{tab.LFVtauB}, we summarize the above-mentioned tree-level LFV decay modes of $\tau$, with the dominant coupling dependence in our RPV3 setup and the model predictions in each of the three cases discussed above, corresponding to the parameters in the overlap regions shown in Figs.~\ref{fig:ANITARdRkgm2CKM},~\ref{fig:ANITARdRkgm2ST} and~\ref{fig:ANITARdRkgm2}. Also shown are the current experimental constraints on each channel. As can be seen, all the three benchmarks are consistent with the current bounds, while some of the predictions might be accessible at future $B$-factories. Note that the tree-level BRs in Case 1 turn out to be much smaller than our naive estimate discussed above, because we have used the value of $|\epsilon|=0.02$ for the overlap region in this case  (cf.~Fig.~\ref{fig:ANITARdRkgm2CKM}), which is a factor of 10 smaller than the simple choice of $|\epsilon|\simeq \lambda \approx 0.23$.
%\textcolor{red}{need to adapt the last paragraph, taking into account that $\epsilon$ in our Case 1 is much smaller than 0.23}

%%%%%%%%%%%%%%%%%%%%%newly added for Prof. Soni

\subsection{LFV via Loop Decays of \texorpdfstring{$\tau$}{}}  
\begin{figure}[tbh!]
		\centering
		\includegraphics[width=0.33\textwidth]{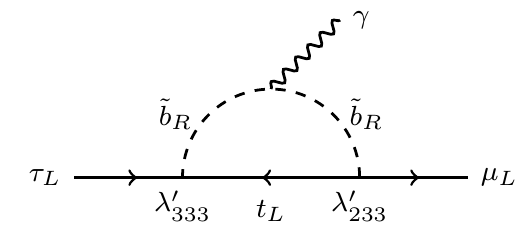}
	\caption{Dominant contribution to $\tau \to \mu \gamma$ in RPV3 at one-loop level. Note that the emitted photon could be attached to all possible charged propagators and external legs and what we show here is just one possible diagram.}
	\label{fig:tautomugamma}
\end{figure}

There are interesting LFV loop decays of $\tau$ that we can estimate quite easily by using existing calculations of $b \to s \gamma$~\cite{Deshpande:1987nr} and of $b\to s \ell^+ \ell^-$~\cite{Hou:1986ug}. These calculations are relevant as the virtual top quark dominates in $b$ decay as well as in $\tau$ decays because of the simple picture of mixing angles that we have adopted. The dominant diagram is shown in Fig.~\ref{fig:tautomugamma}, and we find the decay width for $\tau \to \mu \gamma$ contributed from RPV to be
%%%%%%%%%%%%%%%%%%%%%%%%%%%%%%%%%%%%%%%%%%%%%%%%%%
%%%%%%%%%%%diagram for tau to mu  gamma
\begin{align}
& {\Gamma(\tau \to \mu \gamma) \  \simeq \ \frac{\alpha_{\rm em} m_{\tau}^5 G_F^2}{256 \pi^4} \frac{m_W^4}{g^4}}\times \nonumber \\
& {\left| \sum_k \left( \frac{\lambda^{\prime\,*}_{3k3}\lambda^\prime_{2k3}}{m_{\tilde b_R}^2} + \frac{2\lambda^*_{3k3}\lambda_{2k3}}{3 m_{\tilde \nu_\tau}^2} - \frac{\lambda^*_{k33}\lambda_{k23}}{3 m_{\tilde \tau_R}^2}- \frac{\lambda^*_{k33}\lambda_{k32}}{3 m_{\tilde \tau_L}^2} \right) \right|^2}
\end{align}
which reduce to the following when keeping only the dominant term:
\begin{align}
& {\Gamma(\tau \to \mu \gamma) \ \simeq \ \frac{\alpha_{\rm em} m_{\tau}^5 G_F^2}{256 \pi^4} \frac{m_W^4}{g^4}} {\left| \left( \frac{\lambda^{\prime\,*}_{333}\lambda^\prime_{233}}{m_{\tilde b_R}^2} \right) \right|^2}
\end{align}
% \begin{align}
% \Gamma(\tau \to \mu \gamma) \ \simeq \ & \frac{\alpha_{\rm em} m_{\tau}^5 G_F^2}{288 \pi^4} \frac{m_W^4}{m_{\widetilde b}^4}  \left(\frac{(3\lambda'_{333}\lambda'_{233})^2}{g^4}\right) 
% %\nonumber\\
% %&\times
% \left(1 - \frac{m_{\mu}^2}{m_{\tau}^2}\right)^3,
% \end{align}
% where 
% %for $m_{t} \approx 174$ GeV, we have $\kappa_t = \alpha_s \ln(m_t^2/{m_c^2}) \approx 0.40$ and 
% $I\sim 0.075$ is a loop function~\cite{Lavoura:2003xp} that depends on the ratio of top mass to sbottom mass. 
Thus, with $\lambda'_{233}\sim\lambda'_{333}\lambda$ and $\lambda\sim 0.23$, we estimate that BR$(\tau \to \mu \gamma) \sim 10^{-8}$.

% \textcolor{red}{WA: Here is the general expression that I get (in analogy to g-2 there are also loops with staus and sneutrinos with LLE interactions that can contribute)}
% \begin{align}
% & \textcolor{red}{\Gamma(\tau \to \mu \gamma) \simeq \frac{\alpha_{\rm em} m_{\tau}^5 G_F^2}{256 \pi^4} \frac{m_W^4}{g^4}} \nonumber \\
% & \textcolor{red}{\left| \sum_k \left( \frac{\lambda^{\prime\,*}_{3k3}\lambda^\prime_{2k3}}{m_{\tilde b_R}^2} + \frac{2\lambda^*_{3k3}\lambda_{2k3}}{3 m_{\tilde \nu_\tau}^2} - \frac{\lambda^*_{k33}\lambda_{k23}}{3 m_{\tilde \tau_R}^2}- \frac{\lambda^*_{k33}\lambda_{k32}}{3 m_{\tilde \tau_L}^2} \right) \right|^2}
% \end{align}
% comment: the second and the fourth term vanish:
% the second term, only lambda_313* lambda_213 survive the anti-symmetric rule, this is small in case 1 and case 2 and 0 in case 3. Also, in case 1 and 2, m_{\tilde nu} is much larger
% the fourth term, in all cases, m_{\tilde tau_L} being large, lambda_{133}* lambda_{132} and lambda_{233}* lambda_{232} survive the anti-symmetric rule, but they are 0 for case 3 and 2 and  small in case 1.
%%% as for the third term,
% m_{\tilde tau_R} is of the good range, but couplings vanish for case 2 and 3, and is small for case 1.
In an analogous fashion, in the loop decays $\tau \to \mu \ell^+ \ell^-$ (for $\ell = \mu, e$),
the virtual top-quark dominates as in the case of $b \to s \ell^+ \ell^-$. This leads one to the estimate,
\begin{align}
&\frac{{\rm BR}(\tau \to \mu \ell^+ \ell^-)}{{\rm BR}(\tau \to \mu \gamma)} \ \approx \ 
\frac{{\rm BR}(b \to s \ell^+ \ell^-)}{{\rm BR}(b \to s \gamma)} \  
\approx \ 0.05 \, .
\end{align}
Thus, we conclude that the loop contribution to BR($\tau \to 3 \mu$) is about a hundred times smaller compared to the tree contribution estimated above. 

Another class of loop modes emerges from considering $\tau \to \mu +$ gluon(s). This is difficult to estimate reliably.
Based on gauge invariance the $\tau \to \mu +$ gluon amplitude vanishes and we expect that the amplitude for $\tau \to \mu +$ 2~gluons is suppressed by four powers of sfermion masses. A rough estimate thus gives 
\begin{align}
{\rm BR}(\tau \to \mu gg) \ \sim \  \frac{\alpha_s^2}{4 \pi \alpha_\text{em}} \frac{m_\tau^4}{m_{\tilde b_R}^4} \textrm{BR}(\tau \to \mu \gamma) \, .
\end{align}
Using $\alpha_s \simeq 0.3$ and $m_\tau/m_{\tilde b_R} \simeq 10^{-3}$, we obtain ${\rm BR}(\tau \to \mu gg) \sim 10^{-20}$, which is many orders of magnitude below our expectation for the tree level $\tau \to \mu s \bar s$ branching ratio. 
%We therefore expect exclusive hadronic modes such as $\tau \to \mu + \eta', \pi^+ \pi^-, ...$ with branching ratios that do not exceed the branching ratio $\tau \to \mu K^+ K^-$.

Our RPV3 predictions for the loop-level $\tau$ LFV decays are also summarized in Table~\ref{tab.LFVtauB} for all three cases, along with the corresponding experimental bounds. 

%%%%%%%%%%%%%%%%%%%%%%%%%%%%%%%%%%%%%%%%%%%%

\subsection{\texorpdfstring{LFV Decays of $B$-mesons}{LFV Decays of B-mesons}}
%%%%%%%%%%%diagram for b to s  l l
\begin{figure}[tbh!]
		\centering
		\includegraphics[width=0.33\textwidth]{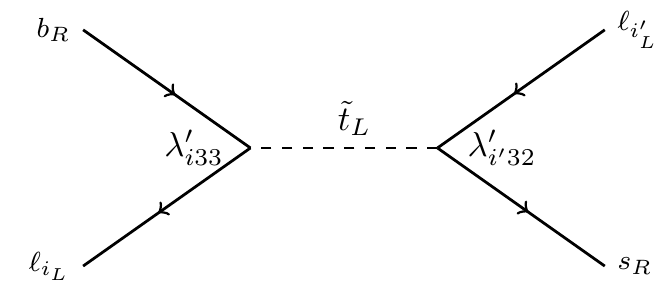}
	\caption{Generic diagram for $b \to s \ell_i \ell_{i'}$ in RPV3 at tree level.}
	\label{fig:btosll}
\end{figure}

We briefly discuss here some illustrative examples of distinctive LFV decays of $B$-mesons that
proceed via tree processes in our RPV3 scenario and can be estimated readily.
%%%%%%%%%%%%%%%%%%%%%%%%%%%%%%%%%%%%%%%%%%%%%%%%%%
First example we want to discuss is $b \to s \tau \mu$, whose general diagram is shown in Fig.~\ref{fig:btosll}. In RPV, because of the presence of leptoquark-type interactions, leptons and quarks should be treated on the same footing when it comes to flavor rotations accompanying RPV interactions. With this in mind,  in our third-generation-centric setup with flavor rotations as explained above, $b \to s\tau \mu$ results in the exclusive modes, $B \to K^{(*)} \mu \tau$~\cite{Choudhury:2017qyt} or $B_s \to \phi \mu \tau$.  In this case the $b \to \tau$ vertex has no suppression but $s \to \mu$,  both being second generation fermions, carry a suppression of $\lambda^2$ at amplitude level, making BR$(b \to s \mu \tau) \sim (\lambda'_{333}/g)^4 \lambda^4$. Thus once again taking the mediator mass of 1.6 TeV and taking the normalizing weak decay
$B \to \ell \nu X_c$ with ${\rm BR}\approx 11\%$ which involves a suppression factor of $V_{cb}^2$ results in
BR$(b \to s \mu \tau) \approx 7 \times 10^{-7}$.  The BRs of the corresponding exclusive manifestations are likely a factor of 10 smaller as indicated in Table~\ref{tab.LFVtauB}. Also notice that for both case 1 and case 2, contributions from $\lambda'_{233}\lambda'_{232}$ and $\lambda'_{332} \lambda_{232}$ dominate due to the smaller stop mass compared to sneutrino mass. As for case 3, loop-level contribution is taken into account due to the tree-level terms being vanishingly small.

Another related extremely interesting example is $B_s \to \tau \mu$.
Let us normalize this to the SM mode of $B \to \tau \nu$. In this case though the LFV BR($B_s \to \tau \mu$) carries a suppression of $\lambda^4$, that is more than compensated by $V_{ub}^2$ factor in the normalizing mode.
Thus, again for a mediator mass of 1.6 TeV, we get BR($B_s \to \tau \mu$) $\approx 8.4 \times 10^{-8}$.

It is to be stressed that these BRs of flavor violations involving $\tau \mu$ final states of $B$ and $B_s$ are rather large and future experiments like Belle II and upgraded LHCb should be able to constrain them quite well.

For completeness, we also list in Table~\ref{tab.LFVtauB} our RPV3 predictions in the lepton flavor conserving FCNC decay modes, such as $B_s\to \ell^+\ell^-$. Although in all our benchmark cases, the model predictions are quite small for these channels, it is conceivable that in a less restrictive setup, the RPV contributions could be within reach of upcoming experiments.

%%%%%%%%%%%%%%%%%%%%%%
\section{\texorpdfstring{High-$p_T$ Predictions at the LHC}{High-pT Predictions at the LHC}}\label{sec:LHC}
%%%%%%%%%%%%%%%%%%%%%%%%%%%%%%%%%%%%%%%%%%%%%%%%%%%%%%%%%
As shown in Ref.~\cite{Altmannshofer:2017poe}, simple crossing symmetry arguments can be used to establish a high-$p_T$ model-independent test of the $R_{D^{(*)}}$ anomaly in CMS and ATLAS experiments; see also Refs.~\cite{Faroughy:2016osc, Greljo:2018tzh, Afik:2019htr}. The basic idea is that the underlying quark-level process for $R_{D^{(*)}}$ is $b\to c\tau \nu$, which by crossing symmetry also implies the existence of the processes like $gc\to b\tau \nu$ and $gb\to c\tau \nu$, which can be searched for in the high-$p_T$ LHC experiments. We do not wish to repeat the same analysis here, but would like to stress the point that similar model-independent tests can be done for the $R_{K^{(*)}}$ anomaly as well; see also Refs.~\cite{Greljo:2017vvb, Afik:2018nlr}. 

Specifically, the underlying parton-level process for $R_{K^{(*)}}$ is $b\to s\ell^+\ell^-$ (with $\ell=e,\mu$), and by crossing symmetry, the following processes must also occur in the $pp$ collisions at the LHC: (i) $bs\to \ell^+\ell^-$, (ii) $gb\to s\ell^+\ell^-$ and (iii) $gs\to b\ell^+\ell^-$ (here $g$ stands for gluon and $q$ generically stands for both quarks and anti-quarks). So if the $R_{K^{(*)}}$ anomaly were true, we must also have an anomaly in these channels, which might be observable depending on the signal to background ratio. 
%\blue{For illustration purpose, the Feynman diagram from the most substantial process, $gs\to b l^+ l^-$, is denoted in Fig.~\ref{fig:gstobll}.}

The signal in each case can be analyzed in the four-fermion setup with the vector operators defined in Eq.~\eqref{eq:Heff} and an effective mass scale of ${\cal O}$(TeV). Scalar and tensor operators do not work here, unlike in the $R_{D^{(*)}}$ case, because of the $B_s\to \mu^+\mu^-$ constraint, which can be helicity-suppressed only in the vector case. Moreover, as the LHC center-of-mass energy is comparable to the mass scale of the effective operator being studied here, it is more accurate to use an explicit mediator. To be concrete, we use the RPV3 model as our benchmark, where one of the squarks serves as the mediator for the processes listed above and couples via the $\lambda'$-type $LQD$ interaction. Note that the $\lambda$-type $LLE$ interactions do not enter here at the leading order, since we must have $b$ and $s$ quarks in the external legs to relate to the $R_{K^{(*)}}$ anomaly.   

\begin{figure}[t!]
		\centering
		\includegraphics[width=0.33\textwidth]{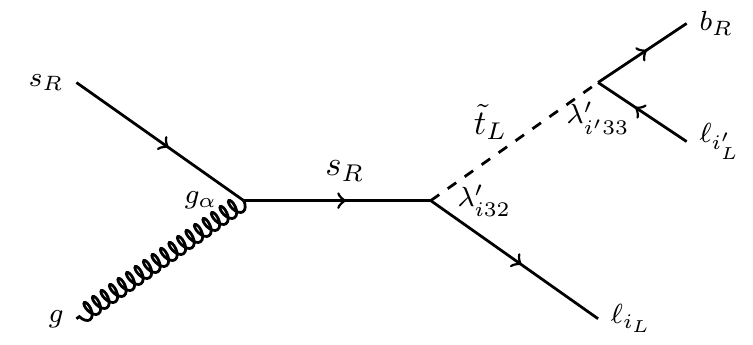}
	\caption{Dominant contribution to the collider process $pp \to b \ell_i \ell_{i'}$ in RPV3 at tree level.}
	\label{fig:gstobll}
\end{figure}

\begin{figure*}[t!]
\includegraphics[width=0.4\textwidth]{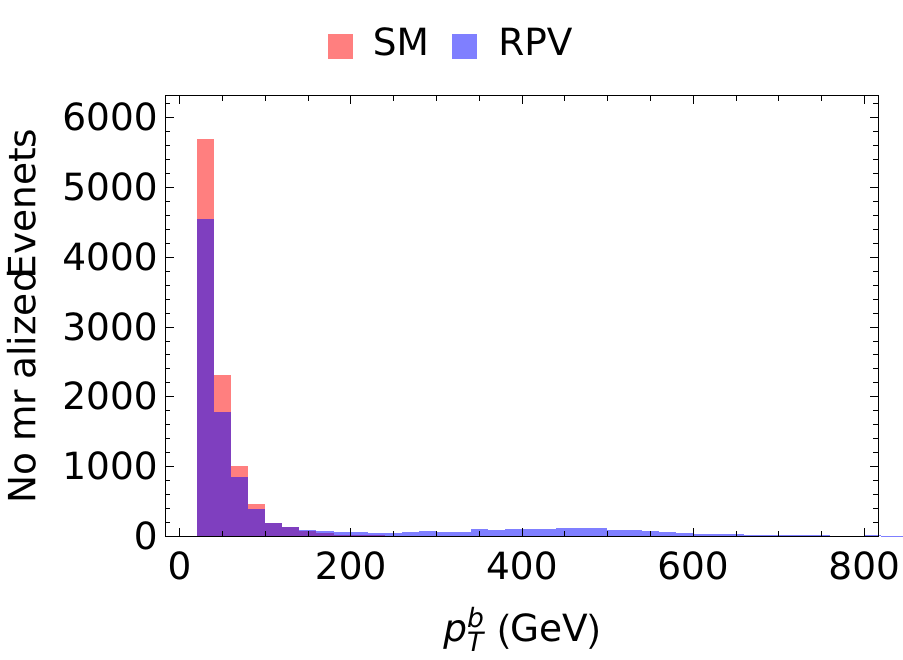}
\includegraphics[width=0.4\textwidth]{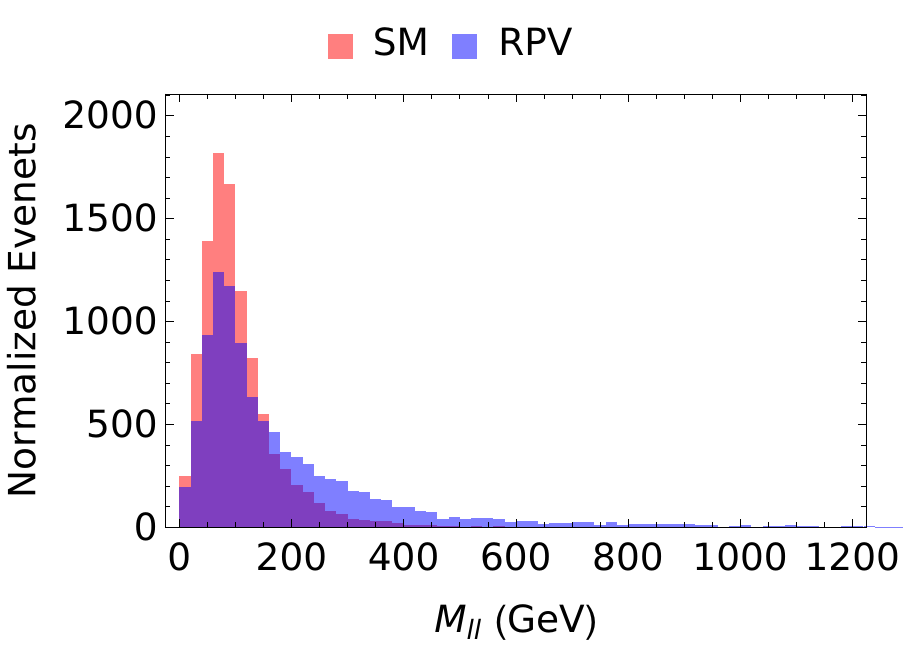}\\
\includegraphics[width=0.4\textwidth]{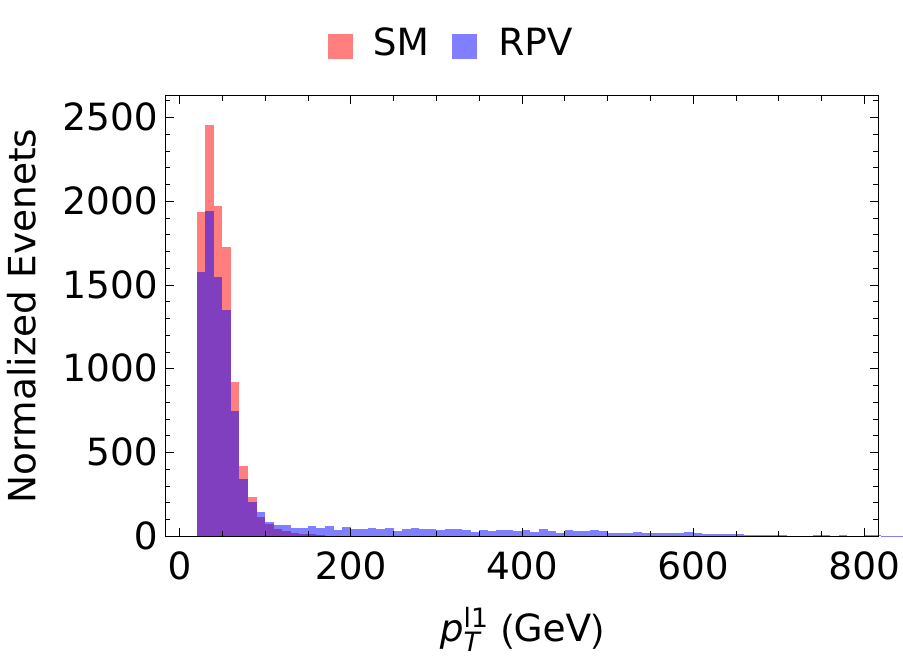}
\includegraphics[width=0.4\textwidth]{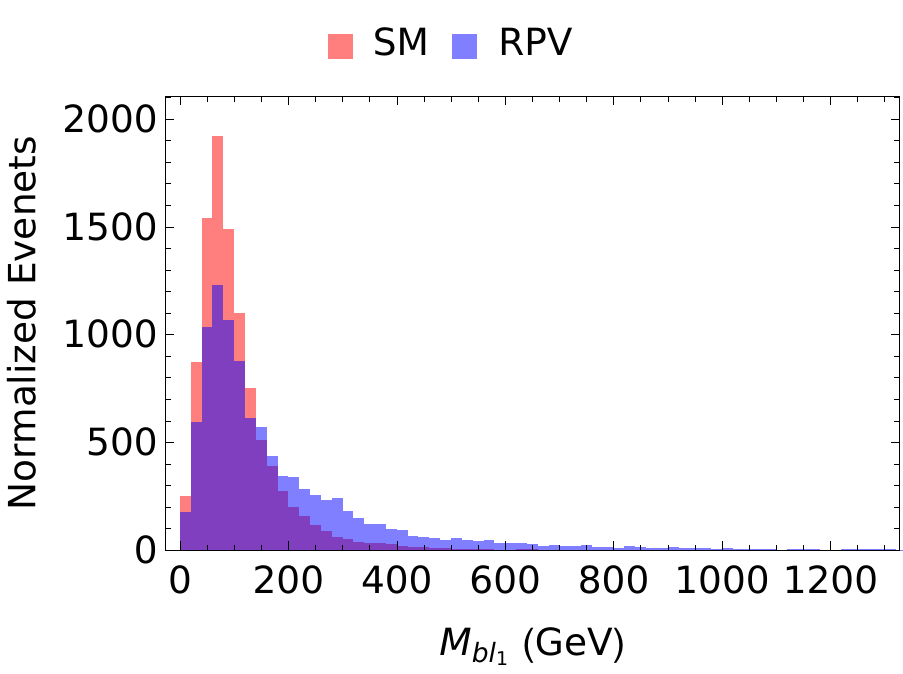}
 \\
\includegraphics[width=0.4\textwidth]{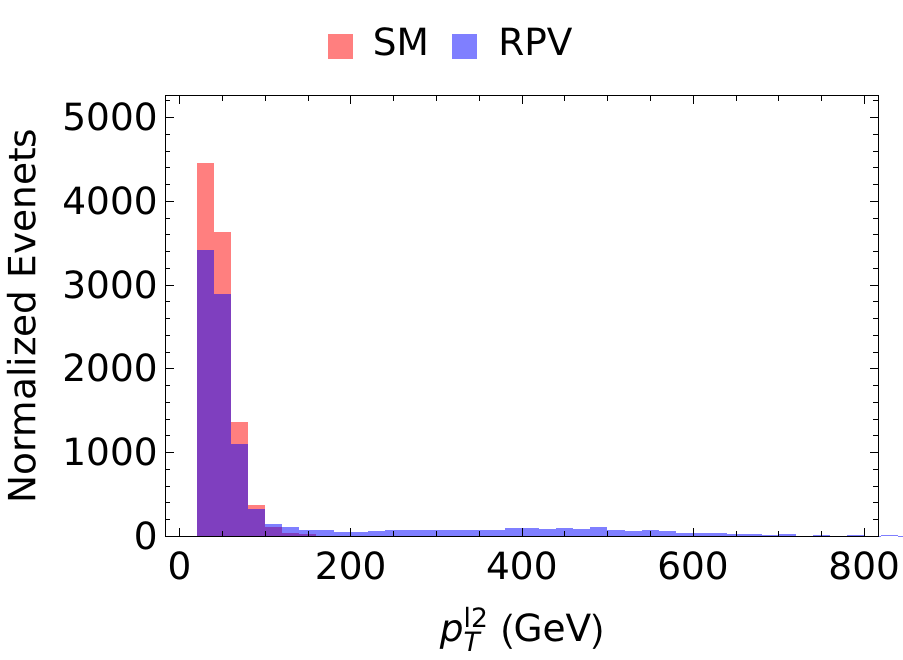}
\includegraphics[width=0.4\textwidth]{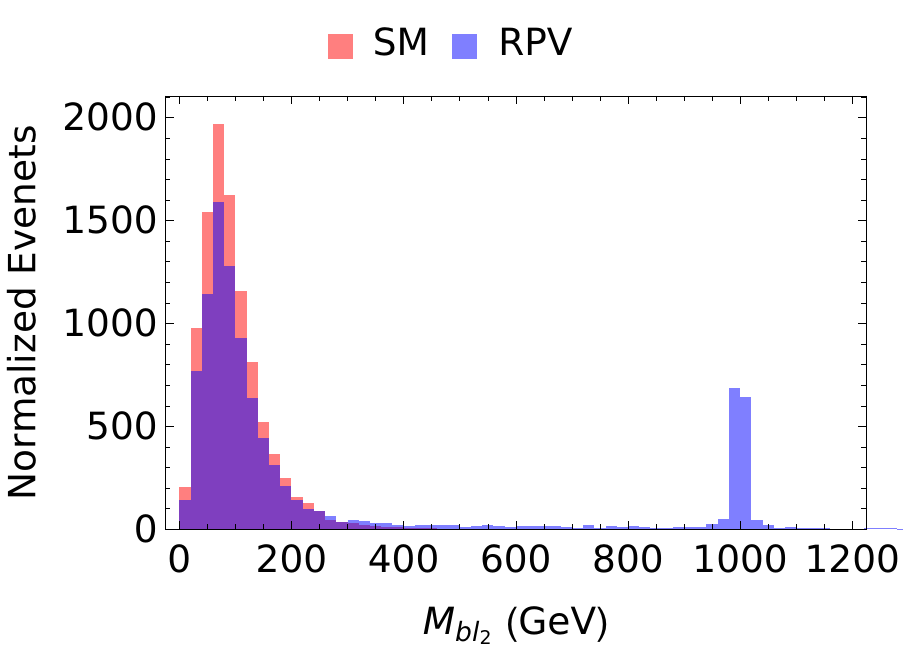}
\caption{Kinematic distributions for the $pp\to b\ell_1\ell_2$ signal in the RPV model (blue) and the corresponding SM background (red). The left panels show the transverse momentum distributions for the bottom quark and the two charged leptons, whereas the right panel shows the invariant mass distributions for the dilepton and the two bottom quark-lepton combinations. In the RPV3 model under consideration, the right combination of $M_{bl}$ gives a peak at the squark mass, as shown in the last plot.}
\label{fig:dist}
\end{figure*}

For case (i), we find that the SM background is overwhelmingly large, mostly coming from $Z\to \ell^+\ell^-$. Imposing an invariant mass cut on $M_{\ell\ell}$ to exclude the $Z$-mass window helps, but still we find it difficult to achieve a signal significance of more than $3\sigma$. Similarly, for case (ii), the signal cross section is suppressed due to the bottom-quark parton distribution function in the proton. So the best case scenario is case (iii), as shown in Fig.~\ref{fig:gstobll}, where the additional $b$ jet in the final state provides a better handle on the signal over background. Some simple kinematic distributions for the corresponding signal and background are shown in Figure~\ref{fig:dist}. Here we have used the minimal trigger cuts: $p_T^{j,b\ell}>20$ GeV, $|\eta^{j,b,\ell}|<2.5$ and $\Delta R^{\ell\ell}>0.4$ and an average $b$-tagging efficiency of 70\%. From the distributions, we find that although the $M_{\ell\ell}$ distribution can distinguish the RPV signal from the SM background to some extent, the striking signature of RPV comes in the distribution of the invariant mass $M_{b\ell}$ with the correct lepton combination. This is because in RPV3, we have the process $gs\to \widetilde{t}_L\mu^-\to b\mu^+\mu^-$ through the $\lambda'$-couplings [cf. the penultimate term in~Eq.~\eqref{Eq.lambda_prime}]. Thus, if kinematically allowed, the stop can be produced on-shell in the $s$-channel, followed by its decay into $b\mu^+$, thereby giving a resonance peak in the $M_{b\mu^+}$ distribution, as shown in the bottom right panel of Fig.~\ref{fig:dist} for a representative stop mass of 1 TeV. Using this resonance feature, it is possible to achieve more than $3\sigma$ signal significance for the overlap region shown in Fig.~\ref{fig:ANITARdRkgm2} at the 14 TeV LHC with an integrated luminosity of 3 ab$^{-1}$. 
%%%%%%%%%%%%%%%%%%%%%%%%%%%%%%%%%%%%%%%%%%%%%%%%%%
%%%%%%%%%%%diagram for g s to b  l l

Also note that in all the benchmark scenarios studied here, the sfermion masses and RPV
couplings are quite well determined to fit all the anomalies. Therefore, we can make concrete predictions for the sfermion production and decay at the LHC. A particularly striking signature in our RPV3 setup would be final states with third-generation fermions, such as $pp\to t\tau^+\tau^-$, via the resonant production of a bottom squark. A detailed analysis of this signal and the corresponding SM backgrounds will be reported in a separate publication. 

Similarly, the RPV3 explanation of the ANITA anomaly can be independently tested at colliders. The key thing to note here is that we require a light, long-lived bino with a rest lifetime of about 10 ns to explain the ANITA anomaly~\cite{Collins:2018jpg} (see Section~\ref{ANITAdetail}). The bino can be produced at the LHC from either gluino or squark decay through gauge interactions, followed by the  3-body decay of bino into two quarks and a lepton (through $LQD$ coupling) or into three leptons (through $LLE$ coupling). For TeV-scale gluinos and squarks, a GeV-scale bino will have a boost factor $\gamma\sim 10^3$ at the LHC and will have a decay length of $\sim 100$ m in the lab frame. This leads to distinct displaced vertex signatures~\cite{Zwane:2015bra, deVries:2015mfw, Dercks:2018eua}, which should be accessible to dedicated long-lived particle searches at the LHC~\cite{Curtin:2018mvb, Alimena:2019zri}.

%\blue{Comment on the comparison of LQ vs RPV explanation.}

%%%%%%%%%%%%%%%%%%%%%%%%%%%%%%%%%%%%%%%%%%%%
\section{Conclusion} \label{sec:conclusions}
%%%%%%%%%%%%%%%%%%%%%%%%%%%%%%%%%%%%%%%%%%%%
Taking the reported $B$-physics anomalies, as well as the muon anomalous magnetic moment and ANITA anomalous upgoing air shower events at face value, we examine the exciting possibility that these anomalies in vastly different systems could actually be connected by a single underlying BSM framework. In particular, we point out that the origin of these anomalies might be related to a third-generation-centric BSM scenario, which could also address the SM Higgs naturalness issue, while preserving all the good features of a generic supersymmetric framework. In a promising minimalist approach, we consider the so-called `RPV3' scenario, wherein only the superpartners of the third-generation SM fermions are relatively light, at (sub)TeV scale, whereas all other sparticles (except the lightest neutralino) are much heavier and do not play a significant role in explaining the anomalies. 

We have considered three benchmark cases for this RPV3 setup and analyzed the reduced parameter space to carve out the regions favored by each of the above mentioned anomalies, while making sure that all relevant experimental constraints are satisfied. We find that some combination of these indication(s) of deviations from the Standard Model can be explained in all three cases, but finding an allowed overlap region between all of them may only be possible in one of the three cases studied here. Nevertheless, it seems remarkable to us that such an overlap region exists at all (see Fig.~\ref{fig:ANITARdRkgm2}), given the stringent experimental constraints from a large number of low and high energy processes on the masses and couplings.  

We have also given a sample of  predictions for various LFV decays of the $\tau$-lepton and of $B$-mesons, which can in principle be used to test the RPV3 hypothesis in the current and upcoming  precision $B$-physics experiments.  Some complementary tests in the high-$p_T$ LHC experiments are also  discussed here. Moreover, improved measurements in the experimental inputs showing the current indications of deviation will likely  have significant consequences for our RPV3 scenario.

\section*{Notes Added} 
\begin{enumerate}
\item While finalizing our paper, we became aware of Ref.~\cite{Hu:2020yvs} which only uses the $\lambda'$-type RPV couplings to simultaneously address $R_{D^{(*)}}$ and $R_{K^{(*)}}$. In our study, we consider both $\lambda$ and $\lambda'$-type couplings to address $R_{D^{(*)}}$ and $R_{K^{(*)}}$, as well as muon $g-2$ and ANITA anomalies, which are not discussed in Ref.~\cite{Hu:2020yvs}.

\item Recently three new lattice calculations of the hadronic vacuum polarization
 contribution to muon $(g-2)$ have appeared~\cite{Aubin:2019usy, 
Borsanyi:2020mff,Lehner:2020crt};  all three use the 
so-called ``staggered" quarks.   One  of these calculations~\cite{Borsanyi:2020mff} by the BMW collaboration claims to have the smallest errors of all lattice calculations 
to date and that its results imply that no new physics is needed  
to explain the BNL experimental result on muon $(g-2)$~\cite{Bennett:2006fi}.  However, as emphasized in Ref.~\cite{Crivellin:2020zul} even if the BMW result is correct, the need for new physics is still there; it just gets shifted from muon $(g-2)$ to the electroweak sector.

\end{enumerate}

%%%%%%%%%%%%%%%%%%%%%%%%%%%%%%%%%%%%%%%%%%%%%%%
\section*{Acknowledgments}
%%%%%%%%%%%%%%%%%%%%%%%%%%%%%%%%%%%%%%%%%%%%%%%

Discussions with Bhubanjyoti Bhattacharya, Angelo Di Canto, Tim Gershon, Hassan Jawahery, Christoph Lehner and Sheldon Stone are gratefully acknowledged. The research of WA is supported by the National Science Foundation under Grant No. PHY-1912719. The work of BD and YS is supported in part by the U.S. Department of Energy under Grant No. DE-SC0017987 and in part by the MCSS funds, as well as by the Neutrino Theory Network Program under Grant No. DE-AC02- 07CH11359. The work of AS was supported in part by the U.S. DOE contract 
\#DE-SC0012704. WA and AS acknowledge support by the Munich Institute for Astro- and Particle Physics (MIAPP) which is funded by the Deutsche Forschungsgemeinschaft (DFG, German Research Foundation) under Germany's Excellence Strategy - EXC-2094 - 390783311. BD and AS would like to thank the organizers of FPCP 2018 at University of Hyderabad for hospitality, where part of this work was done. BD and YS would like to thank the High Energy Theory group at Oklahoma State University for warm hospitality during a summer visit in 2019, where part of this work was done.

\appendix 
\section{\texorpdfstring{$b\to c \tau \widetilde{\chi}_1^0$}{}} \label{app:bino}
\begin{figure}[t!]
	\begin{subfigure}{.23\textwidth}
		\centering
		% include first image
		\includegraphics[width=1\linewidth]{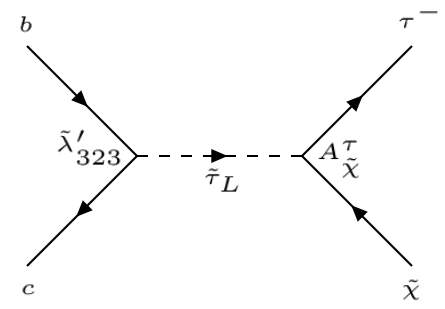}  
		\caption{}
		\label{fig:a}
	\end{subfigure}
	\begin{subfigure}{.23\textwidth}
		\centering
		% include second image
		\includegraphics[width=1\linewidth]{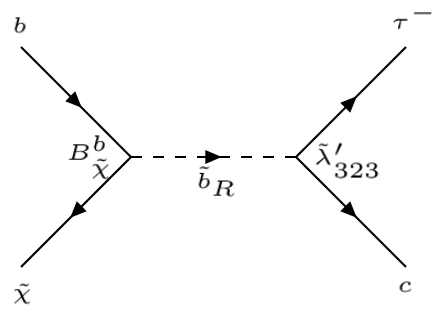}  
		\caption{}
		\label{fig:b}
	\end{subfigure}
	\caption{Contributions to $b\to c\tau {\chi}$ from (a) left-handed stau, and (b) right-handed sbottom.}
	\label{fig:feynman}
\end{figure}

As shown in Fig.~\ref{fig:feynman}, there are two possible diagrams for the $b\to c \tau \widetilde{\chi}_1^0$ decay process. In theory, the vertex with $b\tau\widetilde c$ may also give rise to a third diagram but we only consider the third generation of sparticles to be light, so that diagram will not be considered here.

The effective Hamiltonian for the two diagrams could be written respectively as:
\begin{align}
H_1& \ \simeq \ \frac{1}{m_{\widetilde \tau_L}^2}(-1)\widetilde{\lambda'}_{323}^* (-i(A_{\widetilde \chi}^{\tau})^*) (\bar c_L b_R)  ( \bar \chi L \tau)^{\dagger}\nonumber \\
& \ = \ i\frac{\widetilde{\lambda'}_{323}^* (A_{\widetilde \chi}^{\tau})^*}{m_{\widetilde \tau_L}^2} (\bar c P_R b)  ( \bar \tau P_R  \chi) \, ,\\
H_2& \ \simeq \ \frac{1}{m_{\widetilde b_R}^2}(-1)\widetilde{\lambda'}_{323}^* (iB_{\widetilde \chi}^{b}) (\bar c_L \tau^{c}_L)  ( \bar \chi R b )\nonumber \\
& \ = \ (-i)\frac{\widetilde{\lambda'}_{323}^* B_{\widetilde \chi}^{b}}{m_{\widetilde b_R}^2}  (\bar c_L \tau^{c}_L)  ( \bar \chi P_R b ) \, ,\label{eq:fereq}
\end{align} 
where $\widetilde{\lambda}'_{ijk} = \lambda'_{ilk}V_{jl}$, $A^\tau_{\widetilde\chi}$ is a linear combination of the $SU(2)_L$ and $U(1)_Y$ gauge couplings in the sparticle sector, and $B^b_{\widetilde \chi}$ is related to the $U(1)_Y$ gauge coupling $g'$; for the exact definitions of $A$ and $B$, see Eqs.~(8.88a) and (8.88c) in Ref.~\cite{Baer:2006rs}. 
Via Fierz transformation and using the property of charge-conjugate operator, we could rewrite $H_2$ in a similar fashion as $H_1$:
\begin{align}
H_2& \ \simeq \ (-i)\frac{\widetilde{\lambda'}_{323}^* B_{\widetilde \chi}^{b}}{8m_{\widetilde b_R}^2}  
\left[4(\bar c R b)(\bar \tau R \chi) - (\bar c R \sigma_{\mu\nu} b) (\bar \tau \sigma^{\mu\nu} \chi)
\right] \, .
\end{align}

To compare with the SM contributions, we can rewrite both $H_1$ and $H_2$ into the form of standard Wilson coefficients multiplying the corresponding operators [cf.~Eq.~\eqref{eq:effH}]:
	\begin{align} 
	H_{\rm eff}^{\rm extra}& \ = \ \frac{4 G_F}{\sqrt{2}}V_{cb}[C_{S_R} {\cal O}_{S_R} +  C_{T_R} {\cal O}_{T_R}] \, ,
	\end{align}
with the following Wilson coefficients which are the dominant ones in our RPV3 scenario: 
\begin{align}
C_{S_R} \ = \ &\frac{i\sqrt2 \widetilde{\lambda'}_{323}^*}{4 G_F} \left( \frac{ (A_{\widetilde \chi}^{\tau})^*}{m_{\widetilde \tau_L}^2}  -  \frac{B_{\widetilde \chi}^{b}}{2m_{\widetilde b_R}^2}   \right) \, , \label{eq:CSR} \\
% C_{T_L}=&\frac{\sqrt2}{4 G_F} \left(-\frac{3}{8} \frac{(\widetilde{\lambda'}_{323}^* B_{\widetilde \chi}^{b} (-i))}{m_{\widetilde b_R}^2} \right)\\
C_{T_R} \ = \ &\frac{i\sqrt 2\widetilde{\lambda'}_{323}^*}{4 G_F} \left( \frac{ B_{\widetilde \chi}^{b}}{8m_{\widetilde b_R}^2} \right) \, , \label{eq:CTR}
\end{align}
and the operators defined as:
\begin{align}
{\cal O}_{S_R}& \ = \ (\bar c P_R b)(\bar \tau P_R \widetilde \chi) \, , \\
% O_{T_L}&=(\bar c L \sigma_{\mu\nu} b)(\bar \tau \sigma^{\mu\nu} \chi) \\
O_{T_R}& \ = \ (\bar c P_R \sigma_{\mu\nu} b)(\bar \tau \sigma^{\mu\nu} \widetilde \chi) \, .
\end{align}

With these expressions, we could easily compare the contribution to the decay width from the extra channel and from the SM. We define the following ratio: 
\begin{align}
R^{\rm extra}_{\rm int} \ \equiv \ \frac{\int dq^2 \frac{d\Gamma( B \to D\tau\widetilde\chi)}{d q^2} }{\int dq^2 \frac{d\Gamma( B \to D\tau{{\bar\nu}_{\tau}})_{\rm SM}}{d q^2} } \, ,
\end{align}
where  $\frac{d\Gamma(B \to D\tau{{\bar\nu}_{\tau}})_{\rm SM}}{d q^2}$ and $\frac{d\Gamma(B \to D\tau\widetilde\chi)}{d q^2}$ could be written as~\cite{Murgui:2019czp}:
\begin{align}
&\frac{d\Gamma(B \to D\tau{{\bar\nu}_{\tau}})_{\rm SM}}{d q^2} \ \propto \  \left(1-\frac{m_{\tau}^2}{q^2}\right)^2\nonumber\\
& \quad \times\left[\left(1+\frac{m_{\tau}^2}{2 q^2}\right) \left(H_{V,0}^s\right)^2+ \frac{3}{2}\frac{m_{\tau}^2}{q^2}\left(H_{V,t}^s\right)^2 \right] \, , \label{eq:SMwidth}
\end{align}
and 
\begin{align}
&\frac{d\Gamma(B \to D\tau\widetilde\chi)}{d q^2} \ \propto \   \left(1-\frac{(m_{\tau}+m_{\chi})^2}{q^2}\right)^2 \left[
\frac{3}{2} |C_{S_R}|^2 \left(H_{S}^s\right)^2\right.\nonumber\\
& \qquad \left.
+8 |C_{T_R}|^2 \left(1+\frac{2(m_{\tau}+m_{\chi})^2}{q^2}\right) \left(H_{T}^s\right)^2  
\right] \, , \label{eq:extrawidth}
\end{align}
where $H_{V,0}^s$, $H_{V,t}^s$, $H_{S}^s$ and $H_{T}^s$ are the helicity amplitudes defined in the same way as in Appendix B of Ref.~\cite{Murgui:2019czp}.
Taking the Wilson coefficients from Eqs.~\eqref{eq:CSR} and \eqref{eq:CTR}, masses of the final state particles as in our benchmark cases, helicity amplitudes and form factors for $B$ and $D$ mesons from Refs.~\cite{Sakaki:2013bfa,Murgui:2019czp}, we find that for $m_{\widetilde\chi_1^0}=2$ GeV, $R^{\rm extra}_{\rm int}=0.6\%$, which is insignificant compared to the SM and typical RPV contributions discussed in Section~\ref{sec:IIIA}. 
\section{\texorpdfstring{Bino Mean Free Path}{}} \label{app:anita}

\begin{figure*}[t!]
	\begin{subfigure}{0.48\textwidth}
		\centering
		% include first image
		\includegraphics[width=1.0\textwidth]{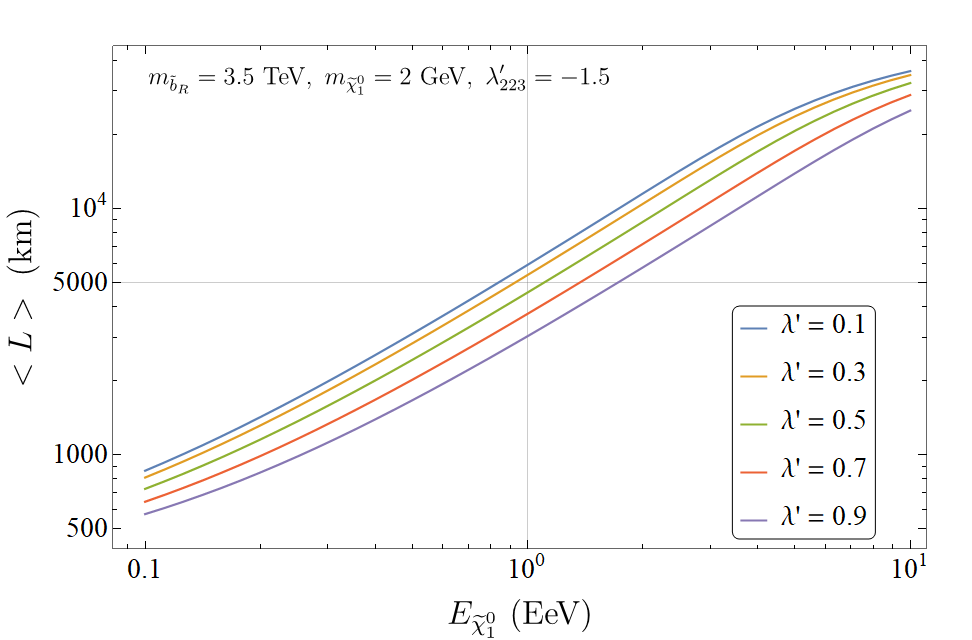}  
		\caption{}
		\label{fig:a}
	\end{subfigure}
	\begin{subfigure}{0.48\textwidth}
		\centering
		% include first image
		\includegraphics[width=1.0\textwidth]{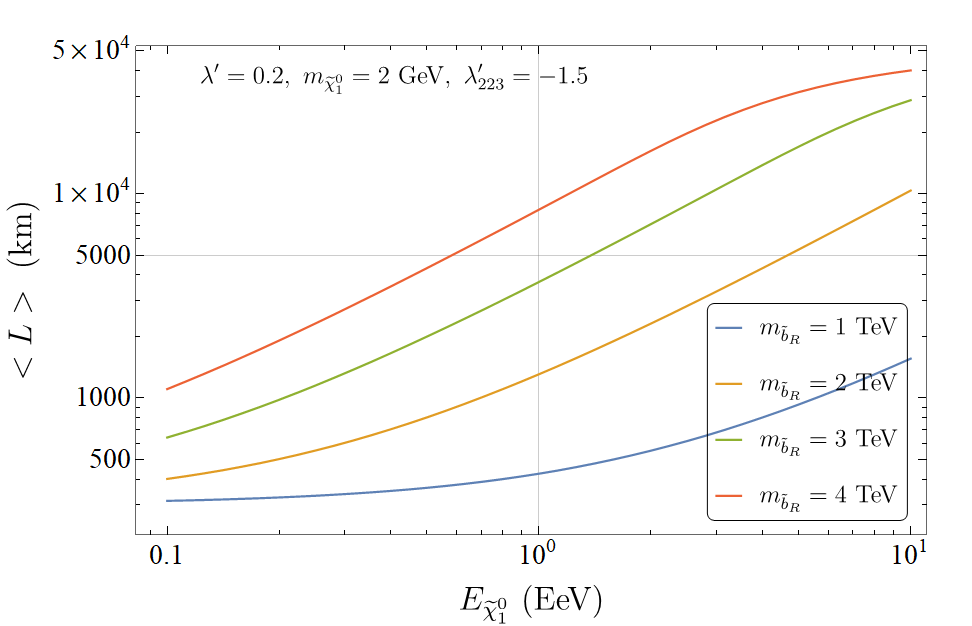}  
		\caption{}
		\label{fig:a}
	\end{subfigure}
	\caption{The bino mean free path in earth as a function of its energy for Case 3 (cf.~Section~\ref{sec:ours}): (a) for a fixed $m_{\widetilde{b}_R}=3.5$ TeV and different values of $\lambda'$, and (b) for a fixed $\lambda'=0.2$ and different values of $m_{\widetilde{b}_R}$. The vertical line is for the bino energy of 1 EeV, while the horizontal line is for its mean free path of 5000 km, which are the ballpark values required to fit the ANITA anomaly.}
	\label{fig:mfp}
\end{figure*}

In Fig.~\ref{fig:mfp}, we show the variation of the bino mean free path $\langle L\rangle$ inside the earth as a function of its energy for Case 3 (cf.~Section~\ref{sec:ours}); the results for Case 1 and 2 are similar. Here we fix $m_{\widetilde \chi_{0}^1} = 2$ GeV and $\lambda'_{223} = -1.5$ [cf.~Eq.~\eqref{eq:bp3}]. In Fig.~\ref{fig:mfp}(a), we show the mean free path for different values of $\lambda'$, keeping the sbottom mass fixed at 3.5 TeV, whereas in Fig.~\ref{fig:mfp}(b), we take different sbottom masses, while keeping $\lambda'$ fixed at 0.2. As we can see from these figures, with bino energy $\sim$ EeV (shown by the vertical line) and for suitable choice of $\lambda'$ and $m_{\widetilde b_R}$, the mean free path can be around $\sim$ 5000 km (shown by the horizontal line), as required to fit the ANITA observation. This calculation is done with the approximation that only the bino decay process matters in the bino propagation. This is valid due to the small bino-nucleon cross section, which gives an effective interaction length of  $\sim 10^9$ m, much larger than its decay length ($\sim 10^6$ m). For more details, including the analytic expression for the mean free path used in this context, see Ref.~\cite{Collins:2018jpg}.

\bibliographystyle{apsrev4-1}
\bibliography{main}
\end{document}